\documentclass{article}
\usepackage{todonotes}
\usepackage{amsmath, amsthm, amsfonts, accents, subfiles, graphicx, subcaption, float}
\usepackage{hyperref}
\usepackage{geometry}
\usepackage{cleveref}

\usepackage[style=numeric,sorting=none]{biblatex}
\numberwithin{equation}{section}
\newcommand{\pleb}[0]{Pleba\'nski}
\newcommand{\urb}[0]{Urbantke}

\newcommand{\so}[0]{\mathfrak{so}}
\newcommand{\E}[0]{\so(3,\mathbb{C})}
\newcommand{\C}[0]{\mathbb{C}}
\newcommand{\R}[0]{\mathbb{R}}
\newcommand{\id}[0]{\mathbb{I}}
\newcommand{\im}{{\rm i}}
\newcommand{\re}{\operatorname{Re}}
\newcommand{\imag}{\operatorname{Im}}

\newcommand{\sd}[0]{\Sigma}
\newcommand{\spasurf}[0]{\mathcal{S}}
\newcommand{\utilde}[1]{\underaccent{\tilde}{#1}}
\newcommand{\asd}[0]{\overline{\Sigma}}
\newcommand{\tsd}[0]{\tilde{\Sigma}}

\newcommand{\teps}[0]{\tilde{\epsilon}}
\newcommand{\uteps}[0]{\utilde{\epsilon}}
\newcommand{\tg}[0]{\tilde{g}}
\newcommand{\utg}[0]{\utilde{g}}
\newcommand{\tE}[0]{\tilde{E}}
\newcommand{\tbE}[0]{\tilde{\bar{E}}}
\newcommand{\utE}[0]{\utilde{E}}
\newcommand{\dlap}[0]{\utilde{N}}

\author{Adam Shaw\footnote{adam.shaw@nottingham.ac.uk}\\\\University of Nottingham, Nottingham, NG7 2RD, UK}

\title{Hyperbolic Chiral Numerical Schemes for General Relativity}
\date{November 2026}

\begin{document}
\maketitle

\begin{abstract}
We develop numerical relativity evolution schemes based on \pleb{}'s chiral formulation of general relativity and its 3+1 form, Ashtekar's variables. At the linear level, we construct a two-parameter family of first-order gauge fixings that dynamically fixes both diffeomorphisms and internal $SO(3,\mathbb{C})$ rotations; after a linear field redefinition, the resulting system splits into independent 12- and 4-component sectors. Lifting this gauge fixing to its minimal nonlinear extension and performing the 3+1 decomposition yields a strongly hyperbolic modification of Ashtekar's evolution equations that remains polynomial in the fields and includes constraint damping. We also introduce a conformal (unimodular) reformulation in which the nonlinear 12-component sector evolves completely independently of the remaining fields. All three systems are tested with gauge-wave convergence tests, robust-stability tests on a noisy Minkowski background, and long-time evolutions. The unmodified Ashtekar system crashes at $T \sim 325$, whereas the gauge-fixed hyperbolic and conformal systems remain stable up to $T = 2000$, with constraint violations and the imaginary part of the metric remaining at the $10^{-10}$ level.
\end{abstract}

\tableofcontents
\newpage

\section{Introduction}

It has been two decades since the first stable evolutions of binary black holes~\cite{Evolution_of_Bi_Pretor_2005,Gravitational_w_Baker_2005}, and numerical relativity is now a mature field~\cite{NumericalRelatBaumga2010,AlcubierreMiguelIntro3+12008}. Accurate simulations nonetheless remain computationally expensive, so schemes that achieve the same or better accuracy at lower cost are highly desirable. One route to such schemes is to start from a different formulation of general relativity (GR). GR admits many equivalent reformulations~\cite{FormGenRelGravity2020}, several of which have been explored in numerical relativity~\cite{Formulations_of_Shinka_2008}, and some may prove better suited to numerical evolution than others. In this paper we focus on chiral descriptions of GR, in particular \pleb{}'s formulation.

In Lorentzian signature, chiral formulations exist only in four dimensions. Their name refers to the use of only one factor in the decomposition of the Lorentz Lie algebra, \cref{eq:lorentz-group-iso}. The basic variables are a triple of complex-valued 2-forms $\sd^i \in \Lambda^2$, which are self-dual, i.e.\ lie in the $+\im$ eigenspace of the Hodge star (the anti-self-dual forms lie in the $-\im$ eigenspace).\footnote{The index $i=1,2,3$ is an $\so(3,\C)$ index; $\so(3,\C)$ is the chiral half of the Lorentz algebra $\so(1,3)$.} When these 2-forms satisfy the metricity constraints, \cref{eq:metricity-cond}, they encode a metric together with a complex frame rotation. The first-order formulation also introduces the self-dual connection $A^i$, defined as the connection compatible with the 2-forms, $d^A \sd^i = 0$; it is the chiral projection of the full spin connection. Its curvature, $F^i = dA^i + \frac{1}{2} \epsilon^{ijk} A^j \wedge A^k$, contains the Ricci tensor, so Einstein's equations become an algebraic condition relating $F^i$ and $\sd^i$. The metric obtained in this way is in general complex, and the reality conditions, \cref{eq:reality-cond}, remove its imaginary part.

Describing geometry in this way has already revealed attractive features of both gravity and Yang-Mills theory~\cite{Kerr_metric_fro_Krasno_2024,Plebanski_compl_Krasno_2025,Kahler_decoupli_Green_2026}. Most relevant here is the pure-connection numerical scheme of~\cite{WeylCurvatureKrasno2022}, developed by the author and collaborators, which explores a first-order formulation in terms of the self-dual connection and the Weyl curvature.

A numerical evolution scheme must satisfy two requirements: the underlying system of partial differential equations (PDEs) must be strongly hyperbolic, and the discretisation must be stable. Strong hyperbolicity is a property of the PDEs themselves; it guarantees a unique solution that depends continuously on the initial data. Stability means that the numerical solution remains bounded in terms of its initial data. For linear systems strong hyperbolicity implies stability, but in general stability must be checked numerically.

Hyperbolicity depends only on the principal part (the highest-derivative terms) of a PDE, which the nonlinear equations share with their linearisation. The linear structure of gravity in our formalism is therefore the natural starting point. Hyperbolicity can be changed only by exploiting additional freedom in the description, such as gauge freedom, or by adding constraint terms. Gauge fixing can address both requirements: it can change the hyperbolicity of the equations and control the constraints, which in turn promotes stability. The linearisation of chiral theories is also interesting in its own right~\cite{Local_rigidity_Fine_2021,ChiralPerturbaKrasno2020,Plebanski_compl_Krasno_2025} and~\cite[sec.~5.9.5]{Krasnov2020-qr}. For instance, a first-order gauge fixing exists that eliminates certain propagators~\cite{ChiralPerturbaKrasno2020} and recasts the linearised theory as an elliptic complex~\cite{Plebanski_compl_Krasno_2025}.

The 3+1 (Hamiltonian) decomposition of \pleb{}'s formulation yields Ashtekar's variables for gravity~\cite{NewVariablesFAshtek1986}. The dynamical fields are the spatial (densitised) triad $\tE^{ia}$ and the spatial connection $A^i_a$,\footnote{Here $a=1,2,3$ is an index on the spatial hypersurface.} while the inverse densitised lapse $\dlap$, the shift $N^a$ and the temporal connection $A^i_0$ act as gauge fields. We call the resulting evolution scheme the original system (\cref{subsec:original-system}). Because of the extra internal symmetry, three Gauss constraints appear alongside the usual Hamiltonian and momentum constraints. All constraints are first-class, which is what allows gauge choices and constraint additions to control constraint growth. A further, potentially useful, property is that all evolution equations and constraints are polynomial in the fields and their derivatives.

Ashtekar variables are central to loop quantum gravity and spin-foam models~\cite{A_Short_Review_Ashtek_2021}. In numerical relativity they were studied in a series of papers by Yoneda and Shinkai~\cite{Constructing_hy_Yoneda_1999,Formulations_of_Shinka_2008,Hyperbolic_form_Shinka_2000,Hyperbolic_form_Yoneda_2001,Will_hyperbolic_Shinka_2001}, who introduced the adjusted-system~\cite{Hyperbolic_form_Yoneda_2001} and $\lambda$-system~\cite{Einstein_s_Equa_Brodbe_1998} modifications to damp constraint growth, working with a non-dynamical gauge. Here we instead consider dynamical gauges, in the spirit of the 1+log, gamma-driver and harmonic gauges, which are known to produce stable long-term simulations~\cite{Gauge_condition_Alcubi_2003,Numerical_Relat_Pretor_2004}; such gauges can also be formulated at the level of the action~\cite{ActionPrinciplBona2011,Action_Principl_Brown_2010,Strongly_Hyperb_Brown_2008}.

The main results of this paper are:

\begin{enumerate}
    \item A two-parameter family of first-order gauge fixings for linearised \pleb{} gravity, giving a 12+4-component split into self-dual/anti-self-dual Maxwell-type systems.
    \item Its minimal nonlinear lift, with a strongly hyperbolic 3+1 system and constraint damping parametrised by four constants $(c_1,c_2,\alpha,\beta)$.
    \item A conformal (unimodular) reformulation in which the 12-component sector evolves independently, parametrised by two constants $(c,\alpha)$.
    \item Numerical robust-stability and long-time tests showing that the gauge-fixed systems survive to $T=2000$, while the original crashes at $T \sim 325$.
\end{enumerate}

First, we extend the first-order gauge fixing of~\cite{ChiralPerturbaKrasno2020} to a two-parameter family. We introduce an internal gauge field $\chi^i \in \C^3$ and two modified linearised connections, $a^i := \delta A^i$ and $\xi_\mu$ (\cref{eq:mod-sd-con-def,eq:mod-first-ord-de-donder}); $\xi_\mu$ plays the role of the Z4 vector $Z_\mu$~\cite{General_covaria_Bona_2003} in controlling constraint growth, and for $\chi^i = 0 = \xi_\mu$ the definitions reduce to the standard linearised self-dual connection and the de Donder gauge on flat space. The linearised Einstein equations then become two first-order equations on these connections, a Z4-like modification of the Ricci-flat condition and a modified Lorenz gauge on $a^i$ (\cref{eq:mod-ricci-flat-cond,eq:mod-lin-lorenz-cond}), which together dynamically fix both diffeomorphisms and $SO(3,\C)$ rotations. Every field then satisfies the wave equation at second order, so the system is strongly hyperbolic.

Moreover, the system separates under the linear transformation in \cref{eq:con_inv_transform,eq:lin-org-frame-to-con}. The equations of motion become two independent systems of size 12 and 4. They contain 12 and 4 components, respectively, of both the frame and the connection. Structurally, they take the same form as four copies of Maxwell's equations in Lorenz gauge: three self-dual and one anti-self-dual. The split also allows for a one-parameter family of Lagrangians that all produce the same gauge-fixed linearised gravity theory; see \cref{eq:lin-two-lagrangians}. This result has been discussed in a different context in~\cite{Plebanski_compl_Krasno_2025}.

Second, we extend the linear gauge fixing to its minimal nonlinear generalisation. Here, minimal means using the fewest nonlinear terms needed to obtain the same linear result upon linearisation. Using the same $\chi^i$ and $\xi_\mu$ to represent the new nonlinear fields, this modification becomes the following two-parameter family,

\begin{gather}
    A^i_\mu = -\im J_1^{-1}(\star d \sd)^i_\mu + c_1 \partial_\mu \chi^i + c_2 \epsilon^{ijk} \sd^j_\mu{}^\nu \partial_\nu \chi^k, \quad \xi_\mu = - \Gamma_\mu - 2 c_2 \sd^i_\mu{}^\nu \partial_\nu \chi^i, \\
    \sd^i_\mu{}^\rho F^i_{\nu\rho} + \partial_{(\mu} \xi_{\nu)} = 0, \quad \partial^\mu A^i_\mu - \frac{1}{2} \sd^{i\mu\nu} \partial_\mu \xi_\nu = 0.
\end{gather}

The constants $c_1,c_2$ are arbitrary complex numbers, subject to $c_2 \neq 0$. The equations are not fully covariant because of the presence of partial derivatives and connections. Since the gauge has been fixed, the above system is invariant under neither diffeomorphisms nor $SO(3,\C)$ frame rotations.

We then perform the 3+1 decomposition of these equations to obtain the hyperbolic evolution system of \cref{subsec:hyperbolic-3+1-system}, adding damping terms controlled by two further complex constants, $\alpha$ and $\beta$. Compared with the original Ashtekar system, the hyperbolic system differs only by terms that vanish on the constraint surface: the physical dynamics are unchanged, but the constraint surface becomes an attractor.

The third, and most surprising result, is the analogue of the 12+4 sector split for this nonlinear system. This is a miraculous property of gravity in four dimensions when using the chiral formulations. After appropriate gauge fixing, Einstein's equations separates into two sectors. The 12-component sector can be solved completely independently of the other. To do so we introduce a unimodular 2-form, $H^i = (-\det(g))^{-\frac{1}{4}} \sd^i$, where unimodular means that $\det(g_H) = -1$, with $g_H$ the metric constructed from $H^i$. The unimodular requirement removes one component from $H^i$, so that it contains $13-1=12$ components. A new set of connections is introduced using the nonlinear version of the linear connection transformation,

\begin{align}
    \Omega^i_\mu &= A^i_\mu - \frac{1}{2} \sd^i_\mu{}^\nu \xi_\nu, \\
    \omega_\mu &= \xi_\mu + 2 \sd^i_\mu{}^\nu A^i_\nu.
\end{align}

In the nonlinear case the separation not as strong as the linear. The 12 components $(H^i,\Omega^i)$ are evolved using initial data for themselves alone, whereas the evolution of $(n,\chi^i)$ requires their initial conditions and knowledge of the 12 sector for the desired integration time. The 3+1 decomposition also suggests new variables, $H^{ia} = \dlap \tE^{ia}$ and $n = \log(\dlap)$, built algebraically from Ashtekar's variables, in which the evolution equations simplify. Together with the connections $\Omega^i_a,\Omega^i_0,\omega_a,\omega_0$ and the unchanged $N^a$ and $\chi^i$, these form the conformal system of \cref{subsec:gauge-fixed-3p1-systems:conformal-system}, which depends on one gauge constant $c$ and one damping constant $\alpha$. Because the conformal triad $H^{ia}$ must be inverted, this system is no longer polynomial. That this separation occurs at all is a consequence of our choice of formulation, no other known description has this property. It is a property that has only been shown in the chiral formulations making it an interesting description to explore.

Finally, we test the systems numerically. Robust stability is assessed by adding random noise to the Minkowski solution and monitoring the constraints, including the reality conditions, over time. After choosing suitable parameters, we also run long-time evolutions. By tuning the damping and gauge constants $\alpha,\beta,c_1,c_2$ and $c$, we obtain systems that control constraint growth far better than the original Ashtekar system.

\subsection{Wave Equation Analogy}

The difference between gauge-fixed ADM-type systems and the Ashtekar-based systems developed here is captured by a simple analogy: the two standard ways of reducing the wave equation to first order. In the first, only some of the first-order variables satisfy the wave equation; in the second, a Dirac-type reduction, all of them do. Our gauge-fixed systems are of the second type.

In four dimensions the Minkowski wave equation is

\begin{align}
        \partial^2_t u = \partial^i \partial^i u,
\end{align}

where $u(t,x^i)$ is a real function on Minkowski space. The solution $u$ represents an Einstein metric in harmonic gauge. Both the ADM and Ashtekar systems recast Einstein's equations as an evolution system that is first order in time. For the wave equation there are two ways of doing this. The first is to introduce a first-order variable for the time and spatial derivatives of $u$,

\begin{align}
    v = \partial_t u, \quad a^i = \partial^i u.
\end{align}

Then the wave equation becomes the following first-order system,

\begin{align}
    \partial_t u = v, \quad \partial_t v = \partial^i a^i, \quad \partial_t a^i = \partial^i v,
\end{align}

along with the constraint $C^i = a^i - \partial^i u$. This is the most common way of producing a first-order system from a second-order one. As presented here, the system is symmetric hyperbolic. This is analogous to the ADM decomposition of Einstein's equations, with $v$ playing the role of the extrinsic curvature and $a^i$ that of the Levi-Civita connection.
The second way of producing a first-order system is to use spinors and gamma matrices. The gamma matrices in four dimensions are a set of four matrices $\gamma^\mu$ such that $\gamma^{(\mu} \gamma^{\nu)} = g^{\mu\nu} \mathbb{I}$. More formally, they generate the matrix representation of the Clifford algebra $Cl_{1,3}(\R)$. The smallest real representation is the $4\times 4$ Majorana representation, for which we can take the solution vector, now denoted $U$, to be a real 4-vector. The first-order reduction of the wave equation is then

\begin{align}
    \gamma^\mu \partial_\mu U = 0.
\end{align}

To make the time evolution explicit, we split $\gamma^{\mu} = (\gamma^0, \gamma^i)$ to obtain

\begin{align}
    \partial_t U = \hat\gamma^i \partial^i U, \quad \hat\gamma^i = -(\gamma^0)^{-1} \gamma^i.
\end{align}

This system is strongly hyperbolic, since the gamma matrices are diagonalisable in any representation and their eigenvalues are all real. This spinorial wave equation is analogous to Ashtekar's formalism, with $U$ representing the collection $(\tE^{ia}, A^i_a)$. In the ADM-like reduction, only $u$ and $v$ satisfy a second-order wave equation, whereas the $a^i$ do not. In contrast, all components of $U$ satisfy the wave equation at second order. Unlike the scalar wave equation, gravity cannot be written using real spinors and one must use complex-valued objects.

\subsection{Organisation}

The paper is organised as follows. \Cref{sec:chiral-gr} reviews the chiral formulation of gravity used throughout this paper. Hamiltonian field theories and their constraint analysis are covered in \cref{sec:constraint-hyperbolicity}, and \cref{sec:ashtekar-formalism} gives an overview of Ashtekar's formalism from the numerical relativity perspective. The main work begins in \cref{sec:linearised-gauge-fixing}, which presents the linearised gauge fixing, its separation into two sectors and its Lagrangian interpretation. \Cref{sec:gauge-fixing} contains the nonlinear extension of the linearised gauge fixing. In \cref{sec:gauge-fixed-3+1-evolution-systems} we present the original, hyperbolic and conformal evolution systems, including damping terms, derived from the 3+1 decomposition of the nonlinear gauge fixings. Their hyperbolicity and constraint evolution are analysed in \cref{sec:hyp-and-con-anal-of-evo-sys}. \Cref{sec:numerical-results} describes the numerical set-up and presents the simulation results, which are discussed in \cref{sec:discussion}. The appendices contain a detailed 3+1 decomposition of the \pleb{} fields (\cref{apx:3+1-decomp-of-chiral-objects}) and of the gauge-fixed equations of motion (\cref{apx:3+1-eom-split}), the analysis of the eigenvalues of the constraint matrix (\cref{apx:eigenvalue-analysis}) and further numerical plots (\cref{apx:numerical-plots}).

\section{Chiral Formulations of Gravity}\label{sec:chiral-gr}

Chiral formulations stem from the fact that in four dimensions the Lorentz Lie algebra is not simple: it splits as

    \begin{align}\label{eq:lorentz-group-iso}
        \so(1,3) = \so(3,\C)  \oplus \overline{\so(3,\C)}.
    \end{align}

General relativity can be described using only one of these factors, and chiral refers to this projection. \pleb{}'s formulation exploits this property. Its basic fields are a triple of self-dual 2-forms $\sd^i \in \Lambda^2(M)$, where $i = 1,2,3$ is an internal $\so(3,\C)$ index; since the internal metric is $\delta^{ij}$, the position of this index carries no meaning. For Lorentzian metrics the 2-forms are necessarily complex, because the Hodge star acting on 2-forms has eigenvalues $\pm \im$. The action is

    \begin{align}\label{eq:plebanski-action}
        S[\sd^i,A^i] = \int \sd^i \wedge F^i, 
    \end{align}

    where $F^i = dA^i + \frac{1}{2} \epsilon^{ijk} A^j \wedge A^k$ is the curvature of the self-dual connection 1-form $A^i \in \Lambda^1(M)$. Additional constraints must be imposed on the fields to ensure that the above action describes a real-valued Einstein metric: the metricity and reality conditions,

    \begin{gather}
        \sd^i \wedge \sd^j \sim \delta^{ij}, \label{eq:metricity-cond} \\
        \re(\sd^i \wedge \sd^i) = 0, \qquad \sd^i \wedge \overline{\sd}^j = 0. \label{eq:reality-cond}
    \end{gather}

    Here the anti-self-dual 2-forms are denoted by $\overline{\sd}^i$ and are related to the self-dual ones through complex conjugation. The conditions \cref{eq:metricity-cond} are called the metricity conditions; they remove the components of the 2-forms that do not encode a metric and an $SO(3,\C)$ rotation. The metric can then be recovered using the \urb{} formula~\cite{OnIntegrabilitUrbant1984},

    \begin{align}\label{eq:urbantke-metric}
        g(u,v) v_\sd = \frac{\im}{6} \epsilon^{ijk} \iota_u \sd^i \wedge \iota_v \sd^j \wedge \sd^k, \quad u,v \in TM.
    \end{align}

    Here $v_\sd = \sd^i \wedge \sd^i \in \Lambda^4$ is a top form. This is the physical metric that is used to raise and lower the spacetime indices. An $SO(3,\C)$ transformation of the 2-forms $\sd^i$ leaves this metric invariant; the other factor, $\overline{SO(3,\C)}$, does not act on $\sd^i$ and therefore also leaves the metric invariant. The $\sd^i$ are then automatically self-dual with respect to the Hodge star operator constructed from \cref{eq:urbantke-metric},

    \begin{align}
        {(\star \sd^i)}_{\mu\nu} = \frac{1}{2} \epsilon_{\mu\nu}{}^{\rho\sigma} \sd^i_{\rho\sigma} = \im \sd^i_{\mu\nu},
    \end{align}

    where $\star : \Lambda^2 \rightarrow \Lambda^2$ is the Hodge star and $\epsilon^{\mu\nu\rho\sigma} = \pm \frac{1}{\sqrt{-g}}$ is the totally antisymmetric tensor. Furthermore, satisfying \cref{eq:reality-cond} implies that the metric, \cref{eq:urbantke-metric}, is real-valued. The reality conditions are only required when working with chiral theories in Lorentzian signature, since the self-dual 2-forms are naturally complex-valued. The 2-forms that satisfy the metricity and reality conditions also satisfy the following algebraic identities:

    \begin{equation} \label{eq:algebraic-identities}
        \begin{aligned}
            \sd^i_\mu{}^\rho \sd^j_\rho{}^\nu & = -\delta_\mu^\nu \delta^{ij} + \epsilon^{ijk} \sd^k_\mu{}^\nu, \\
            \sd^i_{\mu\nu} \sd^i_{\rho\sigma} & = g_{\mu\rho} g_{\nu\sigma} - g_{\mu\sigma} g_{\nu\rho} - \im \epsilon_{\mu\nu\rho\sigma}, \\
            \epsilon^{ijk} \sd^j_{\mu\nu} \sd^k_{\rho\sigma} & = 2 \sd^i_{\rho[\mu} g_{\nu]\sigma} - 2 \sd^i_{\sigma[\mu} g_{\nu]\rho}, \\
            \epsilon^{\mu\nu\rho\sigma} \sd^i_{\alpha\beta} & = 12 \im \delta^{[\mu}_\alpha \delta^\nu_\beta \sd^{i\rho\sigma]}.
        \end{aligned}
    \end{equation}

    These will be used throughout the paper, often without mention. Varying the action with respect to $A^i$ gives

    \begin{align}
        \frac{\delta S}{\delta A^i} = d \sd^i + \epsilon^{ijk} A^j \wedge \sd^k = 0. \label{eq:torsion-free}
    \end{align}

    This associates $A^i$ with the self-dual projection of the Levi-Civita connection. Varying with respect to the self-dual 2-forms, while respecting the constraints, results in the following equation of motion,

    \begin{align}
        \frac{\delta S}{\delta \sd^i} = 0 \qquad \Rightarrow \qquad \sd^i_{\mu}{}^\rho F^i_{\nu\rho} = 0. \label{eq:ricci-flat-cond}
    \end{align}

    When \cref{eq:torsion-free} is satisfied, $F^i$ is the self-dual projection of the Riemann curvature tensor~\cite{Su2StructureBhoja2024}. It contains the self-dual half of the Weyl tensor and all the components of the Ricci tensor. The above condition is then equivalent to the vanishing of the Ricci tensor. The Ricci-flat condition can also be written in a way that does not involve the metric,

    \begin{align}
        \sqrt{-g} R^\mu{}_\nu = \teps^{\mu\rho\sigma\alpha} \sd^i_{\rho\sigma} F^i_{\alpha \nu} = 0, \label{eq:ricci-no-metric}
    \end{align}
    where $\teps^{\mu\nu\rho\sigma} = \pm 1$ is the totally antisymmetric symbol. Since Ricci-flat metrics are vacuum Einstein metrics with $\Lambda=0$, we find that the critical points of the action \cref{eq:plebanski-action} are Einstein spacetimes. Modifications that include a cosmological constant and matter are possible and are described in~\cite{Krasnov2020-qr}, but they are not considered here since we are interested in vacuum spacetimes.
    
\section{Constraint Analysis and Hyperbolicity}\label{sec:constraint-hyperbolicity}
To simulate equations derived from a gauge theory, the gauge must be fixed so that the dynamics are uniquely determined. Here we consider gauge fixing that adds terms to the Lagrangian or modifies the equations of motion directly. This section describes the procedure in a general Hamiltonian setting and then explains how to assess the propagation of constraints, following~\cite{Diagonalizabili_Yoneda_2002}.

\subsection{Hamiltonian Analysis}\label{subsec:hamiltonian-analysis}
A useful example of a gauge fixing procedure, in the metric formulation, is the Z4 formulation~\cite{General_covaria_Bona_2003}. In essence, the choice of gauge enlarges the phase space to include extra fields that control the evolution of the constraints. This type of gauge fixing can be understood within the Hamiltonian framework. We analyse its effect on a generic Hamiltonian system with the same properties as GR.\ To do so, we introduce the generalised coordinates $q^I, q^A$ and corresponding momenta $p_I, p_A$.\footnote{The indices $I,J,K,\ldots$ denote dynamical fields and $A,B,C,\ldots$ denote Lagrange multipliers.} Here we assume that the coordinates $q^A$ appear as Lagrange multipliers enforcing some constraints, and as such their momenta $p_A$ do not appear. The Hamiltonian is therefore a function of only

\begin{align}
	\mathbf{H} = \mathbf{H}(q^I,q^A,p_I).
\end{align}

From the general theory, the conjugate momenta $p_A$ form the primary constraints. The surface in phase space on which the constraints vanish is called the constraint surface. On the constraint surface we have $p_A = 0$. By computing the time derivative of the primary constraints, we obtain the secondary constraints,

\begin{align}
	\partial_t p_A = [p_A,\mathbf{H}] = - \frac{\partial \mathbf{H}}{\partial q^A} := \phi_A,
\end{align}

where $\phi_A$ are the secondary constraints, which are also required to vanish on the constraint surface, i.e.\ $\phi_A = 0$ and $p_A = 0$ now define the constraint surface. The brackets $[\cdot,\cdot]$ are the Poisson brackets. We further assume, as is true for GR, that the secondary constraints are first-class. This means that their time derivatives are again proportional to the constraints,
\begin{align}
	\partial_t \phi_A = [\phi_A,\mathbf{H}] = f_A{}^B \phi_B + \tilde{f}_A{}^B p_B.
\end{align}

If the constraints vanish initially, their time derivatives vanish too, so analytically they remain zero for all time. Numerically, however, errors push the solution off the constraint surface and the constraints can then grow. A second issue is that the theory gives no prescription for the evolution of the Lagrange multipliers $q^A$, which are arbitrary. We therefore consider gauge fixings that make the Lagrange multipliers dynamical and that can also improve the behaviour of solutions near the constraint surface. A general construction adds terms proportional to $p_A$ to the Hamiltonian,

\begin{align}
	\tilde{\mathbf{H}} = \mathbf{H} + \lambda^A p_A,
\end{align}

for arbitrary functions $\lambda^A$ of all the phase-space variables. To see that this change does not affect the dynamics, we write the equations of motion for all the coordinates and momenta,

\begin{align}
	\partial_t q^I &= [q^I,\mathbf{H}]+[q^I,\lambda^A] p_A,\quad \partial_t q^A = \lambda^A + [q^A, \lambda^B] p_B, \\ \partial_t p_I &= [p_I,\mathbf{H}] + [p_I,\lambda^A] p_A, \quad \partial_t p_A = \phi_A + [p_A,\lambda^B] p_B.
\end{align}

The evolution equations for $q^I, p_I$ and $p_A$ are thus modified only by the addition of terms proportional to the constraints. As for $q^A$, since these are Lagrange multipliers, their dynamics are irrelevant to the physical system and can be chosen freely. Therefore, on the constraint surface, this system evolves identically to the original system. To ensure that the system also behaves well numerically, we must require that the constraint surface be stable under perturbations. To do so, we compute the equations of motion for the secondary constraints,

\begin{align}
	\partial_t \phi_A = [\phi_A,\mathbf{H}] + \lambda^B [\phi_A, p_B] + [\phi_A,\lambda^B] p_B.
\end{align}

We can now use the first-class assumption to deduce that $[\phi_A, p_B] = D_{A B}{}^C \phi_C + \tilde{D}_{AB}{}^{C} p_C$. Therefore, the evolution of the secondary constraints becomes

\begin{align}
    \partial_t \phi_A = E_A{}^B \phi_B + F_A{}^B p_B
\end{align}

for some functions $E_A{}^B$ and $F_A{}^B$ of the phase-space variables. By defining $[p_A,\lambda^B] := G_A{}^B$, we can write the evolution of the full constraint system in closed form,

\begin{align} \label{eq:hamiltonian-constraint-evolution-gauge-fixed}
	\partial_t \binom{p_A}{\phi_A} = \begin{pmatrix} G_A{}^B & \delta_A^B \\ F_A{}^B & E_A{}^B \end{pmatrix} \binom{p_B}{\phi_B} = \mathcal{C}_A{}^B \binom{p_B}{\phi_B}.
\end{align}

The functions $G,F,E$ are now fixed by a choice of $\lambda^A$ and the original system, and we call the matrix $\mathcal{C}$ the amplification matrix. In gauge theories the number of symmetries is equal to the number of Lagrange multipliers present; therefore $A=1,\ldots,N$, where $N$ is the total number of symmetries. Choosing $N$ functions $\lambda^A$ then fixes all the gauge degrees of freedom; we call this first-order gauge fixing. We wish for deviations from the constraint surface to be attracted back to the constraint surface during evolution. Therefore, we want to analyse the linearised behaviour of \cref{eq:hamiltonian-constraint-evolution-gauge-fixed}. First, one linearises around the constraint surface, treating the constraints $\phi_A$ and $p_A$ as independent variables. This amounts to evaluating $\mathcal{C}$ on a static background. Next, we move to the spatial frequency domain with a Fourier transformation,
\begin{align}
	\phi_A(t,x) = \int d^3 k e^{\im k_a x^a} \hat{\phi}_A(t,k), \quad p_A(t,x) = \int d^3 k e^{\im k_a x^a} \hat{p}_A(t,k).
\end{align}
The evolution equation for these variables is then a simple matrix product involving no derivatives. To ensure that the constraints are well-behaved, we require that the eigenvalues of the Fourier-transformed amplification matrix have non-positive real parts. A positive real part of an eigenvalue corresponds to exponential growth of the constraints, whereas a zero or negative real part implies conservation or damping, respectively. The imaginary part of the eigenvalue can be arbitrary; however, following the analysis in~\cite{Diagonalizabili_Yoneda_2002}, a non-zero imaginary part is preferred when the real part is zero. The functions $\lambda^A$ should therefore be chosen to respect these guidelines.

The $\lambda^A$ are not uniquely determined by this procedure. Furthermore, one can add arbitrary functions of the constraints to all the evolution equations without changing the effective equations of motion; this can be done in a way that is not equivalent to a gauge fixing. In general, a modification involving both methods should be considered in order to obtain the desired properties.

The adjusted-system and $\lambda$-system modifications~\cite{Hyperbolic_form_Yoneda_2001,Einstein_s_Equa_Brodbe_1998} are of this second type: they differ from gauge fixing but achieve a similar effect, and still require a gauge to be chosen before the system can be evolved. We therefore start from a gauge fixing, so that a single construction addresses both requirements. We also relax the requirement that the gauge fixing derive from an action or Hamiltonian, and allow direct modifications of the equations of motion that share the same first-order structure.

\subsection{Hyperbolicity}
For modern numerical techniques the underlying PDEs should be well-posed, meaning that, given some initial data, the solutions at a later time are uniquely determined and depend continuously on the initial data. For generic PDEs this is difficult to prove; therefore, we restrict our attention to first-order equations of motion. That is, given a solution vector $u = (q^I, q^A, p_I, p_A)$, the equations of motion can be written as
\begin{align} \label{eq:first-order-evolution-system}
	\partial_t u = {M(u,t,x)}^a \partial_a u + S(u,t,x),
\end{align}
where $M$ is a matrix of functions that does not contain any spatial derivatives of $u$, and $S$ is the source vector, which similarly contains no derivatives. To check the well-posedness of this system, we can instead consider a related property known as hyperbolicity. We construct the characteristic matrix $P = M^a n_a$, where $n_a$ is an arbitrary real unit vector. The different levels of hyperbolicity are then
\begin{enumerate}
	\item Weakly Hyperbolic: Eigenvalues of $P$ are all real,
	\item Strongly Hyperbolic: $P$ is diagonalisable with all real eigenvalues,
	\item Symmetric Hyperbolic: $P$ is Hermitian.
\end{enumerate}
The three levels are defined such that $\textrm{Symmetric} \Rightarrow \textrm{Strongly} \Rightarrow \textrm{Weakly}$. If the system is at least strongly hyperbolic, it is well-posed (at least locally). The general discussion of gauge fixing above showed that the equations of motion are modified by a choice of $\lambda^A$; as such, the hyperbolicity of the system can change. We can further restrict the choice of $\lambda^A$ so that the constraints are preserved and the system remains strongly hyperbolic. While this reduces the available gauge choices, it still does not produce a unique choice. In general, one can still add terms proportional to the constraints to the evolution equations, which will affect the hyperbolicity and constraint amplification.

In practice we combine Hamiltonian-type gauge fixing with the addition of constraint terms. Since a Hamiltonian cannot always be constructed for a given modification, we work directly with the equations of motion, and regard the modified system as equivalent to the original provided the constraints remain controlled during evolution. This can ultimately be confirmed only numerically, but the amplification-matrix analysis supplies a necessary condition. The rest of the paper applies these ideas to \pleb{}'s chiral formulation.

\section{Ashtekar's Formalism}\label{sec:ashtekar-formalism}
    Ashtekar's formalism~\cite{NewVariablesFAshtek1986} introduced a new canonical pair for gravity. The position variables are densitised inverse triads, $\tE^{ia} =  \det(e) e^{ia}$ (this form of the triad applies only for certain gauge choices), where $e^I_\mu$ is the tetrad and $e^{ia}$ the inverse of its spatial part. Adding the purely imaginary Holst term to the action turns the conjugate momenta into a complex combination of the spin connection and the extrinsic curvature, $A^i_a = -\frac{1}{2}\epsilon^{ijk} \omega^{jk}_a + \im e^{ib} K_{ab}$. In these variables the Hamiltonian is polynomial and, unlike in the tetrad formalism, there are no second-class constraints. Both features are attractive for numerics: the constraints obey a closed system of evolution equations, which, as discussed in \cref{sec:constraint-hyperbolicity}, allows the equations of motion to be modified so that the constraint surface is attractive. Below we show how the \pleb{} action becomes polynomial in these variables and write down the equations of motion and constraints that define the theory, starting from the 3+1 decomposition of its fields.

    \subsection{3+1 Decomposition}\label{subsec:ashtekar:3+1-decomposition}
    We assume that the manifold is globally hyperbolic, so that it admits a foliation $M = \R \times \spasurf$, where $\spasurf$ is a 3-dimensional spatial hypersurface; any 3+1 decomposition splits the tangent space with respect to this foliation. For us the fundamental fields are the self-dual 2-forms $\sd^i \in \Lambda^+$ and the self-dual connection $A^i \in \Lambda^1$. However, instead of computing their decomposition directly, we compute it for the bivectors $\tE^{i\mu\nu}$ as these appear as conjugate fields to the connections. More details on this process can be found in \cref{apx:3+1-decomp-of-chiral-objects}; here we summarise the splitting.

    We use coordinates $x^\mu = (t,x^a)$, with $t \in \R$ and $x^a \in \spasurf$. In these coordinates the bivector splits into
    \begin{align}\label{eq:3+1-decomp-E}
        \tE^{i0a} = \tE^{ia}, \quad \tE^{iab} = -\im \dlap \epsilon^{ijk} \tE^{ja} \tE^{kb} - N^a \tE^{ib} + N^b \tE^{ia}.
    \end{align}
    The metric is given by
    \begin{align}
        \tg^{00} = -\frac{1}{\dlap}, \quad \tg^{0a} = \frac{N^a}{\dlap}, \quad \tg^{ab} = \dlap \tE^{ia} \tE^{ib} - \frac{N^a N^b}{\dlap}.
    \end{align}
    We see how the 2-forms are used to construct the metric. Indeed, all components of the metric can be written in terms of the 2-forms and the inverse triad,
    \begin{align}
        \dlap = \frac{\im}{6} \epsilon^{ijk} \utE^i_a \utE^j_b \tE^{kab}, \quad N^a = -\frac{1}{2} \utE^i_b \tE^{iab}.
    \end{align}
    Here $\utE^i_a$ is the inverse of $\tE^{ia}$. The decomposition of the inverse of $\tg^{\mu\nu}$ is
    \begin{align}
        \utg_{00} = - \dlap + \utg_{ab} N^a N^b, \quad \utg_{0a} = \utg_{ab} N^b, \quad \utg_{ab} = \frac{1}{\dlap} \utE^i_a \utE^i_b.
    \end{align}
    Combining this result with the bivector decomposition, we find the splitting of $\sd^i_\mu{}^\nu$,
    \begin{equation}\label{eq:ashtekar:trip-endo-3+1}
    \begin{gathered}
        \sd^i_0{}^0 = \im \utE^i_a \frac{N^a}{\dlap}, \quad \sd^i_a{}^0 = \frac{\im}{\dlap} \utE^i_a, \quad \sd^i_0{}^a = \im \dlap \tE^{ia} - \im \utE^i_b \frac{N^aN^b}{\dlap} - \epsilon^{ijk} \utE^j_b N^b \tE^{ka}, \\ \sd^i_a{}^b = -\im \utE^i_a \frac{N^a}{\dlap} - \epsilon^{ijk} \utE^j_a \tE^{kb}.
    \end{gathered}
    \end{equation}
    The decomposition of the connections is much simpler. For $A^i$ and $\xi$ we have
    \begin{align}
        A^i_\mu dx^\mu = A^i_0 dt + A^i_a dx^a, \quad \xi_\mu dx^\mu = \xi_0 dt + \xi_a dx^a.
    \end{align}
    The curvature of the self-dual connection $A^i$ becomes
    \begin{align}
        F^i_{0a} = \partial_t A^i_a - \partial_a^A A^i_0, \quad F^i_{ab} = \partial_a A^i_b - \partial_b A^i_a + \epsilon^{ijk} A^j_a A^k_b.
    \end{align}
    We call $A^i_0$ the temporal connection and $A^i_a$ the spatial connection. The above provides a full dictionary for the decomposition of \pleb{}'s formulation, which we now apply to the Lagrangian.

    \subsection{Ashtekar Lagrangian}
    We now write the \pleb{} Lagrangian in terms of the new fundamental variables, $\tE^{ia}$, $N^a$, $\dlap$, $A^i_0$ and $A^i_a$. Starting from the \pleb{} Lagrangian and moving to components, we find
    \begin{align}
        L = \Sigma^i \wedge F^i /d^4x = \frac{1}{4} \teps^{\mu\nu\rho\sigma} \Sigma^i_{\mu\nu} F^i_{\rho\sigma} = \frac{1}{2} \tE^{i\mu\nu} F^i_{\mu\nu}.
    \end{align}
    Applying the 3+1 decomposition to this action, we find
    \begin{align}
        L = \tE^{ia} \left( \partial_t A^i_a - \partial^A_a A^i_0 \right) -\frac{\im}{2} \dlap \epsilon^{ijk} \tE^{ja} \tE^{kb} F^i_{ab} + \tE^{ia} N^b F^i_{ab}.
    \end{align}
    This is the Lagrangian for gravity in Ashtekar variables. Clearly $\tE^{ia}$ is conjugate to $A^i_a$, and since time derivatives appear only in this pairing, these are the only dynamical variables. The variation with respect to these two fields gives
    \begin{align}
        \frac{\delta L}{\delta \tE^{ia}} &= \partial_t A^i_a - \partial^A_a A^i_0 - \im \dlap \epsilon^{ijk} \tE^{jb} F^k_{ab} + N^b F^i_{ab} = 0, \\
        \frac{\delta L}{\delta A^i_a} &= -\partial_t \tE^{ia} - \epsilon^{ijk} A^j_0 \tE^{ka} - \im \epsilon^{ijk} \partial_b^A \left( \dlap \tE^{ja} \tE^{kb} \right) + 2 \partial_b^A \left( \tE^{i[a} N^{b]} \right) = 0.
    \end{align}
    Here we have used the variational formula $\delta F^i_{ab} = 2\partial^A_{[a} \delta A^i_{b]}$. These variations give two evolution equations, for the triad and the spatial connection,

    \begin{align}\label{eq:orginal-pleb-system}
        \partial_t A^i_a  &= \partial^A_a A^i_0 + \im \dlap \epsilon^{ijk} \tE^{jb} F^k_{ab} + N^b F^i_{ba}, \\
        \partial_t \tE^{ia} &= - \epsilon^{ijk} A^j_0 \tE^{ka} - \im \epsilon^{ijk} \partial_b^A \left( \dlap \tE^{ja} \tE^{kb} \right) + 2 \partial_b^A \left( \tE^{i[a} N^{b]} \right). \label{eq:num-ashtekar-triad-evolution-equation}
    \end{align}

    The remaining fields act as Lagrange multipliers and impose the following constraints,
    \begin{align}
        \frac{\delta L}{\delta \dlap} & = -\frac{\im}{2} \epsilon^{ijk} \tE^{ja} \tE^{kb} F^i_{ab} = 0, \\
        \frac{\delta L}{\delta N^a} & = \tE^{ib} F^i_{ba} = 0, \\
        \frac{\delta L}{\delta A^i_0} & = \partial_a^A \tE^{ia} = 0.
    \end{align}
    Variation with respect to the lapse and shift gives the Hamiltonian and momentum constraints,
    \begin{align}
        \mathcal{H} & = \epsilon^{ijk} \tE^{ja} \tE^{kb} F^i_{ab} = 0, \\
        \mathcal{M}_a & = \tE^{ib} F^i_{ba} = 0.
    \end{align}
    These generate the time and space diffeomorphism gauge transformations. The Ashtekar formalism comes with three extra constraints, known as the Gauss constraints,
    \begin{align}
        \mathcal{G}^i = \partial^A_a \tE^{ia}.
    \end{align}
    They generate the $SO(3,\C)$ gauge transformations. Since this theory has an extra $SO(3,\C)$ symmetry compared with the metric formulation, the existence of these additional constraints is expected. It can be shown that the constraints are first-class and therefore form a closed system under the time derivative~\cite{Hamiltonian_Ana_Buffen_2004}. Moreover, every equation of motion and constraint is polynomial in the fields and their derivatives, with degree at most four in the equations of motion (five in the Lagrangian), which is remarkably simple for GR. This can matter numerically. For some initial data, fields in standard formulations pass through degenerate points at which a division by zero occurs, causing the equations to diverge and the code to crash~\cite[sec.~6.6]{AlcubierreMiguelIntro3+12008}. No divisions appear in Ashtekar's equations, so this failure mode is absent.

    The Ashtekar system is only weakly hyperbolic~\cite{Constructing_hy_Yoneda_1999}, but adding constraint terms makes it symmetric hyperbolic~\cite{Einstein_s_equa_Iriond_1997}. Shinkai and Yoneda~\cite{Constructing_hy_Yoneda_1999,Formulations_of_Shinka_2008,Hyperbolic_form_Shinka_2000,Hyperbolic_form_Yoneda_2001,Will_hyperbolic_Shinka_2001} explored this further, applying the adjusted- and $\lambda$-system techniques to obtain strongly hyperbolic systems in which constraint growth is tamed. Interestingly, they found no striking difference between weakly, strongly and symmetric hyperbolic versions: the constraints remained under control in all cases~\cite{Will_hyperbolic_Shinka_2001}. A similar conclusion was reached for a different formulation in~\cite{Numerical_Relat_Hern_2000}, suggesting that the signs of the constraint-propagation eigenvalues matter more than the level of hyperbolicity. With this in mind, and following~\cite{Diagonalizabili_Yoneda_2002}, we require our systems to be strongly hyperbolic and to have as many eigenvalues of the constraint amplification matrix (see~\cref{subsec:hamiltonian-analysis}) with non-positive real part as possible.
  
\section{Linearised \pleb{} Gauge Fixing}\label{sec:linearised-gauge-fixing}
In this section we write down first-order equations for the linearised Einstein equations about Minkowski space in \pleb{}'s formulation; their principal part controls the local well-posedness of the full nonlinear system. We first show that a suitable gauge fixing makes every field satisfy the wave equation at leading order, so that the first-order system is strongly hyperbolic. We then show that, in suitable variables, the system splits into two independent subsystems of sizes 12 and 4, a split that relies on both this choice of gauge and the chiral formulation.

\subsection{Parametrising \pleb{} Perturbations}
Before we begin computing the perturbation of the equations of motion, we note that the 2-forms $\Sigma^i$ are not completely independent. Indeed, they satisfy the metricity constraints, \cref{eq:metricity-cond}, so the perturbations must be parametrised in a way that respects this condition. In this section $\Sigma^i$ denotes the background Minkowski 2-forms, $\asd^i = -{(\Sigma^*)}^i$, where $*$ is the complex conjugation operator, and $\delta \Sigma^i$ their linearisation. Linearising the metricity condition, we find
\begin{align}
    \delta \sd^{(i} \wedge \Sigma^{j )} \sim \delta^{ij},
\end{align}
for which the solutions are
\begin{align}
    \delta \sd^i = \frac{1}{2}\epsilon^{i j k} h^k \Sigma^j + h \diamond \sd^i,
\end{align}
where $h^i \in \E$ is an internal complex vector and $(h\diamond \sd)^i_{\mu\nu} = h_{[\mu}{}^{\rho} \sd^i_{|\rho|\nu]}$ with $h_{\mu\nu}$ being a complex symmetric rank-2 tensor. This parametrises a complex spacetime metric and a perturbation of the $SO(3,\C)$ frame. To ensure that we only describe a real metric we must impose the linearised versions of the reality conditions, \cref{eq:reality-cond}, given by
\begin{align}
    \re(\delta \Sigma^i \wedge \Sigma^i) = 0, \qquad  \delta \Sigma^i \wedge \asd^j + \Sigma^i \wedge \delta \asd^j = 0.
\end{align}
By making use of the algebra of the (anti-)self-dual 2-forms, \cref{eq:algebraic-identities}, and their complex conjugates, it is possible to show that the linearised reality conditions imply that $h_{\mu\nu}$ is a real tensor.

By solving the linearised metricity and reality conditions, we find that the linearisation of the self-dual 2-forms is parametrised by the metric perturbation $\delta g_{\mu\nu} := h_{\mu\nu}$ and the $SO(3,\C)$ perturbation $h^i$. At times it will be useful to decompose the metric perturbation further into
\begin{align}
    h_{\mu\nu}  = \frac{1}{4} h g_{\mu\nu} + \hat{h}_{\mu\nu},
\end{align}
where $h$ is a real scalar and $\hat{h}_{\mu\nu}$ is a real symmetric and traceless tensor. The parametrisation of the self-dual 2-forms that we use most often is
\begin{align}\label{eq:lin-frame-param}
    \delta \sd^i_{\mu\nu} = \frac{1}{2} \epsilon^{ijk} h^k \sd^j_{\mu\nu} + h_{[\mu}{}^\rho \sd^i_{|\rho|\nu]}.
\end{align}

\subsection{Linearised Connection and First-Order Ricci Tensor}
We now compute the solution for the linearised self-dual connection, $a^i := \delta A^i$, from the linearised torsion-free condition. It is known (see section 5.3.7 of~\cite{Krasnov2020-qr}) that the full torsion-free condition, \cref{eq:torsion-free}, can be uniquely solved to find
\begin{align}\label{eq:sd-con-def}
    A^i_\mu = -\im J_1^{-1}{(\star d\Sigma)}^i_\mu,
\end{align}
where $J_1 : \E \otimes \Lambda^1 \rightarrow \E \otimes \Lambda^1$ is given by $J_1(a)^i_\mu = \epsilon^{ijk} \sd^j_\mu{}^\nu a^k_\nu$. To find a solution for $a^i$, we first linearise \cref{eq:torsion-free} around an arbitrary background to find
\begin{align}
    d \delta \Sigma^i + \epsilon^{ijk} A^j \wedge \delta \Sigma^k  + \epsilon^{ijk} a^j \wedge \Sigma^k = 0.
\end{align}
We combine the first two terms into the gauge-covariant exterior derivative, $d^A \chi^i = d\chi^i + \epsilon^{ijk} A^j \wedge \chi^k$ for any $\chi^i \in \Lambda^n(M)$. This allows us to write the linearised torsion-free condition as
\begin{align}
    d^A \delta \Sigma^i + \epsilon^{ijk} a^j \wedge \Sigma^k = 0.
\end{align}
Comparing this with \cref{eq:torsion-free,eq:sd-con-def}, we can immediately write down the solution for the linear connection,
\begin{align}
    a^i = -\im J_1^{-1}{(\star d^A \delta \Sigma)}^i.
\end{align}
Here, $J_1$ and the Hodge star involve background quantities only. In components this is
\begin{align}
    a^i_\mu = \frac{\im}{2} \left( \epsilon^{ijk} \Sigma^j_\mu{}^\nu - \delta^{i k}\delta_\mu^\nu \right) \frac{1}{2} \epsilon_\nu{}^{\alpha \rho \sigma} \partial^A_\alpha \delta \Sigma^k_{\rho\sigma}.
\end{align}
On the Minkowski background we can choose $\Sigma^i$ such that $d\Sigma^i = 0 = A^i$, and therefore $F^i = 0$, which simplifies the gauge-covariant derivative, allowing us to replace $\partial^A \rightarrow \partial$. Then, using our parametrisation, \cref{eq:lin-frame-param}, the linearised connection is given by
\begin{align}
    a^i_\mu = \frac{1}{2} \partial_\mu h^i -  \frac{1}{2} \Sigma^{i\rho\sigma} \partial_\rho h_{\mu \sigma}.
\end{align}
Thus the perturbation in the $SO(3,\C)$ rotation of the frame produces a gauge transformation on the linearised connection; that is, $a^i = \frac{1}{2} d h^i$ is a linearised gauge transformation of the connection on the Minkowski background.

We are also interested in computing the linearisation of the first-order equation of motion corresponding to the Ricci-flat condition, when written using the self-dual connection. Recall that the full Ricci-flat condition is given by
\begin{align}
    R_{\mu\nu} = \Sigma^i_{(\mu}{}^\rho F^i_{\nu)\rho} = 0.
\end{align}
The antisymmetric part of this expression is purely gauge and imposes no physical equation of motion. Since the pure-gauge contribution does not affect the physical results, we drop the symmetrisation. To linearise this around a Minkowski background, where $F^i = 0$, we only need to consider the perturbation of the curvature, for which we find
\begin{align}
    \delta F^i_{\mu\nu} = 2 \partial_{[\mu} a^i_{\nu]}.
\end{align}
Therefore, the linearisation of the non-symmetric Ricci-flat condition gives
\begin{align}
    \delta R_{\mu\nu} = \sd^i_{\mu}{}^\rho \delta F^i_{\nu\rho} =  2 \sd^i_{\mu}{}^\rho \partial_{[\nu} a^i_{\rho]}  = 0.
\end{align}

\subsection{Linearised First- and Second-Order Equations of Motion}
To summarise, the first-order equations of motion are
\begin{gather}
    \delta \sd^i_{\mu\nu} = \frac{1}{2} \epsilon^{ijk} h^k \sd^j_{\mu\nu} + h_{[\mu}{}^\rho \sd^i_{|\rho|\nu]}, \quad a^i_\mu = \frac{1}{2}\partial_\mu h^i - \frac{1}{2}\sd^{i\rho\sigma} \partial_\rho h_{\mu \sigma}, \\
    \sd^i_\mu{}^\rho \partial_{[\nu} a^i_{\rho]}  = 0.
\end{gather}
This first-order system of equations on the linearised chiral objects constitutes the linearised Ricci-flat condition in \pleb{}'s formulation. A solution $h_{\mu\nu}, a^i_\mu$ to these equations is equivalent to a solution of the linearised Ricci-flat equations in the metric formulation.

To relate this back to the second-order metric formulation, we can simply substitute the solution for $a^i_\mu$ into the Ricci-flat condition. In doing so we find
\begin{align}\label{eq:lin-ricci-on-metric}
    \delta R_{\mu\nu} = 2 \sd^i_\mu{}^\rho \partial_{[\nu} a^i_{\rho]} = -\frac{1}{2} \Box h_{\mu\nu} + \partial_{(\mu} \partial^\rho \left( h_{\nu)\rho} - \frac{1}{2}g_{\nu )\rho} h \right),
\end{align}
where we have used the algebra \cref{eq:algebraic-identities}. The first term is the wave operator $\Box = g^{\mu\nu}\partial_\mu\partial_\nu$, which is hyperbolic. The remaining terms spoil hyperbolicity, and therefore the well-posedness of the linearised Ricci-flat condition; we remove them by a choice of gauge.

\subsection{Linear Gauge Fixing}
Before choosing a gauge we must consider the symmetries that are present in this system that we are able to fix. All standard formulations of general relativity are diffeomorphism invariant; that is, coordinate changes do not affect physical results. Our chiral formulation has an additional $SO(3,\C)$ symmetry, which must also be fixed. We now consider gauge conditions that make our linearised equations of motion well-posed.

\subsubsection*{de Donder Gauge}

To remove the two unwanted terms in \cref{eq:lin-ricci-on-metric}, one can impose the linearised harmonic condition, also known as the de Donder gauge~\cite{Hyperb1+3YvonneTommaso1983,Generalized_har_Alcubi_2005}. It is given by
\begin{align}
    - \partial^\nu h_{\mu\nu} + \frac{1}{2} \partial_\mu h = 0.
\end{align}
The nonlinear harmonic gauge condition is $g^{\rho\sigma} \Gamma^\mu_{\rho\sigma} = 0$, where $\Gamma^\mu_{\rho\sigma}$ are the Christoffel symbols for the metric $g_{\mu\nu}$. Hence, the above condition is the linearisation of the harmonic gauge around the Minkowski background in coordinates where the background Christoffel symbols vanish. Rather than imposing this condition sharply, we introduce a first-order connection $\xi_\mu \in \Lambda^1(M)$, and define
\begin{align}
    \xi_\mu = -\partial^\nu h_{\mu\nu} + \frac{1}{2} \partial_\mu h.
\end{align}
We modify the linearised Ricci-flat condition by introducing the term
\begin{align}
    2 \Sigma^i_\mu{}^\rho \partial_{[\nu} a^i_{\rho]} + \partial_{(\mu} \xi_{\nu )} = 0.
\end{align}
Upon substituting the definitions of the two linearised connections, $a^i$ and $\xi_\mu$, we find
\begin{align}
    2 \Sigma^i_\mu{}^\rho \partial_{[\nu} a^i_{\rho]} + \partial_{(\mu} \xi_{\nu )} = - \frac{1}{2} \Box h_{\mu\nu} = 0.
\end{align}
Only the wave operator acting on the metric perturbation remains, and for this reason we call this choice of gauge hyperbolic. For this to be a genuine gauge fixing, the constraint $\xi_\mu = 0$ must also be imposed, e.g.\ on an initial hypersurface. Fixing the diffeomorphism freedom in this way makes the metric part of the 2-forms evolve hyperbolically, but the perturbation of the frame, $h^i$, is still undetermined.

\subsubsection*{Modified Lorenz Gauge}

Keeping in mind our hyperbolicity requirements, we search for a gauge fixing that provides a wave-like evolution equation for the perturbations of the $SO(3,\C)$ frame. A second-order equation of motion for $h^i$ will be generated by derivatives of $a^i$, i.e.\ $\partial_\mu a^i_\nu$. In particular, we want a condition with only three components, as this is equal to the number of gauge degrees of freedom. It is then natural to consider the Lorenz condition on the connection,
\begin{align}
    \partial^{\mu} a^i_\mu = \frac{1}{2} \Box h^i - \frac{1}{2} \Sigma^{i\rho\sigma} \partial_\rho \partial^\mu h_{\mu\sigma}.
\end{align}
The second term here obstructs the hyperbolic nature of the equations of motion. However, notice that one can add to this condition arbitrary multiples of 
\begin{align}
    \Sigma^{i\rho\sigma} \partial_\rho \partial_\sigma h = 0,
\end{align}
since it vanishes identically. Therefore, by adding this we can form a multiple of $\xi_\mu$,
\begin{align}
    \partial^\mu a^i_\mu = \frac{1}{2} \Box h^i - \frac{1}{2} \Sigma^{i\rho\sigma} \partial_\rho  \left(\partial^\mu  h_{\mu\sigma} - \frac{1}{2} \partial_\sigma h \right) = \frac{1}{2} \Box h^i + \frac{1}{2} \Sigma^{i\rho\sigma} \partial_\rho \xi_\sigma.
\end{align}
Thus, by modifying the Lorenz gauge using the new connection, we can write a first-order gauge condition that implies a second-order hyperbolic equation for $h^i$. We call
\begin{align}
    \partial^\mu a^i_\mu - \frac{1}{2} \Sigma^{i\mu\nu} \partial_\mu \xi_\nu = \frac{1}{2} \Box h^i = 0
\end{align}
the modified Lorenz gauge.

\subsection{Modified Connection Definitions}
For numerical purposes, the connections and the frame variables should have the same number of components. A Dirac-type operator can then be built between the two spaces, its square has the same principal symbol as the wave operator, and the constraints are promoted to evolution equations. The connections $a^i$ and $\xi_\mu$ have $16$ components, but $\delta\Sigma^i$ has only $13$ ($10$ in $h_{\mu\nu}$ and $3$ in $h^i$). We restore the balance by adding a 3-component internal vector $\chi^i$ to the frame sector, whose derivatives contribute to the connections through modified definitions:
\begin{align}
    a^i_\mu &= \frac{1}{2} \partial_\mu h^i - \frac{1}{2} \Sigma^{i\rho\sigma} \partial_\rho h_{\mu \sigma} + c_1 \partial_\mu \chi^i + c_2 \epsilon^{ijk} \Sigma^j_\mu{}^\nu \partial_\nu \chi^k, \\
    \xi_\mu &= - \partial^\nu h_{\mu\nu} + \frac{1}{2} \partial_\mu h + c_3 \Sigma^i_\mu{}^\nu \partial_\nu \chi^i,
\end{align}
where $c_1,c_2,c_3$ are arbitrary complex constants. These are the only additional terms that can be constructed from first derivatives of $\chi^i$, which is not difficult to prove using representation theory. As with the new connection $\xi_\mu$, we require three more constraints, $\chi^i = 0$, for consistency. As we are considering a linear theory, we can focus on the modified terms and compute their contribution alone. In doing so we find
\begin{align}
    2 \Sigma^i_\mu{}^\rho \partial_{[\nu} a^i_{\rho]} + \partial_{(\mu} \xi_{\nu )} = \cdots + c_2 \Sigma^i_{\mu\nu} \Box \chi^i + \left( c_3 + 2 c_2 \right) \Sigma^i_{(\mu}{}^\rho \partial_{\nu)} \partial_\rho \chi^i.
\end{align}
To remove the term that obstructs hyperbolicity, we fix $c_3 = - 2 c_2$. For the Lorenz gauge, with this choice of constants, we have
\begin{align}
    \partial^\mu a^i_\mu - \frac{1}{2} \Sigma^{i\mu\nu} \partial_\mu \xi_\nu = \cdots + (c_1 - c_2) \Box \chi^i.
\end{align}
Note that $c_2 = 0$ is not allowed, since it leads to undefined terms (factors of $1/c_2$) later in the paper.

\subsection{Linearised Gauge-Fixed \pleb{}}\label{subsec:lin-gauge-fix-summary}
Here we summarise the system resulting from the hyperbolic first-order gauge fixing of the linearised \pleb{} formulation. Starting with the linearised self-dual frame $(h^i,h_{\mu\nu})$, we add to this an internal vector $\chi^i$. We then define the modified self-dual connection $a^i_\mu$ and connection 1-form $\xi_\mu$,
\begin{align}
    a^i_\mu &= \frac{1}{2} \partial_\mu h^i - \frac{1}{2} \Sigma^{i\rho\sigma} \partial_\rho h_{\mu\sigma} + c_1 \partial_\mu \chi^i + c_2 \epsilon^{ijk} \Sigma^j_\mu{}^\nu \partial_\nu \chi^k, \label{eq:mod-sd-con-def}\\
    \xi_\mu &= - \partial^\nu h_{\mu\nu} + \frac{1}{2} \partial_\mu h - 2 c_2 \Sigma^i_\mu{}^\nu \partial_\nu \chi^i. \label{eq:mod-first-ord-de-donder}
\end{align}
To impose Einstein's equations plus gauge, we require that the connections satisfy
\begin{align}
    2 \Sigma^i_\mu{}^\rho \partial_{[\nu} a^i_{\rho]} + \partial_{(\mu} \xi_{\nu )} = 0, \label{eq:mod-ricci-flat-cond} \\
    \partial^\mu a^i_\mu - \frac{1}{2} \Sigma^{i\mu\nu} \partial_\mu \xi_\nu = 0. \label{eq:mod-lin-lorenz-cond}
\end{align}
These equations are such that the second-order equations of motion for $\delta \Sigma^i$ and $\chi^i$ have the wave operator as their leading-order term. We call these equations the gauge-fixed linearised \pleb{} system.

\subsection{Linear Separation}\label{subsec:lin-sep}
Before explicitly checking that this system is well-posed, we construct a change of variables that causes the system to separate into two distinct systems. Changing the variables in this way is possible because of the maps between $\Lambda^1$ and $\C^3 \otimes \Lambda^1$; more details on these maps can be found in~\cite{Su2StructureBhoja2024,Plebanski_compl_Krasno_2025}. To display the separation, we introduce the new connection variables
\begin{equation}
\begin{aligned}
    \Omega^i_\mu & = a^i_\mu - \frac{1}{2} \Sigma^i_\mu{}^\nu \xi_\nu, \\
    \omega_\mu & = \xi_\mu + 2 \Sigma^i_\mu{}^\nu a^i_\nu,
\end{aligned}
\end{equation}
with inverse transformation
\begin{equation}\label{eq:con_inv_transform}
\begin{aligned}
    a^i_\mu & = \frac{1}{2} \Omega^i_\mu - \frac{1}{2} \epsilon^{ijk} \Sigma^j_\mu{}^\nu \Omega^k_\nu - \frac{1}{4} \Sigma^i_\mu{}^\nu \omega_\nu, \\
    \xi_\mu & = \Sigma^i_\mu{}^\nu \Omega^i_\nu - \frac{1}{2} \omega_\mu.
\end{aligned}
\end{equation}

Writing their definitions in terms of derivatives of the frame-sector fields, we find

\begin{align}
    \Omega^i_\mu & = \frac{1}{2} \partial_\mu h^i - \frac{1}{2} \Sigma^{i\rho\sigma} \partial_\rho h_{\mu\sigma} + c_1 \partial_\mu \chi^i + c_2 \epsilon^{ijk} \Sigma^j_\mu{}^\nu \partial_\nu \chi^k - \frac{1}{2} \Sigma^i_\mu{}^\nu \left( - \partial^\rho h_{\nu\rho} + \frac{1}{2} \partial_\nu h - 2 c_2 \Sigma^j_\nu{}^\rho \partial_\rho \chi^j \right) \nonumber \\ & = \frac{1}{2} \partial_\mu \left( h^i + 2 (c_1 - c_2) \chi^i \right) + \frac{1}{2} \Sigma^i_\mu{}^\nu \partial^\rho h_{\nu\rho} - \frac{1}{2} \Sigma^{i\rho\sigma} \partial_\rho h_{\mu \sigma} - \frac{1}{4} \Sigma^i_\mu{}^\nu \partial_\nu h,
\end{align}

and

\begin{align}
    \omega_\mu & = - \partial^\nu h_{\mu\nu} + \frac{1}{2} \partial_\mu h - c_2 \Sigma^i_\mu{}^\nu \partial_\nu \chi^i + 2 \Sigma^i_\mu{}^\nu \left( \frac{1}{2} \partial_\nu h^i - \frac{1}{2} \Sigma^{i\rho\sigma} \partial_\rho h_{\nu\sigma} + c_1 \partial_\nu \chi^i + c_2 \epsilon^{ijk} \Sigma^j_\nu{}^\rho \partial_\rho \chi^k \right) \nonumber \\ & = -\frac{1}{2} \partial_\mu h + \Sigma^i_\mu{}^\nu \partial_\nu \left( h^i + 2(c_1 + c_2) \chi^i \right).
\end{align}

We then define another change of variables, this time for the frame sector,
\begin{align}\label{eq:lin-org-frame-to-con}
    h_{\mu\nu} = \hat{h}_{\mu\nu} + \frac{1}{4}g_{\mu\nu} h, \quad \hat{h}^{i} = h^i + 2(c_1 - c_2) \chi^i, \quad \hat{\chi}^i = h^i + 2 (c_1 + c_2) \chi^i,
\end{align}
where $\hat{h}_{\mu\nu}$ is the symmetric tracefree part of $h_{\mu\nu}$. In these variables the connections become
\begin{align}
    \Omega^i_\mu & = \frac{1}{2} \partial_\mu \hat{h}^i + \frac{1}{2} \Sigma^i_\mu{}^\nu \partial^\rho \hat{h}_{\nu\rho} - \frac{1}{2} \Sigma^{i\rho\sigma} \partial_\rho \hat{h}_{\mu\sigma}, \\
    \omega_\mu & = \Sigma^i_\mu{}^\nu \partial_\nu \hat{\chi}^i - \frac{1}{2} \partial_\mu h.
\end{align}
The two connections depend on separate components of the frame sector, so each can be defined independently of the other. Furthermore, we can use \cref{eq:con_inv_transform} on the connection equations of motion to find
\begin{align}
    2 \Sigma^i_\mu{}^\rho \partial_{[\nu} a^i_{\rho]} + \partial_{(\mu} \xi_{\nu)} & = \frac{1}{4} g_{\mu\nu} \partial^\rho \omega_\rho - \frac{1}{4} \Sigma^i_{\mu\nu} \left( \Sigma^{i\rho\sigma} \partial_\rho \omega_\sigma + 2 \partial^\rho \Omega^i_\rho \right) + \Sigma^i_{\langle \mu}{}^\rho \partial_{\nu \rangle} \Omega^i_\rho - \Sigma^i_{\langle \mu}{}^\rho \partial_{|\rho|} \Omega^i_{\nu\rangle} = 0, \\
    \partial^\mu a^i_\mu - \frac{1}{2} \Sigma^{i\mu\nu} \partial_\mu \xi_\nu & = \partial^\mu \Omega^i_\mu = 0.
\end{align}
Here $\langle \rangle$ denotes the symmetric tracefree parts, i.e.\ $T_{\langle \mu \nu \rangle} = \frac{1}{2} T_{\mu\nu} + \frac{1}{2} T_{\nu\mu} - \frac{1}{4} g_{\mu\nu} g^{\rho \sigma} T_{\rho \sigma}$. The independent equations of motion can be read off from the trace, antisymmetric and symmetric tracefree parts of the first equation, along with the second equation. These are
\begin{align}
    \Sigma^i_{\langle \mu}{}^\rho \partial_{\nu \rangle} \Omega^i_\rho - \Sigma^i_{\langle \mu}{}^\rho \partial_{|\rho|} \Omega^i_{\nu\rangle} = 0, \quad \partial^\mu \Omega^i_\mu = 0, \label{eq:lin-12-con-evo}\\
    \Sigma^{i\mu\nu} \partial_\mu \omega_\nu = 0, \quad \partial^\mu \omega_\mu = 0. \label{eq:lin-4-con-evo}
\end{align}
Using the fact that the $\Sigma^i_\mu{}^\rho \asd^j_{\nu\rho}$ form a 9-dimensional basis for symmetric tracefree rank-2 tensors, we can write the first of these equations as
\begin{align}
    \Sigma^i_{\langle \mu}{}^\rho \partial_{\nu \rangle} \Omega^i_\rho - \Sigma^i_{\langle \mu}{}^\rho \partial_{|\rho|} \Omega^i_{\nu\rangle} = 0 \quad \Rightarrow \quad \asd^{j\mu\nu} \partial_\mu \Omega^i_\nu = 0.
\end{align}
Taking all the first-order equations of motion, we can separate the linearised Einstein equations into two sectors, the first of size $12$ (total components $2\times 12 = 24$) and the second of size $4$:
\begin{alignat}{2}
    \Omega^i_\mu & = \frac{1}{2} \left( \partial_\mu \hat{h}^i + \Sigma^i_\mu{}^\nu \partial^\rho \hat{h}_{\nu\rho} - \Sigma^{i\rho\sigma} \partial_\rho \hat{h}_{\mu\sigma} \right), \quad &&\asd^{j\mu\nu} \partial_\mu \Omega^i_\nu = 0, \quad \partial^\mu \Omega^i_\mu = 0, \\
    \omega_\mu & = \Sigma^i_\mu{}^\nu \partial_\nu \hat{\chi}^i - \frac{1}{2} \partial_\mu h, \quad &&\Sigma^{i\mu\nu} \partial_\mu \omega_\nu = 0, \quad \partial^\mu \omega_\mu = 0.
\end{alignat}
This separation is remarkable: it requires both the chiral formulation and this particular family of gauges. It generalises the separation of~\cite{ChiralPerturbaKrasno2020,Plebanski_compl_Krasno_2025} to two parameters and expresses it in spacetime and internal indices rather than spinor indices. The connection equations are the self-dual and anti-self-dual projections of the curvature, each supplemented by a Lorenz gauge, which is exactly the structure of the self-dual and anti-self-dual Yang-Mills equations. The only difference is that here the connections are not fundamental fields but are built from derivatives of the frame-sector fields.

\subsection{Linear Hyperbolicity}
To check that this system is indeed hyperbolic, we first need to define a 3+1 decomposition. In general this assumes that our manifold is globally hyperbolic. We introduce a 1-form basis $n_\mu, e^i_\mu$ adapted to the 3+1 decomposition, where $n_\mu$ is future-pointing and timelike and the $e^i_\mu$ are three mutually orthogonal spatial 1-forms. The index $i=1,2,3$ is an $\so(3,\C)$ index; however, the basis 1-forms are real-valued because we require the metric to be real. The background metric can then be decomposed as
\begin{align}
    g_{\mu \nu} = - n_{\mu} n_{\nu} + e^i_{\mu} e^i_\nu.
\end{align}
The basis vectors are then orthonormal with respect to this metric,
\begin{align}
    n^\mu n_\mu = -1, \quad e^i_\mu n^\mu = 0, \quad e^i_\mu e^{j\mu} = \delta^{ij}.
\end{align}
The self-dual 2-forms have a similar decomposition,
\begin{align}
    \Sigma^i_{\mu\nu} = \im n_\mu e^i_\nu - \im n_\nu e^i_\mu - \epsilon^{ijk}  e^j_\mu e^k_\nu.
\end{align}
For the purposes of this section it is enough to assume that the background basis is constant under the derivative, $\partial_\mu n_\nu = 0 = \partial_\mu e^i_\nu$. We can also decompose the spacetime indices of the linearised fields using this basis,
\begin{align}
    h_{\mu\nu} = f n_\mu n_\nu + 2 n^i n_{(\mu} e^i_{\nu )} + h^{ij} e^i_\mu e^j_\nu, \\
    a^i_\mu = a^i n_\mu + a^{ij} e^j_\mu, \quad \xi_\mu = n_\mu \xi + \xi^i e^i_\mu.
\end{align}
Lastly, we decompose the derivative in the same way,
\begin{align}
    \partial_\mu = n_\mu \partial_t + e^i_\mu \partial^i.
\end{align}
Since the frame is constant, the derivatives $\partial_t$ and $\partial^i$ commute, as ordinary partial derivatives do. With this in mind, one can construct all the equations of motion by substituting the 3+1 decomposition into \cref{eq:mod-sd-con-def,eq:mod-first-ord-de-donder,eq:mod-ricci-flat-cond,eq:mod-lin-lorenz-cond} and solving the resulting linear equations for the time derivatives of all the fields. The result, after some algebra, for the frame sector is
\begin{align}
    \partial_t{h^{i j}} &= \im \partial^{(i}{h^{j )}} + \partial^{(i}{n^{j )}}+\im \epsilon^{(i |k l} \partial^{k}{h^{|j ) l}}+ 2\im (c_{1}-c_2) \partial^{(i}{\chi^{j )}} +2 \im c_{2} \delta^{i j} \partial^{k}{\chi^{k}}, \\
    \partial_t{n^{i}} &= \frac{1}{2}\partial^{j}{h^{i j}}+\frac{1}{2}\partial^{i}{f} + \frac{1}{2} \epsilon^{ijk} \partial^j \left( h^k + \im n^k + 2 (c_1-c_2) \chi^k \right), \\
    \partial_t{h^{i}} & =  -\frac{\im(c_1 + c_2)}{2c_2} \partial^{j}{h^{i j}}+\frac{\im}{2} \partial^{i}{\left( f + \frac{c_1}{c_2} h^{jj} \right)} + \frac{\im}{2c_2} \epsilon^{i j k} \partial^j \left((c_1 - c_2) h^k + 2(c_1^2 - c_2^2) \chi^k + \im (c_1 + c_2) n^k \right), \\
    \partial_t{\chi^{i}} &= \frac{1}{2c_2}\epsilon^{ijk} \partial^j \left( n^k - \im h^k -2 \im (c_1 + c_2) \chi^k \right) + \frac{\im}{4c_2} \partial^{j} h^{ij} - \frac{\im}{4c_2} \partial^{i} h^{jj}, \\
    \partial_t f & = \partial^i \left( n^i - \im h^i - 2 \im c_1 \chi^i \right).
\end{align}
We have dropped all terms of lower than first order in derivatives, as these do not affect the characteristic matrix. For the connections we have
\begin{align}
    \partial_t a^{ij} & = \partial^j a^i - \im \epsilon^{ikl} \partial^j a^{kl} - \im \epsilon^{ikl} \partial^k a^{lj} -\im\partial^{(i} \xi^{j )}, \\
    \partial_t a^i & = \partial^j a^{ij} - \partial^j a^{ji} + \partial^i a^{jj} + \frac{1}{2} \epsilon^{ijk} \partial^j \xi^k - \im \partial^i \xi, \\
    \partial_t \xi^i & = -2\im \partial^i a^{jj} +2\im\partial^j a^{ji} - \partial^i \xi, \\
    \partial_t \xi & = - 2\epsilon^{ijk} \partial^i a^{jk} - \partial^i \xi^i.
\end{align}
The usual way to determine hyperbolicity is to construct the characteristic matrix $P = M^i k^i$, where we assume the above system is written in the form $\partial_t u = M^i \partial^i u$ for some solution vector $u \in \C^{32}$ and matrices $M^i \in {\rm Mat}(32,\C)$. Instead, we take an alternative route: by applying the time derivative twice, one can show that all the fields in the above satisfy
\begin{align}
    \partial_t \partial_t u = \partial^i \partial^i u;
\end{align}
that is, they satisfy the wave equation. Equivalently, $P^2 = k^2 \id$, where $k^2 = k^i k^i$. Therefore, the eigenvalues of $P$ are $\pm \sqrt{k^2}$. All the eigenvalues of $P$ are real, and because of the link to the wave equation the system is guaranteed to be at least strongly hyperbolic. We can also interpret $M^i$ as a $32\times 32$ representation of the gamma matrices, since they satisfy $M^{(i} M^{j)} = \delta^{i j} \id$.

Our linearised system is therefore well-posed. Any nonlinear system with this linearisation shares its principal part and is locally well-posed (see also \cref{sec:hyp-and-con-anal-of-evo-sys}); that is, solutions exist, are unique and depend continuously on the initial data. Since the separated system is merely a reparametrisation of the same equations, it is well-posed as well.

\subsection{Linear Lagrangian}
The first-order linear Lagrangian corresponding to the gauge-fixed linearised \pleb{} system is given by
\begin{equation} \label{eq:lin-lagrangian}
\begin{aligned}
    L = & \left( \Sigma^{i\mu[\rho} h^{\sigma]}_\mu + \frac{1}{2} \epsilon^{ijk}  \Sigma^{j\rho\sigma} h^k + \frac{1}{2} h \Sigma^{i\rho\sigma} \right) \partial_\rho a^i_\sigma + \frac{1}{2} \epsilon^{ijk} \Sigma^{i\mu\nu} a^j_\mu a^k_\nu \\\ &\ - 2 c_2 \chi^i \partial^\mu a^i_\mu - (c_1 + c_2) \epsilon^{ijk} \chi^i \Sigma^{j\mu\nu} \partial_\mu a^k_\nu  \\ &\ - \frac{1}{2}\left( h^{\mu\nu} - \frac{1}{2} h g^{\mu\nu} \right) \partial_\mu \xi_\nu + \frac{1}{4} \xi^\mu \xi_\mu + c_2 \Sigma^{i\mu\nu} \chi^i \partial_\mu \xi_\nu. \end{aligned}
\end{equation}
The first line is obtained by linearising the \pleb{} action $L = \Sigma^i \wedge F^i$ in a way that respects the metricity and reality conditions. The second and third lines are the gauge-fixing terms for the modified de Donder and Lorenz gauges. Varying with respect to each field produces the equations of motion in \cref{eq:mod-sd-con-def,eq:mod-first-ord-de-donder,eq:mod-ricci-flat-cond,eq:mod-lin-lorenz-cond}.

The splitting of the equations of motion into two sectors is also apparent at the level of the Lagrangian. By transforming \cref{eq:lin-lagrangian} into the new fields, we find
\begin{equation}
\begin{aligned}
    L = &\ \Sigma^i_\rho{}^{[\mu} \hat{h}^{\nu]\rho} \partial_\mu \Omega^i_\nu + \frac{1}{2} \hat{h}^i \partial^\mu \Omega^i_\mu + \frac{1}{2} \Omega^{i\mu} \Omega^i_\mu \\ & - \frac{1}{4} \hat{\chi}^i \Sigma^{i\mu\nu} \partial_\mu \omega_\nu + \frac{1}{8} h \partial^\mu \omega_\mu - \frac{1}{8} \omega^\mu \omega_\mu.
\end{aligned}
\end{equation}
The Lagrangian therefore separates into two parts,
\begin{align}
    L_{12} & = \Sigma^i_\rho{}^{[\mu} \hat{h}^{\nu]\rho} \partial_\mu \Omega^i_\nu + \frac{1}{2} \hat{h}^i \partial^\mu \Omega^i_\mu + \frac{1}{2} \Omega^i_\mu \Omega^i_\nu ,\\
    L_{4} & = - \frac{1}{4} \hat{\chi}^i \Sigma^{i\mu\nu} \partial_\mu \omega_\nu + \frac{1}{8} h \partial^\mu \omega_\mu - \frac{1}{8} \omega^\mu \omega_\mu.
\end{align}

Since these Lagrangians share no fields, any linear combination

\begin{align}\label{eq:lin-two-lagrangians}
  L = L_{12} + a L_{4},
\end{align}

for any constant $a$, results in the same dynamics. Physically relevant information resides only in $L_{12}$, through the tracefree metric perturbation $\hat h_{\mu\nu}$, so adding a multiple of $L_4$ changes only pure-gauge terms. At the nonlinear level, as we will see, the separation is incomplete, and this freedom is lost. Therefore, only one Lagrangian is possible, and the dynamical equations remain unique (up to choice of $c_1,c_2$).

\section{Nonlinear \pleb{} Gauge Fixing}\label{sec:gauge-fixing}
We now construct nonlinear analogues of the linear hyperbolic system, which will become our full evolution system. Unlike in the linear case, the self-dual 2-forms can no longer be decomposed into irreducible pieces, so we work with the full 2-form. Recall that the first-order Einstein equations for the self-dual 2-forms $\Sigma^i$ and the connection $A^i$ are
\begin{align}
    A^i = -\im J_1^{-1}(\star d \Sigma)^i, \quad \Sigma^i_\mu{}^\rho F^i_{\nu \rho} = 0.
\end{align}
Here $F^i = dA^i + \frac{1}{2} \epsilon^{ijk} A^j \wedge A^k$ is the curvature of the connection, $J_1$ is defined below \cref{eq:sd-con-def} and $\star$ is the Hodge star of the metric in \cref{eq:urbantke-metric}, which is not an independent field but a convenient repackaging of certain components of $\Sigma^i$. We seek nonlinear equations whose linearisations reproduce \cref{eq:mod-sd-con-def,eq:mod-first-ord-de-donder,eq:mod-ricci-flat-cond,eq:mod-lin-lorenz-cond}.

\subsection{Harmonic Gauge}
We begin with the harmonic condition, a well-known condition that has been used to show that Einstein's equations are well-posed~\cite{Hyperb1+3YvonneTommaso1983}. It is given as a condition on the Christoffel symbols for the metric,
\begin{align}
    \Gamma^\mu := g^{\rho\sigma} \Gamma^\mu_{\rho\sigma} = -\partial_\nu g^{\mu\nu} - \partial^\mu \ln\sqrt{-g} = 0.
\end{align}
Upon perturbing the metric, $g_{\mu\nu} \to g_{\mu\nu} + h_{\mu\nu}$, i.e.\ $\delta g_{\mu\nu} = h_{\mu\nu}$ with $g_{\mu\nu}$ being the Minkowski metric, we find that this becomes $\delta \Gamma^\mu = \partial_\nu h^{\mu\nu} - \frac{1}{2} \partial^\mu h$. Hence the de Donder condition, \cref{eq:mod-first-ord-de-donder}, can be written nonlinearly as
\begin{align}
    \xi_\mu = - \Gamma_\mu - 2 c_2 \Sigma^i_\mu{}^\nu \partial_\nu \chi^i. \label{eq:nonlin-mod-de-donder}
\end{align}
Another useful way to write the harmonic 1-form is as follows:
\begin{align}
    \Gamma_\mu = - \utg_{\mu\nu} \partial_\rho \tg^{\nu\rho},
\end{align}
where $\tg^{\mu\nu} = \sqrt{-g} g^{\mu\nu}$ and $\utg_{\mu\nu} = g_{\mu\nu}/\sqrt{-g}$. The term involving $\chi^i$ in \cref{eq:nonlin-mod-de-donder} takes the same form as in the linear case, since its perturbation is $\delta \chi = 0 + \chi^i$, and therefore it reproduces the linear term exactly. Terms of the form $\chi^i \partial_\mu \Sigma^i$ could also be added without affecting the linearised result, so many potentially useful terms are absent from this gauge fixing. In this section we impose the ``minimal'' number of terms needed to reproduce the linear case, and so ignore terms that do not survive linearisation. As before, the Ricci-flat condition must also be modified; this can simply be written as
\begin{align}
    \Sigma^i_\mu{}^\rho F^i_{\nu \rho} +  \partial_{(\mu} \xi_{\nu )} = 0.
\end{align}
This is constructed so that when $\xi = 0$ the above equation reduces to the original Ricci-flat condition.

\subsection{Lorenz Gauge and Modified Connection}
The Lorenz gauge, \cref{eq:mod-lin-lorenz-cond}, can be lifted directly to a nonlinear condition,
\begin{align}\label{eq:num-plebanski-lorenz-gauge-components}
    \partial^\mu A^i_\mu -\frac{1}{2} \Sigma^{i\mu\nu} \partial_\mu \xi_\nu = 0.
\end{align}
Since the perturbation of the connection 1-form is $ \delta \xi_\mu = 0 + \xi_\mu$, this results in the correct linearised equation up to lower-order terms. Similarly, the definition of the self-dual connection, \cref{eq:sd-con-def}, must be modified to
\begin{align}
    A^i_\mu = -\im J_1^{-1}{(\star d \Sigma)}^i_\mu + c_1 \partial_\mu \chi^i + c_2 \epsilon^{ijk} \Sigma^j_\mu{}^\nu \partial_\nu \chi^k.
\end{align}
By applying $\im \star J_1$ to both sides of this equation, we obtain an alternative way of stating this condition,
\begin{align}
    d^A \Sigma^i + (c_1 + c_2) \epsilon^{ijk} d\chi^j \wedge \Sigma^k + 2 \im c_2 \star d\chi^i = 0.
\end{align}
Written this way, the gauge fixing modifies the torsion-free condition on the connection. Since we require that $\chi^i = 0$ for our equations to be consistent, this condition reduces to the original definition of the self-dual connection.

\subsection{Hyperbolic \pleb{} Gauge System}
Collecting all the gauge-fixed equations of motion we find
\begin{gather}
    A^i_\mu = -\im J_1^{-1}{(\star d \sd)}^i_\mu + c_1 \partial_\mu \chi^i + c_2 \epsilon^{ijk} \Sigma^j_\mu{}^\nu \partial_\nu \chi^k, \quad \xi_\mu = - \Gamma_\mu - 2 c_2 \Sigma^i_\mu{}^\nu \partial_\nu \chi^i, \label{eq:mod-nonlin-con-defs}\\
    \Sigma^i_\mu{}^\rho F^i_{\nu \rho} +   \partial_{(\mu} \xi_{\nu)} = 0, \quad \partial^\mu A^i_\mu - \frac{1}{2} \Sigma^{i\mu\nu} \partial_\mu \xi_\nu = 0.\label{eq:mod-nonlin-ricci-and-lorenz-cond}
\end{gather}
Since the connections are built from derivatives of $\Sigma^i$ and $\chi^i$, we call these fields the frame sector. The connection sector consists of $A^i$ and $\xi$. The number of fields in each sector is $16$, which also matches the number of equations. As we will see in the 3+1 decomposition, this allows for an evolution equation for each component of the frame and connections. Through the link to the linear equations, the implied second-order equations of motion have the wave operator as their leading-order (highest-derivative) term. In the metric formulation it is known that only the harmonic condition, and slight generalisations of it, produce the same leading-order term. Indeed, because of this property, they have been used in various other numerical contexts~\cite{General_covaria_Bona_2003,A_new_approach_Headri_2009}. The Lorenz condition on the connection appears in Yang-Mills theories, where it reduces the equations of motion to the wave operator acting on the connection, e.g.\ see chapter 8 of~\cite{Krasnov2020-qr}. In numerical relativity the Lorenz condition has appeared in the tetrad formalism~\cite{HyperbolicTetrBuchma2003}. 

Setting $\xi_\mu = 0 = \chi^i$ removes all the modifications and leaves Einstein's equations in first-order form together with the Lorenz and harmonic conditions. In Hamiltonian terms, $\xi_\mu$ and $\chi^i$ are the conjugate momenta of the Lagrange multipliers, so the primary constraints require them to vanish. This is also why the modifications are harmless: they preserve the first-class structure of the primary and secondary constraints. We cannot simply discard these fields, however, since the evolution system is not well-posed without them. Instead, we impose $\xi_\mu = 0 = \chi^i$ and $\partial_t\xi_\mu = 0 = \partial_t\chi^i$ initially; by the first-class property, they then remain zero analytically.

Finding exact solutions to the equations with these initial requirements is not always possible in more complicated physical setups, e.g.\ binary black holes. It is then worth considering what happens to these components if they fluctuate away from zero. Since they satisfy the wave equation to leading order, any non-zero fluctuation is expected to propagate away at the speed of light. Numerically, this would force the violating modes to leave the simulation. One problem is that the violating modes can interfere with the physically relevant modes, which also propagate at the speed of light.

For this reason, in later sections we consider further modifications that introduce damping to these components. The damping can be introduced as a form of gauge fixing, without destroying the first-class nature of the constraints.

A second issue is that this gauge fixing is not diffeomorphism covariant. The frame fields and the self-dual connection transform covariantly (the latter involves only exterior derivatives of forms), but $\xi_\mu$ does not, because it is built from Christoffel symbols. This can cause problems, for instance in spherical polar coordinates, where non-tensorial terms with spurious angular dependence appear and can break the symmetry of a configuration. While this term does not always cause problems, it is useful to have a way of restoring covariance, so that different coordinate systems can be related when needed.

To do so we follow~\cite{Covariant_formu_Brown_2009,Formulations_of_Alcubi_2010} and introduce non-dynamical reference 2-forms $\mathring\Sigma^i_{\mu\nu}$, for which there is a corresponding metric $\mathring{g}_{\mu\nu}$. Next, we compute this metric's Christoffel symbols, $\mathring\Gamma^\mu_{\rho\sigma}$, which are in general not zero. We then take the difference with the full Christoffel symbols to construct
\begin{align}
    \Delta \Gamma_\mu = \Gamma_\mu - g^{\rho\sigma} g_{\mu\nu} \mathring{\Gamma}^\nu_{\rho\sigma}.
\end{align}
This is a true tensor, since the difference of two connections is a tensor. In the modified harmonic condition, $\Gamma_\mu$ can be replaced with $\Delta \Gamma_\mu$. If the reference metric is the Minkowski metric, then $\mathring{\Gamma} = 0$ in Cartesian coordinates. However, in spherical polar coordinates one finds that $\mathring{\Gamma} \neq 0$, and it cancels some unwanted terms appearing in $\Gamma$ (see~\cite{Generalized_har_Sorkin_2009}). The other problematic non-covariant terms are the partial derivatives of the connections appearing in \cref{eq:mod-nonlin-ricci-and-lorenz-cond}. To remedy this, one can replace them with metric covariant derivatives, in which case the modification to the Ricci-flat condition resembles the Z4 equations~\cite{General_covaria_Bona_2003}, which have been very successful in numerical relativity. However, as we will see in later sections, leaving the derivatives as partial derivatives does not spoil the numerical solutions for the cases we consider. For this reason we choose to leave them as partial derivatives for the remainder of the paper.

\subsection{Nonlinear Conformal Separation}\label{subsec:nonlin-gauge:-nonlin-conf-sep}
\Cref{subsec:lin-sep} showed that the linear system separates. We now extend this separation, as far as possible, to the nonlinear equations. The result is weaker: only one sector decouples, while the other depends on all the fields. However, it is still surprising that there exists a sector that does decouples nonlinearly. At the linear level, the 4-component sector contains the perturbation of the conformal factor, which suggests a nonlinear conformal decomposition.

We begin by introducing a scalar density, $\lambda = 1/\sqrt{\sqrt{-g}}$, where $g$ is the spacetime metric constructed from $\Sigma^i$. With this we can define new conformal 2-forms,
\begin{align}
    H^i = \lambda \Sigma^i \in \Lambda^+ \subset \Lambda^2.
\end{align}
It is not difficult to check that the metric corresponding to these 2-forms is
\begin{align}\label{eq:conf-metric}
    g_H = \lambda g,
\end{align}
and that $\det(g_H) = -1$. Owing to the conformal invariance of the Hodge star on 2-forms, both $H^i$ and $\Sigma^i$ are self-dual with respect to $g$ and $g_H$. Introducing the components $H^i = \frac{1}{2} H^i_{\mu\nu} dx^\mu \wedge dx^\nu$, and noting that indices on $H^i_{\mu\nu}$ are raised and lowered with $g_H$ and its inverse, we have the following conformal quaternion algebra:
\begin{align}
    H^i_{\mu\rho} H^{j}_\nu{}^\rho = - \delta^{ij} g_{H \mu\nu} + \epsilon^{ijk} H^k_{\mu\nu}.
\end{align}
All the other identities appearing in \cref{eq:algebraic-identities} have conformal counterparts under the replacement $(\Sigma^i,g) \rightarrow (H^i,g_H)$. As in the linear case, we explore how the connection definitions decompose under this transformation. Starting with the self-dual connection, we find
\begin{align}
    A^i & = -\im J_1^{-1} \left( \star d \left(\frac{1}{\lambda} H^i\right) \right) + c_1 d \chi^i + c_2 J_1 {(d\chi)}^i \\
    & = \im J_1^{-1} \left(\frac{1}{\lambda} \star \left(d \log(\lambda) \wedge H^i\right)\right) -\im J_1^{-1} \left(\frac{1}{\lambda} \star d H^i\right) + c_1 d \chi^i + c_2 J_1 {(d\chi)}^i.
\end{align}
Here $\star$ is the Hodge star constructed using $g$. The operator $J_1$ involves the endomorphisms $\Sigma^i_\mu{}^\nu$, but it can be shown that these are equal to $H^i_\mu{}^\nu$ since both are conformally invariant. For this reason we keep the notation $J_1$ without introducing another operator. To proceed, we derive how the Hodge star associated with $g_H$ relates to the original one when acting on 3-forms,
\begin{align}
    {\left(\star T \right)}_\mu = \frac{1}{3!} \epsilon_\mu{}^{\nu\rho\sigma} T_{\nu\rho\sigma} = \frac{1}{3!} g_{\mu\alpha} \epsilon^{\alpha\nu\rho\sigma} T_{\nu\rho\sigma} = \frac{1}{3!} \frac{1}{\lambda} g_{H\mu\nu} \sqrt{-g} \epsilon_H^{\alpha\nu\rho\sigma} T_{\nu\rho\sigma} = \frac{1}{3!} \lambda \epsilon_{H\mu}{}^{\nu\rho\sigma} T_{\nu\rho\sigma} = \lambda {\left(\star_H T\right)}_\mu.
\end{align}
Therefore, the self-dual connection decomposition simplifies slightly,
\begin{align}
    A^i = \im J_1^{-1} \left( \star_H \left(d \log(\lambda) \wedge H^i\right)\right) -\im J_1^{-1} \left(\star_H d H^i\right) + c_1 d \chi^i + c_2 J_1 {(d\chi)}^i.
\end{align}
We can now use the self-duality of $H^i$ to show that
\begin{align}
    \star_H {(d \log(\lambda) \wedge H^i)}_\mu = \frac{1}{2} \epsilon_{H\mu}{}^{\nu\rho\sigma} \partial_\nu \log(\lambda) H^i_{\rho\sigma} = \im H^i_\mu{}^\nu \partial_\nu \log(\lambda) = \im {(H^i \cdot d \log(\lambda))}_\mu.
\end{align}
Here we have introduced the notation ${(H^i \cdot \theta)}_\mu = H^i_\mu{}^\nu \theta_\nu$; the same notation will also be used with $\Sigma^i$. Next, we recall that $J_1^{-1} = \frac{1}{2}(J_1 - \mathbb{I})$ and $J_1 (H^i \cdot \theta) = 2 H^i \cdot \theta$, which implies that $J_1^{-1}(H^i \cdot \theta) = \frac{1}{2} H^i \cdot \theta$. This can be used to simplify the first term in the connection, 
\begin{align}
    A^i = -\frac{1}{2} H^i \cdot d \log(\lambda) - \im J_1^{-1} \left(\star_H d H^i\right) + c_1 d \chi^i + c_2 J_1 {(d\chi)}^i.
\end{align}
A similar conformal decomposition can be performed on the 1-form $\xi$. Using the standard formula for the conformal transformation of the Christoffel symbols we find 
\begin{align}
    \Gamma_\mu = g^{\rho\sigma} g_{\mu\nu} \Gamma^\nu_{\rho\sigma} = g_H^{\rho\sigma} g_{H\mu\nu} \Gamma^\nu_{H \rho\sigma} + \partial_\mu \log(\lambda) = \Gamma_{H\mu} + \partial_\mu \log(\lambda).
\end{align}
Applying this to the connection 1-form, we find that
\begin{align}
    \xi = - \Gamma - 2 c_2 \Sigma^i \cdot d\chi^i = -\Gamma_H - d \log(\lambda) - 2 c_2 H^i \cdot d\chi^i.
\end{align}
As in the linear case, we can introduce two special combinations of these connections,
\begin{align}\label{eq:nonlin-con-transform-Omega-eq-a}
    \Omega^i = A^i - \frac{1}{2} H^i \cdot \xi, \quad \omega = \xi + 2 H^i \cdot A^i. 
\end{align}
Inverting this transformation, we find
\begin{align}\label{eq:nonlin-con-transform-a-eq-Omega}
    A^i = \frac{1}{2} \Omega^i - \frac{1}{2} J_1{{(\Omega)}}^i - \frac{1}{4} H^i \cdot \omega, \quad \xi = H^i \cdot \Omega^i -  \frac{1}{2}\omega. 
\end{align}
Substituting the old connection definitions into the new connections, we find
\begin{align}
    \Omega^i  = & -\frac{1}{2} H^i \cdot d \log(\lambda) - \im J_1^{-1} \left(\star_H d H^i\right) + c_1 d \chi^i + c_2 J_1 {(d\chi)}^i \\ & \quad - \frac{1}{2} H^i \cdot \left( -\Gamma_H - d \log(\lambda) - 2 c_2 H^j \cdot d\chi^j \right) \nonumber\\ = &\ -\frac{1}{2} H^i \cdot d \log(\lambda) - \im J_1^{-1} \left(\star_H d H^i\right) + c_1 d \chi^i + c_2 J_1 {(d\chi)}^i \\ & \quad +  \left(\frac{1}{2} H^i \cdot\Gamma_H + \frac{1}{2} H^i \cdot d \log(\lambda) - c_2 d \chi^i - c_2 J_1{(d\chi)}^i \right) \nonumber\\ = &\ \frac{1}{2} H^i \cdot \Gamma_H - \im J_1^{-1} \left(\star_H d H^i\right) + (c_1 - c_2) d \chi^i.
\end{align}
Doing the same for the other connection, we find
\begin{align}
    \omega = & - \Gamma_H - d\log(\lambda) - 2c_2 H^i \cdot d \chi^i +2 \left( \frac{3}{2} d\log(\lambda) - \frac{\im}{2} H^i \cdot \star_H dH^i + (c_1 + 2 c_2) H^i \cdot d\chi^i  \right) \nonumber \\ = & - \Gamma_H + 2 d \log(\lambda) -\im H^i \cdot \star_H dH^i + 2(c_1 + c_2) H^i \cdot d \chi^i.
\end{align}

The derivatives of the frame fields $H^i, \lambda, \chi^i$ are therefore packaged into two connections:
\begin{align}
    \Omega^i  & = \frac{1}{2} H^i \cdot \Gamma_H - \im J_1^{-1} \left(\star_H d H^i\right) + (c_1 - c_2) d \chi^i, \\
    \omega & = - \Gamma_H + 2 d \log(\lambda) - \im H^i \cdot \star_H dH^i + 2(c_1 + c_2) H^i \cdot d \chi^i.
\end{align}
One choice of $c_1$ and $c_2$ deserves particular attention: when $c_1 = c_2 = c$, the connections simplify to
\begin{align}
    \Omega^i & = \frac{1}{2} H^i \cdot \Gamma_H - \im J_1^{-1} \left(\star_H d H^i\right), \\
    \omega & = - \Gamma_H + 2  d \log(\lambda) - \im H^i \cdot \star_H dH^i + 4 c H^i \cdot d \chi^i.
\end{align}
In this case we find that $\Omega^i$ only depends on $H^i$. Since $H^i$ contains a metric with determinant $-1$ along with a frame, it contains $10 -1 + 3 = 12$ components. The $H^i$ also satisfy the conformal metricity conditions
\begin{align}
    H^i \wedge H^j = 2 \im \delta^{ij} d^4 x,
\end{align}
where, compared with \cref{eq:metricity-cond}, the proportionality factor is now fixed. These are six conditions on $H^i$, leaving $18-6=12$ components, which matches the number of connection components in $\Omega^i$. In contrast, $\omega$ depends on the derivatives of all the fields. The conformal separation is therefore weaker, in that only one sector is independent of the other.

\subsubsection{Conformal Connection Equations of Motion}
We now derive evolution equations for the conformal connections. One might expect that substituting \cref{eq:nonlin-con-transform-a-eq-Omega} into \cref{eq:mod-nonlin-ricci-and-lorenz-cond} would suffice, but a lengthy calculation shows that lower-order terms mix the sectors and no separation occurs. Instead, we impose equations on the conformal connections directly, guided by their linear counterparts, \cref{eq:lin-12-con-evo,eq:lin-4-con-evo}, and then prove that they are equivalent to the original system on the constraint surface:
\begin{align}
    H^i_{\langle \mu}{}^\rho F{{(\Omega)}}^i_{\nu \rangle \rho} = 0, \quad g_H^{\mu\nu} \partial_\mu \Omega^i_\nu = 0, \label{eq:nonlin-12-con-evo-guess}\\
    H^{i\mu\nu} \partial_\mu \omega_\nu = 0, \quad g_H^{\mu\nu} \partial_\mu \omega_\nu = 0. \label{eq:nonlin-4-con-evo-guess}
\end{align}
Here $F{{(\Omega)}}^i = d\Omega^i + \frac{1}{2} \epsilon^{ijk} \Omega^j \wedge \Omega^k$ is the curvature of the new connection. We expect to modify these equations with lower-order terms, involving $\Omega^i$ and $\omega$, in order to recover the original equations of motion. The logic is to show that these equations, when written using $a^i$ and $\xi$, are Einstein's equations with additional terms proportional to the primary constraints $\xi$ or $\chi^i$.

We begin with the first equation in \cref{eq:nonlin-12-con-evo-guess} and use \cref{eq:nonlin-con-transform-Omega-eq-a} to find
\begin{align}
    F{(\Omega)}^i_{\mu\nu} = F{(A)}^i_{\mu\nu} - \partial^A_{[\mu} \left( \Sigma^i_{\nu]}{}^\rho \xi_\rho \right) + \frac{1}{2} \Sigma^i_{[\mu}{}^\rho \xi_{\nu]} \xi_\rho - \frac{1}{4} \Sigma^i_{\mu\nu} \xi^\rho \xi_\rho.
\end{align}
The equation of motion is then 
\begin{align}
    H^i_{\langle \mu}{}^\rho F{(\Omega)}^i_{\nu\rangle \rho} & = \Sigma^i_{\langle \mu}{}^\rho F{(\Omega)}^i_{\nu\rangle \rho} = \Sigma^i_{\langle \mu}{}^\rho F{(A)}^i_{\nu\rangle \rho} - \frac{1}{2} \Sigma^i_{\langle \mu}{}^\rho \partial^A_{\nu\rangle} \left( \Sigma^i_\rho{}^\sigma \xi_\sigma \right) + \frac{1}{2} \Sigma^i_{\langle \mu}{}^\rho \partial^A_\rho \left( \Sigma^i_{\nu \rangle}{}^\sigma \xi_\sigma \right) + \frac{1}{2} \xi_{\langle \mu} \xi_{\nu \rangle} \\ & = \Sigma^i_{\langle \mu}{}^\rho F{(A)}^i_{\nu\rangle \rho} + \partial_{\langle \mu} \xi_{\nu \rangle} + \frac{1}{2} \xi_{\langle \mu} \xi_{\nu \rangle} + \frac{1}{2} \xi_\sigma \left( \Sigma^i_{\langle \mu}{}^\rho \partial^A_{\nu\rangle} \Sigma^i_\rho{}^\sigma - \Sigma^i_{\langle \mu}{}^\rho \partial^A_{|\rho|} \Sigma^i_{\nu \rangle}{}^\sigma \right).
\end{align}
Performing a similar analysis for the second equation in \cref{eq:nonlin-12-con-evo-guess}, we find
\begin{align}
    \lambda g_H^{\mu\nu} \partial_\mu \Omega^i_\nu = g^{\mu\nu} \partial_\mu \Omega^i_\nu = g^{\mu\nu} \partial_\mu A^i_\nu - \frac{1}{2} g^{\mu\nu} \partial_\mu \left( \Sigma^i_\nu{}^\rho \xi_\rho \right) = g^{\mu\nu} \partial_\mu A^i_\nu - \frac{1}{2} \Sigma^{i\mu\nu} \partial_\mu \xi_\nu - \frac{1}{2} \xi_\rho \partial_\mu \Sigma^i_\nu{}^\rho = 0. \label{eq:nonlin-conf-Lorenz-i-gauge-tmp}
\end{align}
In both cases the result is the previously defined gauge condition plus additional terms proportional to the primary constraint, $\xi$. Hence, on the constraint surface $\xi_\mu = 0$, these equations of motion are equivalent. We return to hyperbolicity once all the equations of motion have been shown to be equivalent. So far, we know that the equations for $\Omega^i$ are equivalent to the Lorenz gauge and part of the Ricci-flat condition.

Turning to the equations in \cref{eq:nonlin-4-con-evo-guess}, we find that the first equation becomes
\begin{align}
    \lambda g_H^{\mu\nu} \partial_\mu \omega_\nu & = g^{\mu\nu} \partial_\mu \omega_\nu = g^{\mu\nu} \partial_\mu \xi_\nu + 2 g^{\mu\nu} \partial_\mu \left( \Sigma^i_\nu{}^\rho A^i_\rho \right) \\ & = g^{\mu\nu} \partial_\mu \xi_\nu + 2 \Sigma^{i\mu\nu} \partial_\mu A^i_\nu + 2 A^i_\rho g^{\mu\nu} \partial_\mu \Sigma^i_\nu{}^\rho.
\end{align}
The last term can be rewritten as
\begin{align}
    A^i_\rho g^{\mu\nu} \partial_\mu \Sigma^i_\nu{}^\rho = A^i_\rho g^{\mu\nu} \nabla^A_\mu \Sigma^i_\nu{}^\rho + \epsilon^{ijk} \Sigma^{i\mu\nu} A^j_\mu A^k_\nu + A^i_\rho \Gamma_\sigma \Sigma^{i\sigma\rho} + A^i_\rho \Sigma^{i\mu\nu} \Gamma^\rho_{\mu\nu}.
\end{align}
We now use the facts that $\nabla^A_\mu \Sigma^i_\nu{}^\rho \sim \nabla \chi^i$ and $\Gamma_\mu = - \xi_\mu - 2 c_2 \Sigma^i_\mu{}^\nu \nabla_\nu \chi^i$, so the first and third terms are zero on the constraint surface. The last term is zero by symmetry. We then write
\begin{align}
    A^i_\rho g^{\mu\nu} \partial_\mu \Sigma^i_\nu{}^\rho \simeq \epsilon^{ijk} \Sigma^{i\mu\nu} A^j_\mu A^k_\nu.
\end{align}
The equation of motion is then
\begin{align}
    \lambda g_H^{\mu\nu} \partial_\mu \omega_\nu & \simeq g^{\mu\nu} \partial_\mu \xi_\nu + 2 \Sigma^{i\mu\nu} \partial_\mu A^i_\nu + 2 \epsilon^{ijk} \Sigma^{i\mu\nu} A^j_\mu A^k_\nu \\ &  = g^{\mu\nu} \partial_\mu \xi_\nu + \Sigma^{i\mu\nu} F{(A)}^i_{\mu\nu} + \epsilon^{ijk} \Sigma^{i\mu\nu} A^j_\mu A^k_\nu.
\end{align}
This last term is not present in the original equations of motion and is not zero on the constraint surface; therefore, we must modify the equation of motion for $\omega$. One such modification is to add the following terms:
\begin{align}
    g_H^{\mu\nu} \partial_\mu \omega_\nu + \frac{1}{4} \omega^\mu \omega_\mu - \Omega^i_\mu \Omega^{i\mu} \simeq g^{\mu\nu} \partial_\mu \xi_\nu + \Sigma^{i\mu\nu} F{(A)}^i_{\mu\nu} - \frac{1}{2} \xi^\mu \xi_\mu.
\end{align}
Once again, the last term vanishes on the constraint surface, and what remains is the original equation of motion. Finally, for the second equation in \cref{eq:nonlin-4-con-evo-guess}, we find
\begin{align}
    \lambda H^{i\mu\nu} \partial_\mu \omega_\nu & = \Sigma^{i\mu\nu} \partial_\mu \omega_\nu = \Sigma^{i\mu\nu} \partial_\mu \xi_\nu + 2 \Sigma^{i\mu\nu} \partial_\mu \left( \Sigma^j_\nu{}^\rho A^j_\rho \right) = \Sigma^{i\mu\nu} \partial_\mu \xi_\nu + 2 \Sigma^{i\mu\nu} \partial^A_\mu \left( \Sigma^j_\nu{}^\rho A^j_\rho \right) \nonumber \\ & = \Sigma^{i\mu\nu} \partial_\mu \xi_\nu - 2 g^{\mu\nu} \partial_\mu A^i_\nu - 2\epsilon^{ijk} \Sigma^{j\mu\nu} \partial^A_\mu A^k_\nu + 2 A^j_\rho \Sigma^{i\mu\nu} \partial^A_\mu \Sigma^j_\nu{}^\rho \nonumber \\ & = \Sigma^{i\mu\nu} \partial_\mu \xi_\nu - 2g^{\mu\nu} \partial_\mu A^i_\nu - \epsilon^{ijk}\Sigma^{j\mu\nu} F^k_{\mu\nu} + \epsilon^{ijk} \Sigma^{j\mu\nu} \epsilon^{klm} A^l_\mu A^m_\nu + 2 A^j_\rho \Sigma^{i\mu\nu} \partial^A_\mu \Sigma^j_\nu{}^\rho.
\end{align}
To proceed we derive another useful relation,
\begin{align}
    \Sigma^{i\mu\nu} \partial^A_\mu \Sigma^j_\nu{}^\rho & = \Sigma^{i\mu\nu} \nabla^A_\mu \Sigma^j_\nu{}^\rho + \delta^{ij} g^{\mu\sigma} \Gamma_{\mu\sigma}^\rho \nonumber \\ & \simeq (\partial \chi) + \xi.
\end{align}
This term is therefore proportional to the primary constraints and their derivatives; consequently, it vanishes on the constraint surface. We can now make use of \cref{eq:nonlin-conf-Lorenz-i-gauge-tmp}, which implies that the first two terms vanish on the constraint surface. The result is
\begin{align}
     \lambda H^{i\mu\nu} \partial_\mu \omega_\nu \simeq \epsilon^{ijk} \Sigma^{j\mu\nu} F^k_{\mu\nu} + 2 A^i_\mu \Sigma^{j\mu\nu} A^j_\nu = 0.
\end{align}
The last term here contains the original connection; we therefore remove it by modifying the equation of motion. To do so, we note the identity
\begin{align}
    g^{\mu\nu} \omega_\mu \Omega^i_\nu & = 2 A^i_\mu \Sigma^{j\mu\nu} A^j_\nu + \epsilon^{ijk} \Sigma^{j\mu\nu} A^k_\mu \xi_\nu \\ & \simeq 2 A^i_\mu \Sigma^{j\mu\nu} A^j_\nu.
\end{align}
This affords us a simple way to remove the unwanted term. The result is then
\begin{align}
     \lambda H^{i\mu\nu} \partial_\mu \omega_\nu -g^{\mu\nu}\omega_{\mu} \Omega^i_\nu \simeq \epsilon^{ijk} \Sigma^{j\mu\nu} F^k_{\mu\nu} = 0.
\end{align}
This is equivalent to the antisymmetric part of \cref{eq:mod-nonlin-ricci-and-lorenz-cond}.

Note that no combination of $\omega$ alone can be used to modify \cref{eq:nonlin-4-con-evo-guess} so that it is equivalent on the constraint surface; some terms involving $\Omega^i$ are needed. This is easily checked by constructing all the quadratic terms involving $\omega$ and showing that none of them removes the unwanted terms.

\subsubsection{Conformal Equations of Motion}
The final system of equations of motion, which are equivalent to Einstein's equations on the constraint surface, is given by
\begin{gather}
    \Omega^i = \frac{1}{2} H^i \cdot \Gamma_H - \im J_1^{-1}\left( \star_H d H^i \right) + (c_1 - c_2) d\chi^i, \label{eq:nonlin-12-con-def} \\
    H^i_{\langle \mu}{}^\rho F{(\Omega)}^i_{\nu\rangle \rho} = 0, \quad g_H^{\mu\nu} \partial_\mu \Omega^i_\nu = 0, \label{eq:nonlin-12-con-evo} \\
    \omega = - \Gamma_H + 2 d \log(\lambda) - \im H^i \cdot \star_H d H^i + 2(c_1 + c_2) H^i \cdot d\chi^i, \label{eq:nonlin-4-con-def}\\
    g_H^{\mu\nu} \left( \partial_\mu \omega_\nu + \frac{1}{4} \omega_\mu \omega_\nu - \Omega^i_\mu \Omega^i_\nu \right) = 0, \quad H^{i\mu\nu} \partial_\mu \omega_\nu - g_H^{\mu\nu} \omega_\mu \Omega^i_\nu = 0. \label{eq:nonlin-4-con-evo}
\end{gather}
Remarkably, this system can be written entirely in terms of the new fields. By choosing $c_1=c_2=c$ the first two lines involve only the 12 components of $H^i$ and $\Omega^i$; the last two involve connections from both sectors but, among the frame fields, only $H^i$. The nonlinear separation is therefore one-way: the 12-component sector evolves independently, while the 4-component sector depends on everything. Since this system has the same principal part as the linear system of \cref{eq:lin-12-con-evo,eq:lin-4-con-evo}, it is also locally well-posed.

\subsection{Connection Definitions as Evolution Equations}\label{subsec:con-defs-as-evo-eqs}
In their current form the definitions of the connections do not appear as evolution equations. Here we rewrite them in components so that the corresponding evolution equations are easier to derive. In doing so we introduce the covariant versions of the basic variables in this setup.

\subsubsection{Original Connections}
We begin with the definitions of the original connections in \cref{eq:mod-nonlin-con-defs}. Taking the first equation in this system, we can use the definition of the Hodge star and apply the operator $\im J_1$ to show that
\begin{align}
   \im\epsilon^{ijk} \Sigma^j_\mu{}^\nu A^k_\nu = \frac{1}{2} \epsilon_\mu{}^{\nu \rho\sigma} \partial_\nu \Sigma^i_{\rho\sigma}  + \im (c_1 + c_2) \epsilon^{ijk} \Sigma^j_\mu{}^\nu \partial_\nu \chi^k+ 2 \im c_2 \partial_\mu \chi^i.
\end{align}
It will be useful to raise the $\mu$ index and multiply by $\sqrt{-g}$ to find
\begin{align}
   \im \epsilon^{ijk} \tsd^{i\mu\nu} A^k_\nu = \frac{1}{2}\teps^{\mu\nu\rho\sigma} \partial_\nu \Sigma^i_{\rho\sigma} + \im (c_1 + c_2) \epsilon^{ijk} \tsd^{j\mu\nu} \partial_\nu \chi^k + 2 \im c_2 \tg^{\mu \nu} \partial_\nu \chi^i.
\end{align}
It is now useful to introduce the notation $\tE^{i\mu\nu} = \frac{1}{2} \tilde{\epsilon}^{\mu\nu\rho\sigma} \Sigma^i_{\rho\sigma}$, for which we also find $\tE^{i\mu\nu} = \im \tilde{\Sigma}^{i\mu\nu}$. This, together with the fact that $\teps^{\mu\nu\rho\sigma} = \pm 1$ (so that it commutes with derivatives), simplifies the above to
\begin{align}
    \epsilon^{ijk} \tE^{j\mu\nu} A^k_\nu = \partial_\nu \tE^{i \mu\nu} + (c_1 + c_2) \epsilon^{ijk} \tE^{j\mu\nu} \partial_\nu \chi^k + 2 \im c_2 \tg^{\mu\nu} \partial_\nu \chi^i.   
\end{align}
Using the definition of $\partial^A_\mu$, we can combine the l.h.s.\ with the first term on the r.h.s.\ to find
\begin{align}
    \partial^A_\nu \tE^{i\mu\nu} + (c_1 + c_2) \epsilon^{ijk} \tE^{j\mu\nu} \partial_\nu \chi^k + 2 \im c_2 \tg^{\mu\nu} \partial_\nu \chi^i = 0.\label{eq:nonlin-A-con-def-comps}
\end{align}
For the second equation in \cref{eq:mod-nonlin-con-defs}, we first recall the definition of the harmonic vector in terms of derivatives of $g$,
\begin{align}
    \Gamma_\mu = g^{\rho\sigma} g_{\mu\nu} \Gamma^\nu_{\rho\sigma} = - \utg_{\mu\nu} \partial_\rho \tg^{\nu\rho},
\end{align}
where $\utg_{\mu\nu} = g_{\mu\nu}/\sqrt{-g}$. The connection definition becomes
\begin{align}
    \xi_\mu = \utg_{\mu\nu} \partial_\rho \tg^{\nu\rho} - 2 c_2 \Sigma^i_\mu{}^\nu \partial_\nu \chi^i.
\end{align}
Raising the free index, multiplying by $\sqrt{-g}$ and making use of the definition of $\tE$, we can write this as
\begin{align}
 \partial_\nu \tg^{\mu \nu} - \tg^{\mu\nu} \xi_\nu + 2 \im c_2 \tE^{i \mu\nu} \partial_\nu \chi^i = 0.\label{eq:nonlin-xi-con-def-comps}
\end{align}
In both equations the frame-related terms can be written using only $\tg^{\mu\nu}$, $\tE^{i\mu\nu}$ and $\chi^i$; this will be useful later when we compute the 3+1 decomposition of the equations. The equations in \cref{eq:mod-nonlin-ricci-and-lorenz-cond} can also be written using only these fields by raising an index and/or multiplying by $\sqrt{-g}$ in the appropriate way. The result is
\begin{align}
    \tE^{i\mu\rho} F^i_{\nu\rho} + \im \tg^{\mu\rho} \partial_{(\nu} \xi_{\rho)} = 0, \\
    \tg^{\mu\nu} \partial_\mu A^i_\nu + \frac{\im}{2} \tE^{i\mu\nu} \partial_\mu \xi_\nu = 0.
\end{align}

Placing the equations of motion into this form will make the 3+1 decomposition easier to perform in the later sections.
   
\section{Gauge-Fixed 3+1 Evolution Systems}\label{sec:gauge-fixed-3+1-evolution-systems}

We now apply the spacetime decomposition of \cref{sec:ashtekar-formalism} to the gauge-fixed system, in both its hyperbolic and conformal forms. Since these systems have no Lagrangian, we decompose the equations of motion directly. For reference, the equations derived in \cref{sec:gauge-fixing} are, for the hyperbolic system,

\begin{equation}\label{eq:nonlin-hyperbolic-comp-evo-sys}
    \begin{aligned}
    \partial^A_\nu \tE^{i\mu\nu} + (c_1 + c_2) \epsilon^{ijk} \tE^{j\mu\nu} \partial_\nu \chi^k + 2 \im c_2 \tg^{\mu\nu} \partial_\nu \chi^i = 0, \\
    \partial_\nu \tg^{\mu\nu} - \tg^{\mu\nu} \xi_\nu + 2 \im c_2 \tE^{i\mu\nu} \partial_\nu \chi^i = 0,\\
    \tE^{i\mu\rho} F^i_{\nu\rho} + \im \tg^{\mu\rho} \partial_{(\nu} \xi_{\rho)} = 0, \\
        \tg^{\mu\nu} \partial_\mu A^i_\nu + \frac{\im}{2} \tE^{i\mu\nu} \partial_\mu \xi_\nu = 0,
\end{aligned}
\end{equation}

and, for the conformal system,

\begin{equation}\label{eq:nonlin-conformal-comp-evo-sys}
\begin{aligned}
    g_H^{\rho\sigma} \partial^\Omega_\rho H^{i\mu}_\sigma + (c_1 - c_2) \epsilon^{ijk} H^{j\mu\nu} \partial_\nu \chi^k =0,\\
    2 g_H^{\mu\nu} \partial_\nu \ln(\lambda) + H^{i\mu}_\rho \partial_\sigma H^{i\rho\sigma} + 2(c_1 + c_2) H^{i\mu\nu} \partial_\nu \chi^i - g_H^{\mu\nu} \omega_\nu  = 0, \\
    H^i{}_{\langle \mu}{}^\rho F{(\Omega)}^i_{\nu\rangle \rho} = 0, \quad g_H^{\mu\nu} \partial_\mu \Omega^i_\nu = 0, \\
    g_H^{\mu\nu} \left( \partial_\mu \omega_\nu + \frac{1}{4} \omega_\mu \omega_\nu - \Omega^i_\mu \Omega^i_\nu \right) = 0, \quad H^{i\mu\nu} \partial_\mu \omega_\nu - g_H^{\mu\nu} \omega_\mu \Omega^i_\nu = 0.
\end{aligned}
\end{equation}
     
\subsection{\texorpdfstring{$(\alpha,\beta,c_1,c_2)$}{(ɑ,β,c1,c2)} Hyperbolic System}\label{subsec:hyperbolic-3+1-system}
Recall from \cref{sec:ashtekar-formalism} that, in the foliated coordinates $x^\mu = (t,x^a)$, the tensors decompose as follows:

\begin{align}\label{eq:pleb-E-3+1-decomp}
    \tE^{i0a} &= \tE^{ia}, \quad \tE^{iab} = -\im \epsilon^{ijk} \dlap \tE^{ja} \tE^{kb} - N^a \tE^{ib} + N^b \tE^{ia}, \\
    \tg^{00} &= -\frac{1}{\dlap}, \quad \tg^{0a} = \frac{N^a}{\dlap}, \quad \tg^{ab} = \dlap \tE^{ia} \tE^{ib} - \frac{N^a N^b}{\dlap}.
\end{align}

The decomposition process is described in more detail in \cref{apx:3+1-eom-split}; here we simply state and discuss the results. In addition, we include constraint damping terms for $\chi^i$ and for $\xi_0,\xi_a$, with parameters $\alpha$ and $\beta$, respectively. The resulting evolution equations are

\begin{align}
    (\partial^A_t - N^b \partial^A_b) \tE^{ia} = & - \im \epsilon^{ijk} \tE^{kb} \partial_b^A \left(\dlap \tE^{ja}\right) - \tE^{ib} \partial_b N^a + \tE^{ia} \partial_b N^b + \im (c_1 + c_2) \dlap \tE^{ia} \tE^{jb} \partial_b \chi^j \nonumber \\ & + \frac{c_2-c_1}{2 \im c_2}\left( \epsilon^{ijk} \dlap \tE^{ja} \partial_b^A \tE^{kb} + (c_1+c_2) \dlap \tE^{ja} \tE^{ib} \partial_b \chi^j \right. \nonumber \\ & \left. \hspace{150pt} - (3c_2 + c_1) \dlap \tE^{ja} \tE^{jb} \partial_b \chi^i \right), \label{eq:num-plebanski-triad-evolution-equation} \\
    (\partial_t - N^b \partial_b) N^a = & - \dlap \tE^{ib} \partial_b^A( \dlap \tE^{ia}) - (c_2 - c_1) \epsilon^{ijk} \dlap^2 \tE^{ja} \tE^{kb} \partial_b \chi^i + \dlap^2 \tE^{ia} \tE^{ib} \xi_b, \label{eq:num-plebanski-shift-evolution-equation} \\
    (\partial_t - N^a \partial_a)\dlap = & - \dlap \partial_a N^a - 2 \im c_2 \dlap^2 \tE^{ia} \partial_a \chi^i - \dlap ( \xi_0 - N^a \xi_a), \label{eq:num-plebanski-lapse-evolution-equation} \\
    (\partial_t - N^a \partial_a) \chi^i = & \frac{1}{2 \im c_2} \dlap \partial^A_a \tE^{ia} + \frac{1}{2 \im c_2} (c_1 + c_2) \epsilon^{ijk} \dlap \tE^{ja} \partial_a \chi^k + \alpha \chi^i, \label{eq:num-plebanski-internal-vector-evolution} \\
    (\partial_t - N^b \partial_b) A^i_a & = \partial^A_a A^i_0 - N^b \partial^A_a A^i_b + \im \epsilon^{ijk} \dlap \tE^{jb} F^k_{ab} - \im \dlap \tE^{ib} \partial_{(a} \xi_{b)}, \\
    (\partial_t - N^a \partial_a) A^i_0 & = \ (\dlap^2 \tE^{ja} \tE^{jb} - N^a N^b) \partial_a A^i_b + N^a \partial^A_a A^i_0 - \im \epsilon^{ijk} \dlap \tE^{ja} N^b F^k_{ab} \nonumber \\ &\ + \dlap^2 \tE^{ia} \tE^{jb} F^j_{ab} - \im \dlap \tE^{ia} \partial_a \xi_0 - \im N^a \dlap \tE^{ib} \partial_{[a} \xi_{b]} \nonumber \\ &\ + \frac{1}{2} \epsilon^{ijk} \dlap^2 \tE^{ja} \tE^{kb} \partial_a \xi_b, \\
    (\partial_t - N^a \partial_a)\xi_0 & = - N^a \partial_a \xi_0 - (\dlap^2 \tE^{ia} \tE^{ib} - N^a N^b) \partial_a \xi_b \nonumber \\ &\  -2 \im \dlap N^a \tE^{ib} F^i_{ab} + \epsilon^{ijk} \dlap^2 \tE^{ia} \tE^{jb} F^k_{ab} + \beta \xi_0, \\
    (\partial_t - N^b \partial_b)\xi_a & = 2 \im \dlap \tE^{ib} F^i_{ba} - \partial_a \xi_0 + N^b \partial_a \xi_b + \beta \xi_a.
\end{align}

We now compare the frame equations with the original Ashtekar system, \cref{eq:num-ashtekar-triad-evolution-equation}. For the gauge-fixed system to describe the same physics, the Gauss constraint

\begin{align}
    \mathcal{G}^i = \partial^A_a \tE^{ia} \simeq 0
\end{align}

must vanish during evolution.

The field $\chi^i$ is the conjugate momentum of $A^i_0$, which is the Lagrange multiplier for the Gauss constraint, so $\chi^i$ is a primary constraint and should vanish. When $\chi^i = 0$, \cref{eq:num-plebanski-internal-vector-evolution} implies $\mathcal{G}^i = 0$, and the modifications to the triad equation drop out, recovering Ashtekar's equation. Moreover, $\chi^i$ evolves according to the Gauss constraint and its own derivatives, so if both vanish initially then $\partial_t \chi^i |_{t=0} = 0$ and $\chi^i$ remains zero. When the constraints cannot be satisfied exactly initially, we must check that violations are damped; this is done in \cref{sec:hyp-and-con-anal-of-evo-sys}. Although the system was not derived from a Hamiltonian, this argument still holds.

The lapse and shift are pure gauge, so their evolution equations can be chosen freely. Our choice is a modification of harmonic slicing; like any slicing condition, it changes only the coordinates and has no physical effect.

The last four equations evolve the connections. Before gauge fixing, the connection sector was subject to four constraints, the Hamiltonian and momentum constraints,

    \begin{align}
        \mathcal{H} = \epsilon^{ijk} \tE^{ia} \tE^{jb} F^k_{ab}, \quad \mathcal{M}_a = \tE^{ib} F^i_{ba}.
    \end{align}

The components $\xi_0$ and $\xi_a$ are the conjugate momenta of the lapse and shift, and hence primary constraints that should vanish. The equation for $A^i_a$ differs from Ashtekar's only by derivatives of $\xi_a$, so it reproduces the original dynamics on the constraint surface. The equation for $A^i_0$ is pure gauge: since $A^i_0$ is the Lagrange multiplier for the Gauss constraint, its evolution moves the connection $A^i_\mu$ along its gauge orbit without changing the physical solution. Finally, $\xi_0$ and $\xi_a$ evolve proportionally to the constraints and their derivatives. Every modification is thus either pure gauge or proportional to first-class constraints, so the physical dynamics are unchanged on the constraint surface, and constraints that vanish initially remain zero. Whether the constraint surface is also attractive is examined in \cref{sec:hyp-and-con-anal-of-evo-sys}. Before doing so, we consider the 3+1 decomposition of the conformal system.

The additional constraint terms do not affect the physical dynamics, since they are proportional to first-class constraints that should vanish on the constraint surface. Provided that the system tends towards the constraint surface during numerical evolution, this modification is allowed.

\subsection{3+1 Conformal System}
    We now perform the same computation for the conformal system \cref{eq:nonlin-conformal-comp-evo-sys}. Before tackling the equations of motion, we need to introduce the 3+1 decomposition of the conformal fields.
    \subsubsection{Conformal Fields 3+1 Decomposition}\label{subsec:conf-field-3+1-decomp}
    With the original fields already decomposed, we can now introduce a 3+1 decomposition of the conformal fields. Since these fields are new, we will spend some time explaining how one arrives at their 3+1 decomposition and at the appropriate variables. As discussed in \cref{subsec:nonlin-gauge:-nonlin-conf-sep}, the basic fields that describe the conformal frame are $H^i$ and $\lambda$. Using $\tilde{g}^{\mu\nu} = (\sqrt{-g} g^{-1})^{\mu\nu}$ together with the decomposition in \cref{subsec:ashtekar:3+1-decomposition}, we can, by taking the determinant, identify

    \begin{align}
        \lambda = \frac{1}{\sqrt{\sqrt{-g}}} = \frac{1}{\sqrt{\dlap \det(\tE)}}.
    \end{align}

    As for the original frame, we decompose the bivector $H^{i\mu\nu}$ rather than the 2-form, since it appears more naturally in the equations of motion. We have

    \begin{align}
        H^{i\mu\nu} = \frac{1}{\lambda} \Sigma^{i\mu\nu} = \frac{1}{\lambda} \frac{-\im}{2} \epsilon^{\mu\nu\rho\sigma} \Sigma^i_{\rho\sigma} = \frac{1}{\lambda \sqrt{-g}} \frac{-\im}{2} \teps^{\mu\nu\rho\sigma} \Sigma^i_{\rho\sigma} = -\im \lambda \tE^{i\mu\nu}.
    \end{align}

    Using \cref{eq:3+1-decomp-tE-spacetime}, we split this into

    \begin{align}
        H^{i\mu\nu} = \frac{-\im}{\sqrt{\dlap \det(\tE)}} \left( - \im \dlap \epsilon^{ijk} \tE^{j\mu} \tE^{k\nu} - N^{\mu} \tE^{i\nu} + N^{\nu} \tE^{i\mu} \right).
    \end{align}

    Usually the triple of 2-forms would define a metric and an $SO(3,\C)$ frame, which amounts to $10+3=13$ components. However, since the constructed metric is unimodular, we are left with only $(10-1)+3 = 9 + 3 = 12$ components. Therefore, the 2-forms (bivectors) should encode a 9-component object and a 3-component object. With this in mind, we rewrite the bivector as

    \begin{align}
        H^{i\mu\nu} = \frac{-\im}{\sqrt{\dlap \det(\tE)}} \frac{1}{\dlap} \left( - \im \epsilon^{ijk} \dlap \tE^{j\mu} \dlap \tE^{k\nu} - N^{\mu} \dlap \tE^{i\nu} + N^{\nu} \dlap \tE^{i\mu} \right).
    \end{align}

    Here we have multiplied by $\dlap/\dlap = 1$ to introduce extra factors of the densitised lapse. We now note a common term:

    \begin{align}
        H^{i\mu} = \dlap \tE^{i\mu}.
    \end{align}

    In the foliated coordinates, we find

    \begin{align}
        H^{i0} = 0, \quad H^{ia} = \dlap \tE^{ia}.
    \end{align}

    Its determinant is therefore

    \begin{align}
        \det(H) = \frac{1}{6} \epsilon^{ijk} \utilde{\epsilon}_{abc} H^{ia} H^{jb} H^{kc} = \dlap^3 \det(\tE).
    \end{align}

    The bivector can then be expressed entirely in terms of $H^{i\mu}$ and $N^\mu$:

    \begin{align}
        H^{i\mu\nu} = \frac{1}{\sqrt{\det(H)}} \left( -\epsilon^{ijk} H^{j\mu} H^{k\nu} + \im N^\mu H^{i\nu} - \im N^\nu H^{i\mu} \right).
    \end{align}

    Since $H^{i\mu}$ contains only the 9 components of $H^{ia}$, and $N^\mu = (1,-N^a)$ only 3, $H^{i\mu\nu}$ contains 12 in total, which is the correct number for a unimodular metric and an $SO(3,\C)$ frame. To reinforce this point, we compute the decomposition of the inverse densitised metric,

    \begin{align}
        \sqrt{-g_H} g_H^{\mu\nu} = g_H^{\mu\nu} = \frac{1}{\lambda} g^{\mu\nu} = \lambda \tg^{\mu\nu}.
    \end{align}

    Hence, the decomposition becomes

    \begin{align}
        \sqrt{-g_H} g_H^{\mu\nu} = \frac{1}{\sqrt{\dlap \det(\tE)}} \left( \dlap \tE^{i\mu} \tE^{i\nu} - \frac{N^\mu N^\nu}{\dlap} \right) = \frac{1}{\sqrt{\det(H)}} (H^{i\mu} H^{i\nu} - N^\mu N^\nu).
    \end{align}

    Again, only $H^{ia}$ and $N^a$ appear in the decomposition.

    Overall, the 3+1 decomposition of the conformal frame can be summarised as

    \begin{gather}
        H^{i0a} = \frac{\im H^{ia}}{\sqrt{\det(H)}}, \quad H^{iab} = \frac{1}{\sqrt{\det(H)}} ( -\epsilon^{ijk} H^{ja} H^{kb} - \im N^a H^{ib} + \im N^b H^{ia} ), \\
        g_H^{00} = -\frac{1}{\sqrt{\det(H)}},\quad g_H^{0a} = g_H^{a0} = \frac{N^a}{\sqrt{\det(H)}}, \quad g_H^{ab} = \frac{1}{\sqrt{\det(H)}} ( H^{ia} H^{ib} - N^a N^b).
    \end{gather}

    The endomorphisms $\sd^i_\mu{}^\nu$ and $H^i_\mu{}^\nu$ are equal because they are conformally invariant. Therefore, the decomposition \cref{eq:ashtekar:trip-endo-3+1} is easily converted to the conformal fields, giving

    \begin{gather}
        H^i_0{}^0 = \im H^i_a N^a, \quad H^i_a{}^0 = \im H^i_a, \quad H^i_0{}^a = \im H^{ia} - \im N^a H^i_b N^b - \epsilon^{ijk} H^j_b N^b H^{ka}, \\
        H^i_a{}^b = - \im H^i_a N^b - \epsilon^{ijk} H^j_a H^{kb}.
    \end{gather}

    Here $H^i_a$ is the inverse of $H^{ia}$. For the conformal connections, we can use this frame decomposition together with their definitions \cref{eq:nonlin-con-transform-Omega-eq-a}, which in components read

    \begin{align}
        \Omega^i_\mu = A^i_\mu - \frac{1}{2} \sd^i_\mu{}^\nu \xi_\nu, \quad \omega_\mu = \xi_\mu + 2 \sd^i_\mu{}^\nu A^i_\nu.
    \end{align}

    Decomposing these is relatively simple and gives
    
    \begin{align}\label{eq:gauge-3+1:omega-to-a}
        \Omega^i_N &= A^i_N - \frac{\im}{2} H^{ia} \xi_a, \\
        \Omega^i_a &= A^i_a + \frac{1}{2} \epsilon^{ijk} H^j_a H^{kb} \xi_b - \frac{1}{2} \im H^i_a \xi_N, \\
        \omega_N   &= \xi_N + 2 \im H^{ia} A^i_a, \quad \omega_a = \xi_a - 2\epsilon^{ijk} H^i_a H^{jb} A^k_b + 2 \im H^i_a A^i_N.
    \end{align}

    Here $\Omega^i_N = N^\mu \Omega^i_\mu = \Omega^i_0 - N^a \Omega^i_a$, and similarly for $A^i_N, \xi_N$ and $\omega_N$. The $\mu = 0$ components of the connections are recovered by taking appropriate linear combinations of the above results. The inverse of this change of variables is

    \begin{align}
        A^i_N &=  \frac{1}{2} \Omega^i_N - \frac{\im}{2} \epsilon^{ijk} H^{ja} \Omega^k_a - \frac{\im}{4} H^{ia} \omega_a, \label{eq:gauge-3+1:a-to-omega-1}\\
        A^i_a &= \frac{1}{2} \Omega^i_a - \frac{\im}{2} \epsilon^{ijk} H^j_a \Omega^k_N + \frac{1}{2} H^j_a H^{ib} \Omega^j_b - \frac{1}{2} H^i_a H^{jb} \Omega^j_b - \frac{\im}{4} H^i_a \omega_N + \frac{1}{4} \epsilon^{ijk} H^j_a H^{kb} \omega_b,\label{eq:gauge-3+1:a-to-omega-2}\\ 
        \xi_N &= \im H^{ia} \Omega^i_a - \frac{1}{2} \omega_N, \quad \xi_a = \im H^i_a \Omega^i_N -\epsilon^{ijk} H^j_a H^{kb} \Omega^i_b - \frac{1}{2} \omega_a.
\label{eq:gauge-3+1:a-to-omega-3}
    \end{align}

    The curvature of $\Omega^i$ decomposes in the same way as that of $A^i$:

    \begin{gather}
        F{(\Omega)}^i_{0a} = \partial_t \Omega^i_a - \partial^\Omega_a \Omega^i_0, \quad F(\Omega)^i_{ab} = \partial_a \Omega^i_b - \partial_b \Omega^i_a + \epsilon^{ijk} \Omega^j_a \Omega^k_b.
    \end{gather}

    Instead of the densitised lapse $\dlap$, we use the variable $n = \log(\dlap)$, in terms of which the subsequent equations are simpler. The complete set of conformal variables is then

    \begin{align}\label{eq:nonlin-conf-variables-set}
        H^{ia},\quad N^a, \quad \chi^i, \quad n, \\
        \Omega^i_a, \quad \Omega^i_0,\quad \omega_a, \quad \omega_0.
    \end{align}

    With these variables, we can now derive the evolution equations in the conformal case.

    \subsubsection{Frame Evolution Equations}

        To produce evolution equations for the conformal frame we take the previous 3+1 evolution equations \cref{eq:num-plebanski-triad-evolution-equation,eq:num-plebanski-shift-evolution-equation,eq:num-plebanski-lapse-evolution-equation,eq:num-plebanski-internal-vector-evolution} and translate them using the results of the previous section. The connection definitions in \cref{eq:nonlin-12-con-def,eq:nonlin-4-con-def} then become the frame evolution equations

        \begin{align}
            (\partial_t^\Omega - N^b \partial^\Omega_b) H^{ia} = & -\im \epsilon^{ijk} H^{kb} \partial_b^\Omega H^{ja} - H^{ib} \partial_b N^a \nonumber \\ & + \frac{c_2 - c_1}{2 \im c_2} \left( \epsilon^{ijk} H^{ja} \partial_b^A H^{kb} - \epsilon^{ijk} H^{ja} H^{kb} \partial_b n + (c_1 + c_2) H^{ja} H^{ib} \partial_b \chi^j \right. \nonumber \\ 
             & \quad\quad \left.- (3 c_2 + c_1) H^{ja} H^{jb} \partial_b \chi^i  + 2 c_2 H^{ia} H^{jb} \partial_b \chi^j \right), \\
            (\partial_t - N^b \partial_b) N^a = & - H^{ib} \partial_b^\Omega H^{ia} + (c_1 - c_2) \epsilon^{ijk} H^{ja} H^{kb} \partial_b \chi^i, \\
            (\partial_t - N^a \partial_a) \chi^i = & \frac{1}{2\im c_2} \partial^\Omega_a H^{ia} - \frac{1}{2\im c_2} H^{ia} \partial_a n + \frac{c_1 + c_2}{2 \im c_2} \epsilon^{ijk} H^{ja} \partial_a \chi^k, \\
            (\partial_t - N^a \partial_a) n = & - \partial_a N^a - 2\im c_2 H^{ia} \partial_a \chi^i - \im H^{ia} \Omega^i_a + \frac{1}{2}( \omega_0 - N^a \omega_a).
        \end{align}

        Here the covariant derivatives involving the original connection $A$, i.e.\ in $\partial^A_b H^{ia}$, should be replaced by their expressions in terms of the conformal connections using \cref{eq:gauge-3+1:a-to-omega-1,eq:gauge-3+1:a-to-omega-2,eq:gauge-3+1:a-to-omega-3}. As expected from the previous analysis, the system does not separate into sectors of sizes 12 and 4. For the separation to occur we must choose $c_1 = c_2 = c$, in which case the system simplifies considerably to

        \begin{align}
            (\partial_t^\Omega - N^b \partial^\Omega_b) H^{ia} = & -\im \epsilon^{ijk} H^{kb} \partial_b^\Omega H^{ja} - H^{ib} \partial_b N^a, \\
            (\partial_t - N^b \partial_b) N^a = & - H^{ib} \partial_b^\Omega H^{ia}, \\
            (\partial_t - N^a \partial_a) \chi^i = & \frac{1}{2\im c} \partial^\Omega_a H^{ia} - \frac{1}{2\im c} H^{ia} \partial_a n - \im \epsilon^{ijk} H^{ja} \partial_a \chi^k, \\
            (\partial_t - N^a \partial_a) n = & - \partial_a N^a - 2\im c H^{ia} \partial_a \chi^i - \im H^{ia} \Omega^i_a + \frac{1}{2}( \omega_0 - N^a \omega_a).
        \end{align}

        The result is much simpler than before; we also recover a 12-component system ($H^{ia}, N^a$) that evolves independently of the remaining components ($\chi^i, n$). The first line closely resembles the chiral Yang-Mills evolution equations; see eq.~(19) of~\cite{WeylCurvatureKrasno2022}.

    \subsubsection{Connection Evolution Equations}

    The last set of evolution equations is that for the conformal connections. As with the previous fields, a detailed decomposition can be found in \cref{apx:3+1-eom-split}. The result is

    \begin{align}
        (\partial_t - N^b \partial_b) \Omega^i_a =\ & \partial^\Omega_a \Omega^i_0 - N^b \partial^\Omega_a \Omega^i_b - \frac{\im}{2} H^j_a \epsilon^{jkl} H^{kb} H^{lc} F{(\Omega)}^i_{bc}, \\
        (\partial_t - N^a \partial_a) \Omega^i_0  =\ & N^a \partial_a \Omega^i_0 + ( H^{ja} H^{jb} - N^a N^b ) \partial_a \Omega^i_b \nonumber \\ & - \frac{\im}{2} N^a H^j_a \epsilon^{jkl} H^{kb} H^{lc} F{(\Omega)}^i_{bc} + \epsilon^{ijk} N^a \Omega^j_a \Omega^k_0, \\
    (\partial_t - N^b \partial_b) \omega_a =\ & \partial_a \omega_0 + N^b \partial_a \omega_b - \im \epsilon^{ijk} H^i_a H^{jb} H^{kc} \partial_b \omega_c + \im H^i_a \omega_N \Omega^i_N - \im H^i_a H^{jb} H^{jc} \omega_b \Omega^i_c, \\
    (\partial_t - N^a \partial_a )\omega_0 =\ & N^a \partial_a \omega_0 + N^a N^b \partial_a \omega_b - \im \epsilon^{ijk} N^a H^i_a H^{jb} H^{kc} \partial_b \omega_c + H^{ja} H^{jb} \partial_a \omega_b \nonumber \\ & \ + H^{ja} H^{jb} \left( \frac{1}{4} \omega_a \omega_b - \Omega^i_a \Omega^i_b \right) + \im N^a H^i_a \omega_N \Omega^i_N - \im N^a H^i_a H^{jb} H^{jc} \omega_b \Omega^i_c - \frac{1}{4} \omega_N \omega_N + \Omega^i_N \Omega^i_N.
    \end{align}

    Here the partial separation is available independently of the choice of $c_1, c_2$. Indeed, the evolution equations for $\Omega^i_a$ and $\Omega^i_0$ depend only on themselves and on the 12 independent components of the frame. The evolution equations for the four components of $\omega$ depend on $\Omega^i$ only through terms quadratic in the connections.

    \newpage
    \subsection{\texorpdfstring{$(c,\alpha)$}{(c, α)} Conformal System}\label{subsec:gauge-fixed-3p1-systems:conformal-system}

    In this subsection we collect the evolution equations of the conformal system that we use in the numerical simulations. The system is

    \begin{align}
        (\partial_t^\Omega - N^b \partial^\Omega_b) H^{ia} = & -\im \epsilon^{ijk} H^{kb} \partial_b^\Omega H^{ja} - H^{ib} \partial_b N^a, \\
        (\partial_t - N^b \partial_b) N^a = & - H^{ib} \partial_b^\Omega H^{ia}, \\
        (\partial_t - N^b \partial_b) \Omega^i_a =\ & \partial^\Omega_a \Omega^i_0 - N^b \partial^\Omega_a \Omega^i_b - \frac{\im}{2} H^j_a \epsilon^{jkl} H^{kb} H^{lc} F{(\Omega)}^i_{bc}, \\
        (\partial_t - N^a \partial_a) \Omega^i_0  =\ & N^a \partial_a \Omega^i_0 + ( H^{ja} H^{jb} - N^a N^b ) \partial_a \Omega^i_b \nonumber \\ & - \frac{\im}{2} N^a H^j_a \epsilon^{jkl} H^{kb} H^{lc} F{(\Omega)}^i_{bc} + \epsilon^{ijk} N^a \Omega^j_a \Omega^k_0, \\[10pt]
        (\partial_t - N^a \partial_a) \chi^i = & \frac{1}{2\im c} \partial^\Omega_a H^{ia} - \frac{1}{2\im c} H^{ia} \partial_a n - \im \epsilon^{ijk} H^{ja} \partial_a \chi^k + \alpha \chi^i, \\
        (\partial_t - N^a \partial_a) n = & - \partial_a N^a - 2\im c H^{ia} \partial_a \chi^i - \im H^{ia} \Omega^i_a + \frac{1}{2}( \omega_0 - N^a \omega_a), \\
        (\partial_t - N^b \partial_b) \omega_a =\ & \partial_a \omega_0 + N^b \partial_a \omega_b - \im \epsilon^{ijk} H^i_a H^{jb} H^{kc} \partial_b \omega_c + \im H^i_a \omega_N \Omega^i_N - \im H^i_a H^{jb} H^{jc} \omega_b \Omega^i_c, \\
        (\partial_t - N^a \partial_a )\omega_0 =\ & N^a \partial_a \omega_0 + N^a N^b \partial_a \omega_b - \im \epsilon^{ijk} N^a H^i_a H^{jb} H^{kc} \partial_b \omega_c + H^{ja} H^{jb} \partial_a \omega_b \nonumber \\ & \ + H^{ja} H^{jb} \left( \frac{1}{4} \omega_a \omega_b - \Omega^i_a \Omega^i_b \right) + \im N^a H^i_a \omega_N \Omega^i_N - \im N^a H^i_a H^{jb} H^{jc} \omega_b \Omega^i_c - \frac{1}{4} \omega_N \omega_N + \Omega^i_N \Omega^i_N.
    \end{align}

    The two sectors are now manifestly separate: one contains 12 components of the frame and connection, the other 4. Only the $\chi^i$ equation carries a damping term. Converting the damping terms of the hyperbolic system to conformal variables would introduce $\omega$ into the $\Omega^i$ equations and destroy the separation, while adding damping terms built directly from the new connections would generate $A^i$-dependent terms that break the equivalence with Einstein's equations. Hence, we take the above system as our conformal system.

    \subsection{Original System}\label{subsec:original-system}

    To finish this section and complete the list of evolution equations, we summarise what we call the original system. Its equations are simply the evolution equations derived in \cref{sec:ashtekar-formalism}:

    \begin{align}
        \partial_t \tE^{ia} &= - \epsilon^{ijk} A^j_0 \tE^{ka} - \im \epsilon^{ijk} \partial_b^A \left( \dlap \tE^{ja} \tE^{kb} \right) + 2 \partial_b^A \left( \tE^{i[a} N^{b]} \right), \\
        \partial_t A^i_a  &= \partial^A_a A^i_0 + \im \dlap \epsilon^{ijk} \tE^{jb} F^k_{ab} + N^b F^i_{ba}.
    \end{align}

    The right-hand sides of these equations contain constraint terms that vanish on the constraint surface. We could remove them without changing the dynamics; however, it has long been known that such constraint terms affect stability and can improve the dynamics. We therefore keep them. For the remaining fields $\dlap, N^a, A^i_0$, we enforce the gauge in which they are constant,

    \begin{align}
        \partial_t \dlap = 0, \quad \partial_t N^a = 0, \quad \partial_t A^i_0 = 0,
    \end{align}

    or, when known, we set them to their exact analytic values.

    \subsection{Constraints}

    The final ingredient of the evolution system is the constraints. These are purely spatial conditions that must vanish weakly for a physical set-up. Owing to the extra $SO(3,\C)$ symmetry, there are three additional Gauss constraints on top of the four arising from diffeomorphism freedom. In the standard Ashtekar variables, they are

    \begin{align}
        \mathcal{H}   &= \epsilon^{ijk} \tE^{ja} \tE^{kb} F^i_{ab}, \\
        \mathcal{M}_a &= \tE^{ib} F^i_{ba}, \\
        \mathcal{G}^i &= \partial^A_a \tE^{ia}.
    \end{align}

    These constraints are the same for the original and hyperbolic systems and are required to vanish (or to be as close to zero as is numerically possible). For the conformal system, we do not have direct access to these variables; instead, we use the transformations in \cref{subsec:conf-field-3+1-decomp} to convert the conformal fields into the original ones and then use these to compute the constraints. This requires only one extra algebraic step compared with the original system.

    Another important set of constraints is that arising from the reality conditions. Indeed, for our results to describe a real metric, the following imaginary parts must vanish:

    \begin{align}\label{eq:org-3+1-reality-conds}
        \imag(\tE^{ia} \tE^{ib}) = 0, \quad \imag(N^a) = 0, \quad  \imag(\dlap) = 0.
    \end{align}

    These $6+3+1 = 10$ conditions imply that the imaginary part of the metric vanishes. For the conformal system, we compute

    \begin{align}
        \imag(H^{ia} H^{ib}) = \imag(\dlap^2 \tE^{ia} \tE^{ib}) = 0,
    \end{align}

    instead of the first condition in \cref{eq:org-3+1-reality-conds}. Together with the last two conditions in \cref{eq:org-3+1-reality-conds}, this implies that the metric is real.

    The final set of constraints we could consider is the gauge variables $\chi^i$ and $\xi_\mu$. These should also vanish on the constraint surface, but their dynamics are linked with the constraints above and should vanish, or become constant in time, when the constraints vanish.

\section{Hyperbolicity and Constraint Analysis of Evolution Systems}\label{sec:hyp-and-con-anal-of-evo-sys}

    In this section we analyse the hyperbolicity and constraint propagation of both the hyperbolic and conformal evolution systems from the previous section. We also consider damping terms and their effect on constraint propagation.

    \subsection{Hyperbolicity of the Original and Conformal Systems}

    Strong hyperbolicity is needed for the numerical solutions to converge. In \cref{sec:linearised-gauge-fixing} we found that every field of the linearised system satisfies the wave equation. Writing the system as $\partial_t u = M^a \partial_a u + S(u)$, with characteristic matrix $P = M^a k_a$ for an arbitrary wave vector $k_a$, this property means that

    \begin{align}
        M^a = (\gamma^0)^{-1} \gamma^a
    \end{align}

    for some $32\times 32$ representation of the four-dimensional gamma matrices $\gamma^\mu = (\gamma^0, \gamma^a)$. The Clifford algebra then gives

    \begin{align}
        P^2 + 2 N^a k_a P + (N^a N^b - N^2 g^{ab}) k_a k_b \mathbb{I} = 0.
    \end{align}

    Here $N$ is the usual lapse, defined by $g^{00} = -\frac{1}{N^2}$. This implies that the eigenvalues are

    \begin{align}\label{eq:amp-matrix-eigenvals}
        \lambda_P &= -N^a k_a \pm \sqrt{N^2 g^{ab} k_a k_b}, \\ &= -N^a k_a \pm \sqrt{\dlap^2 \tE^{ia} \tE^{ib}k_a k_b}.
    \end{align}

    Here we have used $N^2 g^{ab} = \dlap \tg^{ab} = \dlap^2 \tE^{ia} \tE^{ib}$, where $g^{ab}$ denotes the inverse spatial metric. Together with ${\rm Tr}(P) = -32 N^a k_a$ (where the 32 is the dimension of the representation), this implies that 16 of the eigenvalues take the plus sign and the remaining 16 the minus sign. Hence $P$ has real eigenvalues and is diagonalisable, so the system is strongly hyperbolic. Alternatively, this is an immediate consequence of the system squaring to the wave equation. None of the fields have zero-speed modes; in~\cite{Towards_an_unde_Alcubi_1999} this was argued to be useful for avoiding singularities.
    
    \subsection{Original System Constraint Analysis}
    Next, we check that the constraint surface is attractive, so that constraint violations are damped rather than amplified. We linearise about Minkowski space, $\dlap = 1$, $N^a = 0$, $\tE^{ia} = \delta^{ia}$, $A^i_\mu = 0$, $\xi_\mu = 0$, $\chi^i = 0$, and Fourier transform; the signs of the real parts of the amplification-matrix eigenvalues then tell us whether the constraints grow or decay. The linearised constraints are

    \begin{align}
        \delta G^i = \partial_a \delta \tE^{ia} + \epsilon^{ijk} \delta A^j_a \tE^{ka}, \quad \delta M^a = \partial^j \delta A^{ja} - \partial^a \delta A^i_i, \quad \delta H = 2 \epsilon^{iak} \partial^i \delta A^k_a.
    \end{align}

    Here, and throughout this section, the background $\tE^{ia} = \delta^{ia}$ is used to convert between spatial and internal indices. Since the frame and connection equations are decoupled, we can treat each block separately, starting with the frame sector. Around this background, the equation of motion for the linearisation $\delta \tE^{ia}$ becomes

    \begin{align}
        \partial_t \delta \tE^{ia} = - \im \epsilon^{ijk} \partial^k \delta \tE^{ja} -\im\epsilon^{iak} \partial^k \delta \dlap - \partial^i \delta N^a + \delta^{ia} \partial_b \delta N^b + \im (c_1 +c_2) \delta^{ia} \partial^j \delta \chi^j \nonumber \\ + \frac{c_2 - c_1}{2 \im c_2} \left( \epsilon^{iak} \delta \mathcal{G}^k + (c_1+c_2) \partial^i \delta \chi^a - (3c_2 + c_1) \partial^a \delta \chi^i \right).
    \end{align}

    The prefix $\delta$ denotes a linearised variable. With this, we can compute the time derivative of the linearised Gauss constraint,

    \begin{align}
        \partial_t \delta \mathcal{G}^i = -\frac{c_1 + c_2}{2 \im c_2} \epsilon^{ijk} \partial^k \delta \mathcal{G}^j - \frac{(c_1+c_2)^2}{2 \im c_2} \partial^i \partial^j \delta \chi^j + \frac{(c_1-c_2)(3c_2 + c_1)}{2 \im c_2} \partial^j \partial^j \delta \chi^i.
    \end{align}

    In deriving this result we have ignored the linearisation of the connection $A^i_\mu$, since it does not affect the eigenvalues. The linearised equation for the internal vector $\chi^i$ is then

    \begin{align}
        \partial_t \delta \chi^i = \frac{1}{2 \im c_2} \delta \mathcal{G}^i + \frac{c_1+c_2}{2 \im c_2} \epsilon^{ijk} \partial^k \delta \chi^j + \alpha \delta \chi^i.
    \end{align}
    Hence $(\chi^i,\mathcal{G}^i)$ form a closed system. Taking the Fourier transform of these equations and writing them in matrix form gives
    \begin{align}
        \partial_t \binom{\delta \hat{\mathcal{G}}^i}{\delta \hat{\chi}^i} = \begin{pmatrix}
            -\frac{c_1 + c_2}{2 c_2} \epsilon^{ijk} k^k & \frac{(c_1+c_2)^2}{2 \im c_2} k^i k^j + \frac{(c_2 - c_1)(3c_2+c_1)}{2 \im c_2} k^2 \delta^{ij} \\
             \frac{1}{2 \im c_2} \delta^{ij}                & \frac{c_1+c_2}{2c_2} \epsilon^{ijk} k^k + \alpha \delta^{ij}
        \end{pmatrix} \begin{pmatrix}\delta \hat{\mathcal{G}}^i \\ \delta \hat{\chi}^i \end{pmatrix}.
    \end{align}
    Here $k^i$ is the wave vector of the Fourier transform and $k^2 = k^i k^i$. Computing the eigenvalues of this amplification matrix, we find
    \begin{align}
        \lambda_\pm = \frac{\alpha}{2} \pm \sqrt{\left(\frac{\alpha}{2}\right)^2 - k^2}, \quad \lambda_{\pm\pm} = \frac{\alpha}{2} \pm \sqrt{\left(\frac{\alpha}{2}\right)^2 \pm \im \left(1 + \frac{c_1}{c_2}\right)\sqrt{k^2} \frac{\alpha}{2} - k^2},
    \end{align}
    where each $\pm$ is independent, giving six eigenvalues in total. When $\alpha = 0$, the eigenvalues reduce to $\pm \im \sqrt{k^2}$, three of each sign, because all the fields satisfy the wave equation. For the constraints to decay, we want the real parts of the eigenvalues to be negative. A vanishing real part is also acceptable, but is expected to handle constraint violations less well. When $\frac{c_1}{c_2} \in \R$, the analysis in \cref{apx:eigenvalue-analysis} shows that the real parts of all eigenvalues are non-positive provided that

    \begin{align}
        \re(\alpha) \leq 0, \quad -3 \leq \frac{c_1}{c_2} \leq 1,
    \end{align}

    except when $\alpha = 0$, in which case $c_1$ and $c_2$ are unconstrained.

    On the connection side of the evolution system, skipping the details, the evolution equations for the linearised Hamiltonian and momentum constraints are

    \begin{align}
        \partial_t \delta \mathcal{M}^i & = -\frac{\im}{2} \partial^i \delta \mathcal{H} + \frac{\im}{2} \partial^i \partial^j \delta \xi^j - \frac{\im}{2} \partial^j \partial^j \delta \xi^i, \\
        \partial_t \delta \mathcal{H} & = 2 \im \partial^i \delta \mathcal{M}^i.
    \end{align}
    The linearised equations of motion for $\xi_0$ and $\xi^i$ are
    \begin{align}
        \partial_t \delta \xi^i &= 2 \im \delta \mathcal{M}^i - \partial^i \delta \xi_0 + \beta \delta \xi^i, \\
        \partial_t \delta \xi_0 &= \delta \mathcal{H} - \partial^i \delta \xi^i + \beta \delta \xi_0.
    \end{align}
    Taking the Fourier transform and choosing the solution vector $u = (\hat{\mathcal{M}}^i,\hat{\mathcal{H}},\hat{\xi^i},\hat{\xi_0})$, we obtain the amplification matrix
    \begin{align}
        \begin{pmatrix}
            0 & \frac{1}{2} k^i & - \frac{\im}{2} k^i k^j + \frac{\im}{2} k^k k^k \delta^{ij} & 0 \\
            -2 k^j & 0 & 0 & 0 \\
            2 \im \delta^{ij} & 0 & \beta \delta^{ij} & -\im k^i \\
            0 & 1 & -\im k^j & \beta
        \end{pmatrix}.
    \end{align}

    The eigenvalues of the connection amplification matrix are 

    \begin{align}\label{eq:hyp-con-amp-matrix}
        \lambda_\pm^1 = \pm \im \sqrt{k^2}, \quad \lambda_\pm^2 = \beta \pm \im \sqrt{k^2}, \quad \lambda^3_\pm = \frac{\beta}{2} \pm \sqrt{\left(\frac{\beta}{2}\right)^2 - k^2},
    \end{align}

    with $\lambda^3$ appearing twice for each sign. Once again, when $\beta = 0$, the eigenvalues all reduce to $\pm \im \sqrt{k^2}$, as expected. By inspection, $\re(\beta) < 0$ gives 6 of the 8 eigenvalues negative real parts.

    Combining both results, we conclude that the best choice for constraint damping is

    \begin{align}\label{eq:hyp-system-constraint-conds}
        \re(\alpha) \leq 0, \quad \re(\beta) \leq 0, \quad c_1 = \hat{\rho} c_2, \quad \hat\rho \in \R, \quad -3 \leq \hat\rho \leq 1,
    \end{align}

    which ensures that the largest number of eigenvalues have negative real parts.

    \subsection{Conformal Systems}

    We now turn our attention to the conformal system. Here, the hyperbolicity analysis can be carried out in exactly the same way. Since the two systems are equivalent (up to a choice of $\alpha,\beta,c_1,c_2$ which do not affect hyperbolicity) they share the same characteristic matrix and strong hyperbolicity. This is in agreement with the linear structure displayed in \cref{sec:linearised-gauge-fixing}.

    Secondly, we consider the constraint damping of the system in \cref{subsec:gauge-fixed-3p1-systems:conformal-system}. Since this system and the hyperbolic system are related by an invertible map, their constraint amplification matrices are related by a similarity transformation. Since similar matrices have the same eigenvalues, the two systems share the same constraint eigenvalues. However, compared with the hyperbolic system, the conformal system corresponds to $\beta = 0$ and $c_1 = c_2 = c$ (i.e.\ $\hat\rho = 1$). From the previous analysis this lies within the acceptable range; however, we do not expect any damping of the Hamiltonian and momentum constraints. In summary, the parameters in the conformal theory must satisfy

    \begin{align}
        \re(\alpha) \leq 0, \quad c \in \C \setminus\{0\}.
    \end{align}

    \section{Numerical Results}\label{sec:numerical-results}

    We test the schemes of \cref{sec:gauge-fixed-3+1-evolution-systems} using standard tests from~\cite{Apples_with_App_Daveri_2018,Towards_standar_Alcubi_2004}. All simulations use Cartesian coordinates $(t,x,y,z)$ on a cubic grid of $N\times N \times N$ points with side length $L = 1.0$ and spacing $\Delta x = \Delta y = \Delta z = \frac{L}{N}$. Time integration uses the classical fourth-order Runge-Kutta method with Courant factor $0.0833$, and spatial derivatives are approximated by fourth-order central finite differences; no Kreiss-Oliger dissipation is applied. We impose periodic boundary conditions. This is a worst case for constraint control, since errors and constraint violations that would otherwise leave through the boundaries are fed back into the domain. The plots on which the following discussion is based are collected in \cref{apx:numerical-plots}.

    \subsection{Convergence Tests}

    Before testing the systems, we perform convergence tests to check that the results can be trusted. To do so, we initialise the systems with a gauge wave; for the original and hyperbolic systems this amounts to imposing

    \begin{gather}
        \tE^{ia} = \begin{pmatrix} 1 & 0 & 0 \\ 0 & W & 0 \\ 0 & 0 & W \end{pmatrix}, \quad \dlap = 1, \quad N^a = 0, \quad \chi^i = 0, \\
         A^i_a =  \begin{pmatrix} \frac{\im \partial_t W}{W} & 0 & 0 \\ 0 & 0 & 0 \\ 0 & 0 & 0 \end{pmatrix}, \quad A^i_0 = \begin{pmatrix} \frac{\im \partial_x W}{W} \\ 0 \\ 0 \end{pmatrix}, \quad \xi_a = 0,\quad \xi_0 = 0.
   \end{gather}

    For the conformal variables, the initial data are

    \begin{gather}
        H^{ia} = \begin{pmatrix} 1 & 0 & 0 \\ 0 & W & 0 \\ 0 & 0 & W \end{pmatrix}, \quad \dlap = 1, \quad N^a = 0, \quad \chi^i = 0, \\
        \Omega^i_a = \begin{pmatrix} \frac{\im \partial_t W}{W} & 0 & 0 \\ 0 & 0 & 0 \\ 0 & 0 & 0 \end{pmatrix}, \quad \Omega^i_0 = \begin{pmatrix} \frac{\im \partial_x W}{W} \\ 0 \\ 0 \end{pmatrix}, \quad  \omega_a = \begin{pmatrix} -2 \frac{\partial_x W}{W} & 0 & 0 \end{pmatrix}, \quad \omega_0 = -2 \frac{\partial_t W}{W},
    \end{gather}

    where we choose $W = 1.0 + \mathcal{A} \cos((t+x)k)$, $\mathcal{A} = 1.0\times10^{-3}$ and $k = \frac{2\pi}{L}$. These are exact periodic solutions that contain only pure gauge modes. By performing a coordinate transformation, each of the resulting metrics can be written as the standard Minkowski metric in Cartesian coordinates. For the original \pleb{} system, \cref{eq:orginal-pleb-system}, since the temporal connection $A^i_0$ has no evolution equation, we set it to its exact value at each time step.

\begin{figure}[H]
        \centering
        \begin{subfigure}{0.31\textwidth}
            \includegraphics[width=\textwidth]{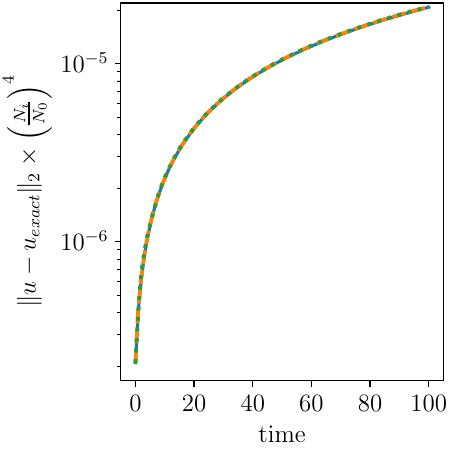}
            \caption{Original}
        \end{subfigure}
        \hspace{-5pt}
        \begin{subfigure}{0.31\textwidth}
            \includegraphics[width=\textwidth]{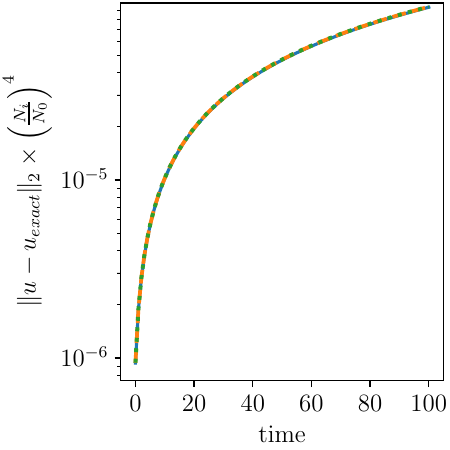}
            \caption{Hyperbolic}
        \end{subfigure}
        \hspace{-5pt}
        \begin{subfigure}{0.31\textwidth}
            \includegraphics[width=\textwidth]{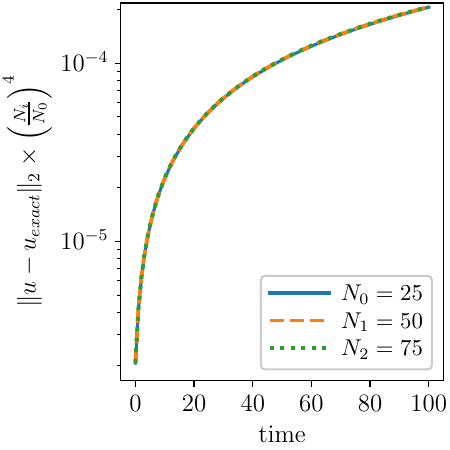}
            \caption{Conformal}
        \end{subfigure}
        \vspace{-8pt}
        \caption{$L^2$ norm of the difference between the exact and numerical solutions for the gauge-wave initial data, for the original, $(0,0,1,1)$ hyperbolic and $(0,1)$ conformal systems with $N_i = 25, 50, 75$ grid points ($i=0,1,2$). The expected error should grow as $N^4$ hence the scaling by $\left(\frac{N_i}{N_0}\right)^4$ should and does undo the convergence. As expected, the errors overlap after scaling, confirming fourth-order convergence for the code in each system.}\label{fig:convergence-plot}
    \end{figure}

    To measure convergence, we plot the $L^2$ norm of the difference between the numerical and exact solutions against time. Here the $L^2$ norm is the sum of the squared moduli of all components over all grid points. For fourth-order methods in space and time, the error should scale as $N^{-4}$, so going from resolution $N_j$ to $N_i$ scales the error in the solution by $\left(\frac{N_j}{N_i}\right)^4$. \Cref{fig:convergence-plot} shows the results for $N = 25, 50, 75$: the rescaled errors overlap for all three systems, confirming fourth-order convergence. In this case the parameters chosen in the hyperbolic and conformal systems were $(\alpha,\beta,c_1,c_2) = (0,0,1,1)$ and $(\alpha,c) = (0,1)$ respectively. We use the intermediate resolution $N_1 = 50$ for the remaining simulations. We do not simply increase the resolution as far as possible, firstly because of the computational cost of large grids, and secondly because at sufficiently large grid sizes systematic numerical errors (e.g.\ accumulated round-off error) begin to emerge.
    
    \subsection{Robust Stability: Minkowski with Noise}

    With convergence confirmed, we now examine the behaviour of the constraints under evolution for different parameters of the schemes. For this test we use uniformly distributed random noise on a Minkowski background. Following~\cite{Apples_with_App_Daveri_2018}, the initial data for the hyperbolic system are

    \begin{gather}
        \tE^{ia} = \delta^{ia} + \varepsilon^{ia}, \quad N^a = \varepsilon^a, \quad \dlap = 1 + \varepsilon, \quad \chi^i = \varepsilon^i, \\
        A^i_\mu = \varepsilon^i_\mu, \quad \xi_\mu = \varepsilon_\mu,
    \end{gather}

    and for the conformal system they are

    \begin{gather}
        H^{ia} = \delta^{ia} + \varepsilon^{ia}, \quad N^a = \varepsilon^a, \quad n = \varepsilon, \quad \chi^i = \varepsilon^i, \\
        \Omega^i_\mu = \varepsilon^i_\mu, \quad \omega_\mu = \varepsilon_\mu,
    \end{gather}
    where $\varepsilon = u + \im v \in \C$ and $u,v \in [-10^{-10}/16,+10^{-10}/16]$ are random real numbers chosen uniformly within the given range. Since the fields in our formalism are complex-valued, we also populate the imaginary part of each component with random noise. Each component at each grid point is assigned a different random complex number. We then evolve these initial data up to $T = 100.0$ or until the simulation crashes.
  
    \paragraph{Original System}

    For the original system, we evolve the scheme as given, without removing the constraint terms that appear in the equations of motion. This provides a baseline for the behaviour of our modified systems: the other systems can be compared against this one to determine whether the gauge-fixing methods described in the previous sections yield an improvement. The evolution of the constraints is plotted in \cref{fig:org-normal-plot}; over this time interval the reality conditions and Gauss constraint grow, while the Hamiltonian and momentum constraints are damped.

    \paragraph{Hyperbolic System}

    For the hyperbolic system we partially explore the parameter space in two stages. First, we vary the ratio $\hat\rho = \frac{c_1}{c_2}$, since the constraint propagation analysis depends only on this combination and not on $c_1, c_2$ individually. Secondly, we take the preferred values of $c_1, c_2$ from this search and explore values of $\alpha, \beta$. For the first stage, we fix $c_2 = 1$ and take $c_1 = -4,-2,-1,0,1,2$; these values are chosen so that $\hat\rho$ lies both inside and outside the range found in the constraint analysis, \cref{eq:hyp-system-constraint-conds}.

    Plots of the $L^2$ norm of the constraints versus time are shown in \cref{fig:hyp-c12}; they indicate that $c_1 = 1$ gives the best constraint conservation, in the sense that the $L^2$ norm of the Gauss constraint is smallest for most of the simulation. For the Hamiltonian and momentum constraints, the performance seems largely independent of $c_1$, consistent with the constraint analysis in \cref{sec:hyp-and-con-anal-of-evo-sys}. For the reality condition, the choices $c_1 = 0$ and $c_1 = -1$ perform better than $c_1 = 1$. Since $c_1 = 1$ gives the best result for the Gauss constraint, and the reality condition is not significantly worse, we adopt this value from here on.

    \Cref{fig:hyp-ab-div,fig:hyp-ab-con} show the effect of the damping parameters. A positive value of either parameter, $(\alpha,\beta) = (1,1), (1,0), (1,-1), (-1,1), (0,1)$, makes the simulation diverge and crash shortly after $T = 25$, well before the designated end time $T = 100$, whatever the sign of the other parameter (\cref{fig:hyp-ab-div}). This is a clear sign of the constraints blowing up, consistent with the analysis in \cref{sec:hyp-and-con-anal-of-evo-sys}. Non-positive values, $(\alpha,\beta) = (-1,0), (-1,-1), (0,-1)$, remove the divergence, and all constraints stay bounded until the end of the run (\cref{fig:hyp-ab-con}); $(\alpha,\beta) = (-1,-1)$ gives the smallest norms overall. Interestingly, the Hamiltonian constraint is insensitive to the choice of $\alpha$ and $\beta$, which suggests that it propagates with the undamped eigenvalues $\pm \im \sqrt{k^2}$ of \cref{eq:hyp-con-amp-matrix}.

    From these tests we conclude that $(\alpha,\beta,c_1,c_2) = (-1,-1,1,1)$ is the best choice among the parameters explored.

    \paragraph{Conformal System} Compared with the hyperbolic system, some parameters of the conformal system are already fixed, leaving $c$ and $\alpha$ to explore. We first explore different values of $c$ with $\alpha = 0$; \cref{fig:conf-c-plot} shows the constraint evolution for $c = -2,-1,1,2$. Changing this value does not affect the constraints in any meaningful way, and the simulations survive the allotted duration without crashing. This is consistent with the constraint analysis: for the conformal system, $\hat\rho = 1$ in \cref{eq:hyp-system-constraint-conds} independently of $c$, so the constraints should behave in the same way.

    From here on we fix $c = 1$, the simplest admissible value ($c = 0$ is excluded because factors of $1/c$ appear). We then explore the damping parameter, simulating $\alpha = 1, 0, -1$. In this case, a positive damping coefficient, $\alpha = 1$ (\cref{fig:conf-a-div}), does not crash the simulation; instead, the reality condition and Gauss constraint grow by roughly $30$ orders of magnitude. This is still considered unstable behaviour, because a longer simulation would be expected to crash eventually. Surprisingly, the momentum and Hamiltonian constraints decrease over time and finish below their initial values. This is explained by the fact that $\alpha$ only affects the evolution of the Gauss constraint, leaving the others unchanged. Lastly, \cref{fig:conf-a-con} shows that a negative damping parameter reduces both the Gauss and reality constraint norms relative to the undamped case.

    The conformal tests show that $\alpha = -1$ and $c = 1$ are the best parameter choices among the cases explored.

    \subsection{Long-Time Simulations}

    As a final test of long-term stability, we evolve each system, using the best parameters determined above, up to $T = 2000$. For the original system no parameters are needed, so we simply run the simulation for longer. The hyperbolic system is run with $(\alpha,\beta,c_1,c_2)=(-1,-1,1,1)$ and the conformal system with $(\alpha,c)=(-1,1)$.

\begin{figure}[H]
        \centering
        \includegraphics[width=0.6\textwidth]{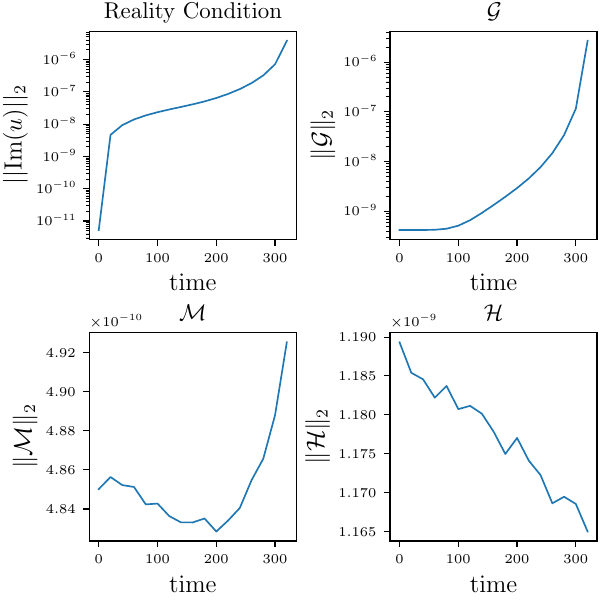}
        \caption{$L^2$ norm of the constraints versus simulation time for the original evolution scheme found in \cref{subsec:original-system} with initial conditions of Minkowski plus random noise on a periodic domain. The reality condition is the distance from  The simulation is run up to $T=2000$, however, here the simulation crashes at $T \sim 325$. The least conserved constraint is the reality condition and it grows 5 orders of magnitude from $10^{-11}$ to $10^{-6}$. The original system in this case is not stable even around Minkowski.}\label{fig:org-long}
    \end{figure}

\begin{figure}[H]
        \centering
        \includegraphics[width=0.6\textwidth]{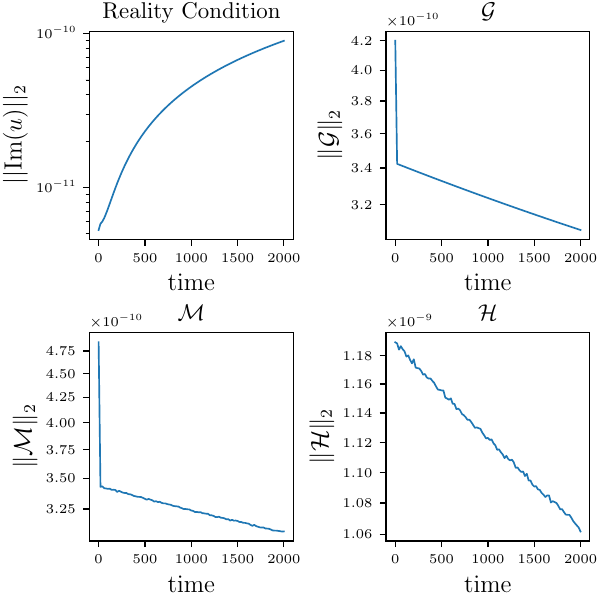}
        \caption{The same plot as \cref{fig:org-long} but for the hyperbolic evolution system in \cref{subsec:hyperbolic-3+1-system} with parameters $(\alpha,\beta,c_1,c_2)=(-1,-1,1,1)$. In this case the simulations are stable up to the end of the simulation at $T=2000$. The reality condition grows by only 1 order of magnitude from $\sim 10^{-11}$ to $\sim 10^{-10}$, the remaining constraints all decrease by less than an order of magnitude. This shows long term stability for the hyperbolic system with the chosen parameters.}\label{fig:hyp-long}
    \end{figure}

\begin{figure}[H]
        \centering
        \includegraphics[width=0.6\textwidth]{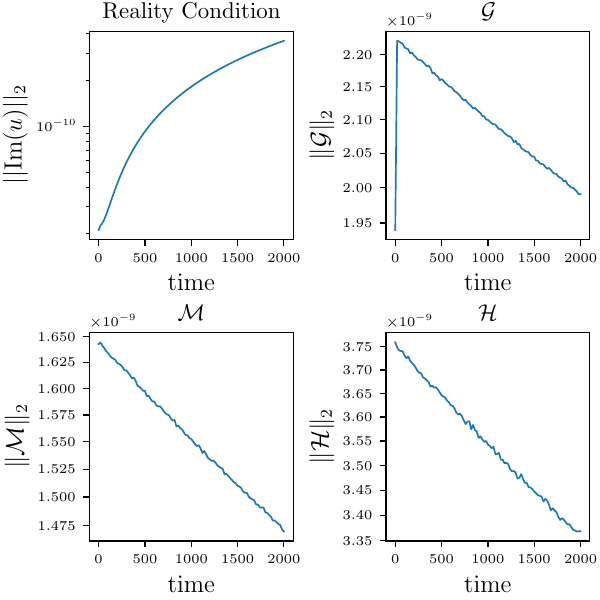}
        \caption{The same plot as \cref{fig:org-long} but for the conformal evolution system in \cref{subsec:gauge-fixed-3p1-systems:conformal-system} with parameters $(\alpha,c)=(-1,1)$. In this case the simulations are stable up to the end of the simulation at $T=2000$. All of the constraints stay within the same order of magnitude and none grow above $10^{-8}$. This shows long term stability for the conformal system with the chosen parameters.}\label{fig:conf-long}
    \end{figure}

    \cref{fig:org-long,fig:hyp-long,fig:conf-long} shows the long-time results. The original system crashes at around $T \sim 325$: its constraint norms grow from $\sim 10^{-10}$ to $\sim 10^{-6}$ shortly beforehand, with the Gauss, momentum and reality constraints all blowing up. (The limited temporal resolution of the output snapshots does not allow the crash time to be pinned down more precisely.) Such behaviour is typical of unmodified numerical relativity formulations and confirms that some modification is required for long-term stability.

    In contrast, with the chosen parameters, both the hyperbolic and conformal systems survive to $T = 2000$ and show no sign of divergence. The only quantity that grows (slowly) is the imaginary part of the metric, which nevertheless remains small enough to be negligible: its $L^2$ norm is at most of order $10^{-10}$, compared with the real part, which is of order $10^0$. Owing to the damping, all other constraints decrease as the simulation continues. This gives us reasonable confidence that these simulations would run until systematic errors take over. 

    \subsection{Summary of Evolution Systems}

    \Cref{tab:evolution-systems-summary} summarises the three numerical evolution schemes, their parameters and the best parameter values found.

    \begin{table}[H]
        \centering
        \begin{tabular}{| c | c | c | c | c | c | c |}
            \hline
            Name & Vars & Comps & Params & Poly & Best Parameters & Eqns \\
            \hline
            Original & $\tE^{ia}, N^a, \dlap, A^i_a, A^i_0$ & $25 \C$ &  N/A & Yes & N/A & \cref{subsec:original-system} \\
            \hline
            Hyperbolic & $\tE^{ia}, N^a, \dlap, \chi^i, A^i_a, A^i_0, \xi_a, \xi_0$ & $32 \C$ & $\alpha,\beta,c_1,c_2$ & Yes & $-1,-1,1,1$ & \cref{subsec:hyperbolic-3+1-system} \\
            \hline
            Conformal & $H^{ia}, N^a, n, \chi^i, \Omega^i_a, \Omega^i_0, \omega_a, \omega_0$ & $32 \C$ & $\alpha, c$ & No & $-1,1$ & \cref{subsec:gauge-fixed-3p1-systems:conformal-system} \\
            \hline
        \end{tabular}
        \caption{Table of the three evolution systems, the variables that they use, the number of components, parameters in the system, whether they are polynomial in their fields, the best parameter choices found and the location of the full evolution systems.}\label{tab:evolution-systems-summary}
    \end{table}

\section{Discussion}\label{sec:discussion}
We have presented two new evolution systems for general relativity based on Ashtekar's variables: the four-parameter hyperbolic system (\cref{subsec:hyperbolic-3+1-system}) and the two-parameter conformal system (\cref{subsec:gauge-fixed-3p1-systems:conformal-system}). Both are minimal nonlinear extensions of a new two-parameter gauge fixing of linearised gravity (\cref{subsec:lin-gauge-fix-summary}). Surprisingly, the conformal system splits the nonlinear Einstein equations into two sectors, of sizes twelve and four, and the resulting equations form a Dirac-like operator acting on the fields. With parameters selected using robust-stability tests, both systems evolved stably to $T = 2000$ (roughly 1.2 million time steps at $\Delta t = 1.66 \times 10^{-3}$), with constraint norms between $\sim 10^{-11}$ and $10^{-10}$ and reality violations of order $10^{-10}$, i.e.\ 10 to 11 orders of magnitude smaller than the physical solution. By contrast, the original system crashes at $T \sim 325$, when constraint growth spoils the solution. For these tests, therefore, the gauge-fixed systems clearly improve constraint control and stability.

Several caveats should be noted.
\begin{itemize}
    \item The numerical tests were performed only around Minkowski space, and the schemes may perform less well for more complicated spacetimes. In particular, black hole spacetimes, which contain singularities, would provide a more stringent test of these schemes in the strongly nonlinear regime.
    \item The simulations used periodic boundaries. Absorbing boundary conditions could be explored instead; these would allow constraint violations propagating as waves to leave the computational domain, possibly improving performance.
    \item Harmonic gauges are prone to gauge shocks~\cite{The_appearance_Alcubi_1996}, coordinate pathologies that spoil the solution. These can be avoided by modifying the gauge~\cite{Hyperbolic_slic_Alcubi_2002}.
    \item The systems are not covariant, owing to the partial derivatives and the explicit appearance of Christoffel symbols. Around a Minkowski solution this is not expected to cause problems, but modifications will be needed in more strongly nonlinear set-ups. Replacing the Christoffel symbols with their difference from those of a reference background metric partly resolves this issue. Dealing with the partial derivatives would require replacing them with covariant derivatives; however, this would introduce explicit metric terms and complicate the resulting evolution equations, which would probably no longer be polynomial.
    \item The reality constraints are evolved freely, so their growth is not controlled. In the tests considered here this growth was not significant, but it is likely to cause problems in other situations. Damping terms could be added, but they would break the holomorphic structure of the systems.
\end{itemize}

Comparisons with other numerical schemes, such as BSSN or Z4, can only be made qualitatively. Here we compare our constraint norms with the ones obtained in~\cite{Apples_with_App_Daveri_2018} using the BSSN and Z4 systems, the initial conditions used are the same in our and their systems. Their constraints, in their best cases, and have a maximum value of around $10^{-10}$. In comparison for our systems, with damping, the norms decrease but stay at roughly the same value of around $10^{-10}$. A more accurate comparison would require running BSSN and Z4 directly on the same tests.

It is also worth comparing the expected computational cost of the hyperbolic and conformal systems with that of the standard real metric systems. The gauge-fixed systems evolve 32 complex (64 real) fields: $16 = 9 + 3 + 1 + 3$ for the triad, shift, lapse and gauge field, and $16 = 9 + 3 + 4$ for the spatial, temporal and gauge connections. Compared with the 25 complex fields of the original Ashtekar system, the 7 extra fields are pure gauge. (In the original system only 18 of the 25 components obey evolution equations, but all 25 must still be stored.) By contrast, the Z4 and BSSN systems of~\cite{Apples_with_App_Daveri_2018} use 24 real fields: $24 = 11 + 7 + 4$, with $11 = 10 + 1$ for the metric and conformal factor, $7 = 6 + 1$ for the extrinsic curvature and its trace, and $4$ additional gauge fields. The additional components in the metric and extrinsic curvature are artefacts of the way these tensors are stored. Our formulation clearly contains more components, because it is both complex and first-order. On the other hand, the metric formulations involve second-order derivatives, so they require more computation per step but less memory. This trade-off between first- and second-order formulations is difficult to assess from the equations and components alone: more components generally means more memory and computation time, but this can be offset by the reduced cost of using a lower-order in derivative scheme.

The conformal system partly offsets this cost. Its 12-component sector can be evolved on its own, so only this sector is needed when the unimodular metric suffices; to the best of our knowledge, this is unique among formulations of gravity for generic set-ups. It is a remarkable property of four-dimensional gravity in the chiral formulation, and is not expected to occur in any other dimension or description due to the unique gauge fixing required. It is natural to ask whether this property helps when extracting physical data, such as gravitational waves. For gravitational wave extraction one wants to compute the Weyl scalars $\Psi^{ij}$ or some linear combination of these. Using only the 12-component sector we can compute

\begin{align}
    \hat{\Psi}^{ij} = \frac{1}{4} H^{\langle i|\mu\nu} F(\Omega)^{|j\rangle}_{\mu\nu}.
\end{align}

Assuming that the simulation is eventually close to the constraint surface, i.e.\ $\xi_\mu = 0$, then we have that $F(\Omega)^i_{\mu\nu} = F^i_{\mu\nu}$ meaning that the 12-component curvature coincides with the full physical curvature. Our conformal Weyl scalars are then related to the physical ones through

\begin{align}
    \hat{\Psi}^{ij} = \frac{1}{4} \lambda^{-1} \sd^{\langle i|\mu\nu} F^{|j\rangle}_{\mu\nu} = \lambda^{-1} \Psi^{ij},
\end{align}

where the $\lambda$ factor appears because the conformal metric, $g_H$, is used to raise the indices of $H^i$. Furthermore, at the spatial boundary, assuming it is far enough away, in Cartesian coordinates the conformal factor goes as $\lambda^{-1} \sim 1 + \mathcal{O}(M/r)$ where $M$ is the mass of the effective Schwarzschild black hole and $r$ is the distance from its center. Therefore we can roughly equate conformal and physical Weyl scalars in this setup $\hat{\Psi}^{ij} \cong \Psi^{ij}$. The more useful quantity to compute, as in~\cite{Extraction_of_g_Bishop_2016}, is $r \Psi^{ij}$. For this we simply multiply $\hat{\Psi}^{ij}$ by $r = \sqrt{x^2+y^2+z^2}$ at each point on the boundary. The usual Weyl extraction techniques then can be applied as usual. This makes it clear that, given enough accuracy, only the 12-component system is needed to extract physical information.

Even when both sectors are required, they could be evolved with different numerical methods. For example, the 12-component sector could be evolved using methods similar to those used here, while the 4-component sector is integrated with coarser grid and time stepping allowing for quicker integration. Provided such a hybrid scheme is stable and convergent, it would recover the same solution at lower cost. Many other combinations of numerical methods are possible, and the performance could vary considerably with the choice; some combinations may significantly outperform current techniques.

Several directions for future research follow from this work.
\begin{itemize}
    \item \emph{Strong-field tests.} Evolving Schwarzschild and Kerr initial data would probe nonlinear stability and singularity avoidance far better than perturbations of Minkowski space.
    \item \emph{Improved gauge conditions.} Modifications of the harmonic and Lorenz gauge conditions, analogous to the 1+log and Gamma-driver conditions, should be explored to prevent gauge shocks and coordinate singularities.
    \item \emph{Benchmarking.} Direct comparisons with BSSN and Z4 on the same tests, including memory usage and run time, are needed to establish whether the schemes presented here offer practical advantages.
    \item \emph{Reality conditions.} The reality conditions can be written as first-order differential constraints. Adding them to the evolution equations should therefore modify the growth of reality violations, and further analysis is needed to determine whether they can then be evolved freely. Alternatively, one could decompose $\tE^{ia} = m^{ij}\tilde{B}^{ja}$ into a real triad $\tilde{B}^{ia}$ and a complex group element $m^{ij} \in SO(3,\C)$. The spatial metric would then satisfy the reality condition automatically, and the lapse and shift could be taken to be real, reducing the number of evolved components from $64\,\R$ to $51\,\R$.
    Implementing this requires a better understanding of how the evolution equations and connections decompose with respect to the reality condition.
    \item \emph{Structure-preserving integrators.} Structure-preserving integrators (SPIs) preserve constraints or other geometric structure exactly at the discrete level, and have been partially explored for gravity~\cite{NumericalRelatFarr2010,Constructing_of_Tsuchi_2016}. Differential forms play a central role in constructing SPIs, and the formulation used here is naturally expressed in terms of them. An SPI based on \pleb{}'s formulation would therefore be of great interest, though technically challenging.
    \item \emph{Wave extraction from the 12-component sector.} Extracting gravitational waves from the 12-component sector in strongly nonlinear set-ups, such as binary black holes, and comparing with standard extraction techniques would test whether the conformal split offers a practical advantage.
\end{itemize}

\section*{Acknowledgements}

The author is supported by the Science and Technology Facilities Council through grant ST/X000672/1. The author also thanks Kirill Krasnov and Miguel Bezares for useful discussions and comments. The author is grateful for access to the University of Nottingham's Ada HPC service, which was used to perform the numerical simulations. These results are partly based on similar work in the author's PhD Thesis~\cite{adam:shaw_2026}.
\appendix

\crefalias{section}{appendix}

\section{3+1 Decomposition of Chiral Objects}\label{apx:3+1-decomp-of-chiral-objects}
    Here we derive the decomposition of the self-dual 2-forms, $\sd^i$, which respect the metricity and reality conditions \cref{eq:metricity-cond,eq:reality-cond}.
    The standard decomposition separates the spacetime into $M = \R \times \spasurf$, where $\R$ is parametrised by the time coordinate $t : M \rightarrow \R$ and we denote coordinates on the spatial hypersurface by $x^a \in \spasurf$. To decompose the tangent space, we define a 1-form in the timelike direction, $dt \in \Lambda^1$. We then introduce a vector such that
    \begin{align}\label{eq:num-n-and-dt-are-dot-product}
    n^\mu {(dt)}_\mu = 1.
    \end{align}
    In coordinates $x^\mu = (t,x^a)$ the components of the 1-form and vector are $dt_\mu = (1,0)$ and $n^\mu = (1,-N^a)$, where $N^a$ is an arbitrary 3-vector.
\subsection{Bivector Decomposition}
    Instead of using $\Sigma^i$ as the basic variables, we use the already introduced bivectors
    \begin{align}
        \tE^{i\mu\nu} = \frac{1}{2}\teps^{\mu\nu\rho\sigma} \Sigma^i_{\rho\sigma} \in \Lambda^2(TM), \label{eq:num-tilde-E-in-eps-S}
    \end{align}
    which are dual to the 2-forms $\Sigma^i$. A tilde over a field denotes a density weight of $+1$; similarly, a tilde under a field denotes a weight of $-1$. The triple of 2-forms only defines a metric up to a conformal factor; we therefore fix the physical metric by fixing the conformal factor in the \urb{} formula, i.e.\ we define the metric via~\cref{eq:urbantke-metric}. When a physical metric is chosen, we can equate the density weight to the determinant of the metric. Keeping track of the number of upper and lower tildes is useful for error checking, as the number of upper tildes minus the number of lower tildes must be the same for every term in an equation. This parametrisation in terms of the bivectors is chosen because they are in a sense the conjugate momenta for the connections $A^i$, meaning that the Lagrangian has the leading-order kinetic term $L = \tE^{i\mu\nu} \partial_\mu A^i_\nu$. The Levi-Civita symbol $\teps^{\mu\nu\rho\sigma}$ takes the values $0, \pm 1$ and is defined without need for the metric. In this way we can take $\tE^{i\mu\nu}$ to be the basic variables of the theory without the need to introduce a metric. To decompose the bivectors, we contract them with the 1-form $dt$,
    \begin{align}
        \tE^{i\mu} = (dt)_\nu \tE^{i\nu\mu}, \label{eq:num-4-variant-triad-definition}
    \end{align}
    giving a triple of vectors. By computing $(dt)_\mu \tE^{i\mu} = (dt)_{[\mu} (dt)_{\nu]} \tE^{i\mu\nu} = 0$, we see that these are tangent to the hypersurfaces. In 3+1 coordinates this means we can parametrise them as $\tE^{i\mu} = (0,\tE^{ia})$, where we refer to $\tE^{ia}$ as triads for brevity (they are really densitised inverse triads). We now have a basis of vectors $n^\mu, \tE^{ia} \in TM$, which we can use to construct a basis for $\Lambda^2(TM)$:
    \begin{align}
        n^{[\mu} \tE^{i\nu]} \in \Lambda^2(TM), \quad \epsilon^{ijk} \tE^{j\mu} \tE^{k\nu} \in \Lambda^2(TM).
    \end{align}
    This basis is 6-dimensional, as expected for the space of bivectors in four dimensions. The full bivector can now be decomposed in this basis as
    \begin{align}
        \tE^{i\mu\nu} = \utilde{X}^{ij} \epsilon^{jkl} \tE^{k\mu} \tE^{l\nu} + M^{ij} n^{[\mu} \tE^{j\nu]}
    \end{align}
    for arbitrary matrices $\utilde{X}^{ij}$ and $M^{ij}$. By contracting this with $(dt)_\mu$ and making use of \cref{eq:num-n-and-dt-are-dot-product,eq:num-4-variant-triad-definition}, we find that $M^{ij} = 2\delta^{ij}$, and therefore the decomposition is
    \begin{align}
        \tE^{i\mu\nu} = \utilde{X}^{ij} \epsilon^{jkl} \tE^{k\mu} \tE^{l\nu} + 2n^{[\mu} \tE^{i\nu]}.
    \end{align}
    For the next step we use the fact that any $3\times 3$ matrix can be written as the sum of its symmetric and antisymmetric parts, $\utilde{X}^{ij} = \utilde{X}^{(ij)} + \epsilon^{ijk} \utilde{X}^k$, where in three dimensions the antisymmetric part can be parametrised by an internal 3-vector $\utilde{X}^i$. Substituting this into the above decomposition, we find that
    \begin{align}
        \tE^{i\mu\nu} & = \utilde{X}^{(ij)} \epsilon^{jkl} \tE^{k\mu} \tE^{l\nu} + \epsilon^{ijm} \utilde{X}^m \epsilon^{jkl} \tE^{k\mu} \tE^{l\nu} + 2 n^{[\mu} \tE^{i\nu]} \\
        & = \utilde{X}^{(ij)} \epsilon^{jkl} \tE^{k\mu} \tE^{l\nu} + 2(n^{[\mu} + \utilde{X}^j \tE^{j[\mu}) \tE^{i\nu]}.
    \end{align}
    We define a new complex vector $N^\mu = n^{\mu} + \utilde{X}^j \tE^{j\mu}$, which in components we write as $N^\mu = (1,-Y^a)$, where $Y^a = N^a - \utilde{X}^j \tE^{ja}$ is a complex 3-vector. The decomposition is then
    \begin{align}
        \tE^{i\mu\nu} = \utilde{X}^{(ij)} \epsilon^{jkl} \tE^{k\mu} \tE^{l\nu} + 2 N^{[\mu} \tE^{i\nu]}. \label{eq:num-bivector-complex-decomposition}
    \end{align}
    Counting the number of components at this point gives $6\C + 3\C + 9\C = 18\C$ for $\utilde{X}^{(ij)}, X^a$ and $\tE^{ia}$ respectively. This is equal to the number of components in a generic triple of complex 2-forms, as expected. We want the bivectors to encode only a real-valued metric and an $SO(3,\C)$ frame; as we have seen, reducing the number of components requires imposing the metricity and reality conditions.
    
    \subsection{Decomposing the Metricity Condition}
    First we focus on the metricity condition,
    \begin{align}
        \Sigma^i \wedge \Sigma^j \sim \delta^{ij}.
    \end{align}
    Writing this out in components gives 
    \begin{align}
        \Sigma^i \wedge \Sigma^j = \frac{1}{4}\Sigma^i_{\mu\nu} \Sigma^j_{\rho\sigma} \teps^{\mu\nu\rho\sigma} d^4x.
    \end{align}
    Using $\Sigma^i_{\mu\nu} = -\frac{1}{2} \uteps_{\mu\nu\rho\sigma} \tE^{i\rho\sigma}$, we can write this condition as 
    \begin{align}
        \tE^{i\mu\nu} \tE^{j\rho\sigma} \uteps_{\mu\nu\rho\sigma} \sim \delta^{ij}.
    \end{align}
    Here $\uteps_{\mu\nu\rho\sigma}$ is the totally antisymmetric symbol with lower indices. Substituting \cref{eq:num-bivector-complex-decomposition} into this and expanding, we find
    \begin{align}
        \tE^{i\mu\nu} \tE^{j\rho\sigma} \uteps_{\mu\nu\rho\sigma} = 4\utilde{X}^{(i|k} \epsilon^{klm} \tE^{l\mu} \tE^{m\nu} N^\rho \tE^{|j)\sigma} \utilde{\epsilon}_{\mu\nu\rho\sigma},
    \end{align}
    where we now assume that $\utilde{X}^{ij}$ is symmetric. To continue we note the identity 
    \begin{align}
        \uteps_{\mu\nu\rho\sigma} N^\mu \tE^{i\nu} \tE^{j\rho} \tE^{k\sigma} = \epsilon^{ijk} \det(\tE) \label{eq:num-epsilon-N-E-E-to-eps-det}
    \end{align}
    where $\det(\tE) = \frac{1}{6} \epsilon^{ijk} \uteps_{abc} \tE^{ia} \tE^{jb} \tE^{kc}$, and we have fixed the orientation to be $\teps^{0abc} = -\teps^{abc}$ or $\uteps_{0abc} = +\uteps_{abc}$. To see that this holds, note that the left-hand side is totally antisymmetric in $ijk$ and must therefore be a multiple of $\epsilon^{ijk}$. Contracting both sides with $\epsilon^{ijk}$ and splitting into 3+1 coordinates, we find that both sides agree. Applying this identity to the metricity condition gives
    \begin{align}
        \tE^{i\mu\nu} \tE^{j\rho\sigma} \uteps_{\mu\nu\rho\sigma} \sim 4 \utilde{X}^{(i|k} \epsilon^{klm} \epsilon^{|j)lm} = 8 \utilde{X}^{(ij)} \sim \delta^{ij}.
    \end{align}
    As $\utilde{X}^{ij}$ is symmetric, this implies that there is a scalar function $\utilde{X}$ such that $\utilde{X}^{ij} = \utilde{X} \delta^{ij}$. This leaves us with the decomposition 
    \begin{align}
        \tE^{i\mu\nu} = \utilde{X} \epsilon^{ijk} \tE^{j\mu} \tE^{k\nu} + 2 N^{[\mu} \tE^{i\nu]}.
    \end{align}
    Counting the remaining components gives $13 \C = 10\C + 3\C$, enough to form a complex-valued metric and an $SO(3,\C)$ frame. Thus the metricity condition removes the unwanted components that do not belong in the usual theory of gravity.

    \subsection{Decomposing the Reality Conditions}
    To restrict ourselves to a real-valued metric, we need to impose the reality conditions,
    \begin{align}
        \Sigma^i \wedge \asd^j = 0, \quad \re(\Sigma^i \wedge \Sigma^i) = 0.
    \end{align}
    The second condition imposes
    \begin{align}
        \utilde{X} \det(\tE) \sim \im V, \label{eq:num-plebanski-1R-reality-condition}
    \end{align}
    where $V$ is some real-valued function to be determined later. The first reality condition requires the anti-self-dual 2-forms. In Lorentzian signature these are related to the complex conjugates of the self-dual 2-forms, so their decomposition follows from that of the self-dual 2-forms:
    \begin{align}
        \bar{\tE}^{i\mu\nu} = \bar{\utilde{X}}  \epsilon^{ijk} \bar{\tE}^{j\mu} \bar{\tE}^{k\nu} - 2 \bar{N}^{[\mu} \bar{\tE}^{i\nu]}.
    \end{align}
    Here the bar above a term denotes the negative complex conjugate, i.e.\ $\bar{X} = -X^*$ where $^*$ denotes complex conjugation. Inserting this into the first reality condition and expanding, we find
    \begin{equation} \label{eq:pleb-num-reality-condition-step-1}
        \begin{aligned}
                \utilde{X} \bar{\utilde{X}} \epsilon^{ikl} \epsilon^{jmn} \tE^{k\mu} \tE^{l\nu} \bar{\tE}^{m\rho} \bar{\tE}^{n\sigma} \uteps_{\mu\nu\rho\sigma} - 2\utilde{X} \epsilon^{ikl} \tE^{k\mu} \tE^{l\nu} \bar{N}^\rho \bar{\tE}^{j\sigma} \uteps_{\mu\nu\rho\sigma} \\ +  2 \bar{\utilde{X}} \epsilon^{jmn} \bar{\tE}^{m\rho} \bar{\tE}^{n\sigma} N^\mu \tE^{i\nu} \uteps_{\mu\nu\rho\sigma} - 4 N^\mu \bar{N}^\rho \tE^{i\nu} \bar{\tE}^{j\sigma} \uteps_{\mu\nu\rho\sigma} = 0.
        \end{aligned}
    \end{equation}
    To simplify this we note that the complex conjugates $\bar{\tE}^{i\mu}$ are still purely spatial with respect to $dt$ and therefore span the same space as $\tE^{i\mu}$. The first term is thus an antisymmetrised product of four 3-vectors, which vanishes. As for the middle two terms, we introduce $\utE^i_\mu$ as the inverse of the triad $\tE^{i\mu}$, i.e.
    \begin{align}
        \tE^{i\mu} \utE^j_\mu = \delta^{ij}, \quad \tE^{i\mu} \utE^i_\nu = \delta^\mu_\nu - N^\mu (dt)_\nu.
    \end{align}
    This implies that its components are $\utE^i_\mu = (-N^b \utE^i_b, \utE^i_a)$, where $\utE^i_a$ are the inverses of the spatial triads,
    \begin{align}
        \tE^{ia} \utE^j_a = \delta^{ij}, \quad \tE^{ia} \utE^i_{b} = \delta^a_b.
    \end{align}
    Using the inverse, we can simplify the second term in the first line of \cref{eq:pleb-num-reality-condition-step-1} to find that
    \begin{align}
        \utilde{X} \epsilon^{ikl} \tE^{k\mu} \tE^{l\nu} \bar{N}^\rho \bar{\tE}^{j\sigma} \uteps_{\mu\nu\rho\sigma} = - 2 \utE^i_\mu \bar{\tE}^{j\mu} \utilde{X} \det(\tE).
    \end{align}
    Splitting both sides into 3+1 components shows that they agree. Taking the complex conjugate of this and using the definition of the barred variables, we obtain
    \begin{align}
        \bar{\utilde{X}} \epsilon^{ikl} \bar{\tE}^{k\mu} \bar{\tE}^{l\nu} N^\rho \tE^{j\sigma} \uteps_{\mu\nu\rho\sigma} & = 2 \bar{\utE}^i_\mu \tE^{j\mu} \bar{\utilde{X}} \det(\bar{\tE}) \\
        & = - 2 \bar{\utE}^i_\mu \tE^{j\mu} \utilde{X} \det(\tE).
    \end{align}
    In the last equality we have used \cref{eq:num-plebanski-1R-reality-condition} to show that $\bar{\utilde{X}} \det(\bar{\tE}) = - \utilde{X} \det(\tE)$. The reality condition then becomes
    \begin{align}
        \utE^i_\mu \bar{\tE}^{j\mu} \utilde{X} \det(\tE) - \bar{\utE}^j_\mu \tE^{i\mu} \utilde{X} \det(\tE) - N^\mu \bar{N}^\rho \tE^{i\nu} \bar{\tE}^{j\sigma} \uteps_{\mu\nu\rho\sigma} = 0.
    \end{align}
    To compute the last term, we split the vector $N^\mu$ as $N^\mu = t^\mu - \utilde{N}^i \tE^{i\mu}$, where $t^\mu = (1,0)$ is purely timelike. A similar split is available for $\bar{N}^\mu = -t^\mu + \bar{\utilde{N}}^i \bar{\tE}^{i\mu}$, where the bar operator is defined the same way with respect to the unbarred fields. The only terms that survive in the last term of the equation above are 
    \begin{align}
        N^\mu \bar{N}^\rho \tE^{i\nu} \bar{\tE}^{j\sigma} \uteps_{\mu\nu\rho\sigma} = -\utilde{N}^k \tE^{k\mu} \bar{N}^\rho \tE^{i\nu} \bar{\tE}^{j\sigma} \uteps_{\mu\nu\rho\sigma} + \bar{\utilde{N}}^k \bar{\tE}^{k\mu} N^\rho \tE^{i\nu} \bar{\tE}^{j\sigma} \uteps_{\mu\nu\rho\sigma}.
    \end{align}
    Using the previous results, this becomes
    \begin{align}
        N^\mu \bar{N}^\rho \tE^{i\nu} \bar{\tE}^{j\sigma} \uteps_{\mu\nu\rho\sigma} = -\epsilon^{ikl} \utilde{N}^k \utE^l_\mu  \bar{\tE}^{j\mu} \det(\tE) - \epsilon^{jkl}  \bar{\utilde{N}}^k \bar{\utE}^l_\mu \tE^{i\mu} \det(\bar{\tE}).
    \end{align}
    Collecting all the results, the reality condition becomes
    \begin{align}
        \utE^i_\mu \bar{\tE}^{j\mu} \utilde{X} \det(\tE) - \bar{\utE}^j_\mu \tE^{i\mu} \utilde{X} \det(\tE) + \epsilon^{ikl} \utilde{N}^k \utE^l_\mu  \bar{\tE}^{j\mu} \det(\tE) + \epsilon^{jkl}  \bar{\utilde{N}}^k \bar{\utE}^l_\mu \tE^{i\mu} \det(\bar{\tE}) = 0.
    \end{align}
    To understand this condition we introduce the matrix $K^{ij} = \utE^i_\mu \bar{\tE}^{j\mu}$ along with the vector $k^i = \frac{\utilde{N}^i}{\utilde{X}}$, such that the above becomes
    \begin{align}
        K^{ij} + \bar{K}^{ji} + \epsilon^{ikl} k^k K^{lj} - \epsilon^{jkl} \bar{k}^k \bar{K}^{li} = 0,
    \end{align}
    where we have made use of \cref{eq:num-plebanski-1R-reality-condition} and removed the factors of $\utilde{X} \det(\tE)$. We note that $-\bar{K}^{ij}$ is the inverse of $K^{ij}$; indeed, computing
    \begin{align}
        - K^{ik} \bar{K}^{kj} = \utE^i_\mu \bar{\tE}^{k\mu} \bar{\utE}^k_\nu \tE^{j\nu} = \utE^i_\mu (\delta^\mu_\nu - N^\mu (dt)_\nu) \tE^{j\nu} = \utE^i_\mu \tE^{j\mu} = \delta^{ij}
    \end{align}
    confirms this. We can then multiply the reality condition by the appropriate factor of $-\bar{K}$ to find
    \begin{align}
        \delta^{ij} - \bar{K}^{ki} \bar{K}^{kj} + \epsilon^{ijk} k^k + \epsilon^{kml} \bar{k}^l \bar{K}^{mi} \bar{K}^{kj} = 0.
    \end{align}
    The first two terms are symmetric in $ij$ and the last two are antisymmetric, so they can be treated separately. The first two terms then imply that $\bar{K}^T = \bar{K}^{-1}$, or equivalently
    \begin{align}
        K^{ij} = -\bar{K}^{ji},
    \end{align}
    which, after multiplying by $\tE^{i\mu}$ and $\bar{\tE}^{j\nu}$, implies that
    \begin{align}
        \tE^{i\mu} \tE^{i\nu} = \bar{\tE}^{i\mu} \bar{\tE}^{i\nu}.
    \end{align}
    Hence the tensor
    \begin{align}
        \tilde{\tilde{\gamma}}^{\mu\nu} = \tE^{i\mu} \tE^{i\nu}
    \end{align}
    is real-valued. An immediate consequence of this is that $\det(\tE) = \pm \det(\bar{\tE})$ and therefore $\det(\bar{K}) = \pm 1$. Without loss of generality, we choose the branch $\det(\tE) = -\det(\bar{\tE})$, which implies that $\det(\tE) = \det(\tE^*)$, $\det(\bar{K}) = 1$ and $\utilde{X} = \bar{\utilde{X}}$. This allows us to identify $\bar{K} \in SO(3,\C)$ and make use of the identity
    \begin{align}
        \epsilon^{kml} \bar{K}^{mi} \bar{K}^{kj}= \epsilon^{jik} \bar{K}^{lk},
    \end{align}
    such that the antisymmetric terms in the reality condition become
    \begin{align}
        \epsilon^{ijk} k^k + \epsilon^{kml} \bar{k}^l \bar{K}^{mi} \bar{K}^{kj} = \epsilon^{ijk} k^k - \epsilon^{ijk} \bar{K}^{lk} \bar{k}^l = 0.
    \end{align}
    The only independent condition is therefore
    \begin{align}
        k^i - \bar{K}^{ji} \bar{k}^j = 0.
    \end{align}
    Substituting the definitions of $K$ and $k$, we find
    \begin{align}
        \frac{\utilde{N}^i}{\utilde{X}} - \bar{\utE}^j_\mu \tE^{i\mu} \frac{\bar{\utilde{N}}^j}{\bar{\utilde{X}}} = 0.
    \end{align}
    Using $\utilde{X} = \bar{\utilde{X}}$ and multiplying by $\tE^{i\mu}$, we find
    \begin{align}
        \bar{\utilde{N}}^i \bar{\tE}^{i\mu} - \utilde{N}^i \tE^{i\mu} = 0.
    \end{align}
    Finally, using the definitions of $N^\mu$ and $\bar{N}^\mu$, this is equivalent to
    \begin{align}
        N^\mu + \bar{N}^\mu = 0, \quad \textrm{ or } \quad N^\mu = (N^\mu)^*.
    \end{align}
    Collecting all the results of the reality conditions, we find
    \begin{align}
        \utilde{X} = \pm \im \dlap \in \im \R, \quad N^\mu \in \R, \quad \tE^{i\mu} \tE^{i\nu} \in \R.
    \end{align}
    Since $N^\mu$ is real, the 3-vector $N^a$ is also real-valued. The decomposition of the bivector that satisfies the reality and metricity conditions is then
    \begin{align}
        \tE^{i\mu\nu} = \pm \im \dlap \epsilon^{ijk} \tE^{j\mu} \tE^{k\nu} + 2 N^{[\mu} \tE^{i\nu]}. \label{eq:num-plebanski-bivector-real-decomposition}
    \end{align}
    Counting the number of components, we see that there are $1\R + 3\R + 6\R + 3\C = 10\R + 3\C$, which make up the real metric and the choice of $SO(3,\C)$ frame.

    \subsection{Recovering the Metric}
    Now that we have the decomposition of $\Sigma^i$, we can use it together with the \urb{} formula to compare the \pleb{} description with the usual ADM metric language. To do so, we make use of the inverse \urb{} formula
    \begin{align}
        \sqrt{-g} \tg^{\mu\nu} = -\frac{1}{12} \epsilon^{ijk} \tE^{i\mu\rho} \tE^{j\alpha\beta} \tE^{k\sigma\nu} \uteps_{\rho\sigma\alpha\beta}.
    \end{align}
    Replacing $\tE^{i\mu\nu}$ with $\Sigma^i$ and using the quaternion algebra, we find that this is indeed the inverse metric.
    Here $g_{\mu\nu}$ is the physical metric, $\sqrt{-g} = \sqrt{-\det(g)}$ and $\tg^{\mu\nu} = \sqrt{-g} g^{\mu\nu}$. Substituting \cref{eq:num-plebanski-bivector-real-decomposition} into this, expanding all the terms and cancelling terms by symmetry, we find
    \begin{align}
        \sqrt{-g} \tg^{\mu\nu} = & \left(\frac{1}{12} \dlap^2 \tE^{l\mu} \tE^{l\nu} - \frac{1}{6} N^\mu N^\nu \right) \epsilon^{ijk} \uteps_{\alpha\beta\rho\sigma} N^\alpha \tE^{i\beta} \tE^{j\rho} \tE^{k\sigma} \\ & + \dlap^2 \tE^{i\mu} \tE^{i\nu} \left(\frac{1}{3} \epsilon^{jkl} \tE^{j\beta} - \frac{1}{12} \epsilon^{ikl} \tE^{i\beta} \right) \uteps_{\alpha\beta\rho\sigma} N^\alpha \tE^{k\rho} \tE^{l\sigma}.
    \end{align}
    Making use of \cref{eq:num-epsilon-N-E-E-to-eps-det} and expanding the product of two epsilons, we find
    \begin{align}
        \sqrt{-g} \tg^{\mu\nu} = \dlap \det(\tE) \left( \dlap \tE^{i\mu} \tE^{i\nu} - \frac{N^\mu N^\nu}{\dlap} \right).
    \end{align}
    We note that the sign in front of the first term in \cref{eq:num-plebanski-bivector-real-decomposition} does not change the definition of the metric. To remove the factors of the determinant of the metric, we compute the determinant of $\sqrt{-g} \tg^{\mu\nu}$ to find
    \begin{align}
        \det(\sqrt{-g} \tg) = \det(g)^3 = - \dlap^6 \det(\tE)^6, \quad \Rightarrow \sqrt{-g} = \pm \dlap \det(\tE).
    \end{align}
    We fix the sign by requiring that $\dlap \det(\tE) > 0$, so that the square root of the determinant of the metric is 
    \begin{align}
        \sqrt{-g} = \dlap \det(\tE).
    \end{align}
    Therefore, the inverse densitised metric is 
    \begin{align}\label{eq:3+1-inverse-densitised-metric-decomp}
        \tg^{\mu\nu} = \dlap \tE^{i\mu} \tE^{i\nu} - \frac{N^\mu N^\nu}{\dlap},
    \end{align}
    and for the inverse metric we find 
    \begin{align} 
        g^{\mu\nu} = \frac{\tE^{i\mu} \tE^{i\nu}}{\det(\tE)} - \frac{N^\mu N^\nu}{\dlap^2 \det(\tE)}.
    \end{align}
    The lapse is
    \begin{align}
        N = (-(dt)_\mu (dt)_\nu g^{\mu\nu})^{-\frac{1}{2}} = \dlap \sqrt{\det(\tE)}.
    \end{align}
    The shift vector, $N^a$, is identified with the 3-vector inside $N^\mu = (1,-N^a)$. The inverse spatial metric is 
    \begin{align}
        \gamma^{\mu\nu} = \frac{\tE^{i\mu} \tE^{i\nu}}{\det(\tE)}.
    \end{align}
    This gives a dictionary between the metric and \pleb{} components.

    \subsection{Enforcing Self-Duality}
    The final step in this decomposition is to fix the sign in \cref{eq:num-plebanski-bivector-real-decomposition}; to do so, we require these bivectors to be self-dual. We compute
    \begin{align}
        \frac{1}{2} \tg^{\mu\rho} \tg^{\nu\sigma} \uteps_{\rho\sigma\alpha\beta} \tE^{i\alpha\beta} = \mp \im \dlap \det(\tE) \left( \pm \im \dlap \epsilon^{ijk} \tE^{j\mu} \tE^{k\nu} + 2 N^{[\mu} \tE^{i\nu]} \right) = \mp \im \dlap \det(\tE) \tE^{i\mu\nu},
    \end{align}
    where the $\pm$ takes the same sign as the one in \cref{eq:num-plebanski-bivector-real-decomposition}. We have used the decomposition of $\tg^{\mu\nu}$ and $\tE^{i\mu\nu}$ into $\tE^{i\mu}, N^\mu, \dlap$ and then the previous identities to obtain the right-hand side. The calculation is lengthy, so we state only the result. The self-dual bivectors are then given by
    \begin{align} \label{eq:3+1-decomp-tE-spacetime}
        \tE^{i\mu\nu} = - \im \dlap \epsilon^{ijk} \tE^{j\mu} \tE^{k\nu} + 2 N^{[\mu} \tE^{i\nu]}.
    \end{align}

\section{3+1 Equation of Motion Splitting}\label{apx:3+1-eom-split}

Here we display the 3+1 decomposition of the equations of motion.

\subsection{Original Frame Equations of Motion}

Beginning with the first equation in \cref{eq:nonlin-hyperbolic-comp-evo-sys}, we substitute $\mu = 0$ and expand to find

\begin{align}
    \partial^A_a \tE^{ia} + (c_1 + c_2) \epsilon^{ijk} \tE^{ja} \partial_a \chi^k - 2\im c_2 \frac{1}{\dlap}(\partial_t - N^a \partial_a)\chi^i = 0.
\end{align}

This can be rearranged as

\begin{align} \label{eq:apx-3+1-eom:chi-eom}
    (\partial_t - N^a \partial_a) \chi^i = \frac{\dlap}{2\im c_2} \partial^A_a \tE^{ia} + \frac{c_1+c_2}{2\im c_2} \epsilon^{ijk} \dlap \tE^{ja} \partial_a \chi^k.
\end{align}

This is the evolution equation for $\chi^i$. Next, we take the $\mu = a$ component:

\begin{align}
    \partial_t^A ( - \tE^{ia} ) + \partial_b^A ( -\im \epsilon^{ijk} \dlap \tE^{ja} \tE^{kb} - N^a \tE^{ib} + N^b \tE^{ia} ) + (c_1 + c_2) \epsilon^{ijk} (-\tE^{ja}) \partial_t \chi^k \nonumber \\ + (c_1 + c_2) \epsilon^{ijk} (-\im \epsilon^{jlm} \dlap \tE^{la} \tE^{mb} - N^a \tE^{jb} + N^b \tE^{ja}) \partial_b \chi^k + 2\im c_2 N^a \partial_t \chi^i + 2\im c_2 (\tE^{ja} \tE^{jb} - N^a N^b) \partial_b \chi^i = 0.
\end{align}

Using \cref{eq:apx-3+1-eom:chi-eom} to replace the time derivatives of $\chi^i$ and then solving for $(\partial^A_t - N^b \partial^A_b)\tE^{ia}$ gives

\begin{align}
 (\partial^A_t - N^b \partial^A_b) \tE^{ia} = & - \im \epsilon^{ijk} \tE^{kb} \partial_b^A \left(\dlap \tE^{ja}\right) - \tE^{ib} \partial_b N^a + \tE^{ia} \partial_b N^b + \im (c_1 + c_2) \dlap \tE^{ia} \tE^{jb} \partial_b \chi^j \nonumber \\ & + \frac{c_2-c_1}{2 \im c_2}\left( \epsilon^{ijk} \dlap \tE^{ja} \partial_b^A \tE^{kb} + (c_1+c_2) \dlap \tE^{ja} \tE^{ib} \partial_b \chi^j \right. \nonumber \\ & \left. \hspace{150pt} - (3c_2 + c_1) \dlap \tE^{ja} \tE^{jb} \partial_b \chi^i \right).
\end{align}

We now move on to the second equation in \cref{eq:nonlin-hyperbolic-comp-evo-sys}, for which we substitute $\mu = 0$ to find

\begin{align}
    \partial_t \left( -\frac{1}{\dlap} \right) + \partial_a \left( \frac{N^a}{\dlap} \right) - \frac{1}{\dlap} ( \xi_0 - N^a \xi_a) + 2 \im c_2 \tE^{ia} \partial_a \chi^i = 0.
\end{align}

This can be solved to give an evolution equation for $\dlap$,

\begin{align}
    (\partial_t - N^a \partial_a)\dlap = & - \dlap \partial_a N^a - 2 \im c_2 \dlap^2 \tE^{ia} \partial_a \chi^i - \dlap ( \xi_0 - N^a \xi_a).
\end{align}

Choosing instead the $\mu = a$ component gives

\begin{align}
    \partial_t \left( \frac{N^a}{\dlap} \right) + \partial_b \left( \dlap \tE^{ia} \tE^{ib} - \frac{N^a N^b}{\dlap} \right) - \frac{N^a}{\dlap} \xi_0 - \left(\dlap \tE^{ia} \tE^{ib} - \frac{N^a N^b}{\dlap} \right) \xi_b - 2 \im c_2  \tE^{ia} \partial_t \chi^i \nonumber \\ + 2 \im c_2 ( -\im \epsilon^{ijk} \dlap \tE^{ja} \tE^{kb} - N^a \tE^{ib} + N^b \tE^{ia} ) \partial_b \chi^i = 0.
\end{align}

As before, we substitute the evolution equations for $\chi^i$ and $\dlap$ to find an evolution equation for $N^a$,

\begin{align}
    (\partial_t - N^b \partial_b) N^a = & - \dlap \tE^{ib} \partial_b^A( \dlap \tE^{ia}) - (c_2 - c_1) \epsilon^{ijk} \dlap^2 \tE^{ja} \tE^{kb} \partial_b \chi^i + \dlap^2 \tE^{ia} \tE^{ib} \xi_b.
\end{align}

This exhausts the equations of motion for the frame sector.

\subsection{Original Connection Equations of Motion}

Having derived the evolution system for the frame sector fields, we now do the same for the connection sector. We begin with the $\mu,\nu = 0,a$ component of the third equation in \cref{eq:nonlin-hyperbolic-comp-evo-sys} to find 

\begin{align}
 \tE^{i0b} F^i_{ab} + \frac{\im}{2} \tg^{00} \left(\partial_t \xi_a + \partial_a \xi_0\right) + \frac{\im}{2} \tg^{0b} \left(\partial_a \xi_b + \partial_b \xi_a\right) = 0.
\end{align}

Using the decomposition and solving for $(\partial_t - N^b \partial_b) \xi_a$, we find

\begin{align}
    (\partial_t - N^b \partial_b) \xi_a = -\partial_a \xi_0 + N^b \partial_a \xi_b - 2 \im \dlap \tE^{ib} F^i_{ab}.
\end{align}

Next, we look at the $\mu,\nu = a,b$ components of the same equation,

\begin{align}
 \tE^{ia0} F^i_{b0} + \tE^{iac} F^i_{bc} + \frac{\im}{2} \tg^{a0} \left(\partial_b \xi_0 + \partial_t \xi_a \right) + \frac{\im}{2} \tg^{ac} \left(\partial_b \xi_c + \partial_c \xi_b\right) = 0.
\end{align}

Using the decomposition gives 

\begin{align}
    \tE^{ia} F^i_{0b} + \left(- \im \epsilon^{ijk} \dlap \tE^{ja} \tE^{kc} + \tE^{ia} N^c - \tE^{ic} N^a\right) F^i_{bc} \nonumber \\ + \frac{\im}{2} N^a (\partial_b \xi_0 + \partial_t \xi_b) + \frac{\im}{2}(\dlap^2 \tE^{ia} \tE^{ic} - N^a N^c)(\partial_b \xi_c + \partial_c \xi_b) = 0.
\end{align}

Collecting into useful factors, we have

\begin{align}
    \tE^{ia}( F^i_{0b} + N^c F^i_{bc} - \im \epsilon^{ijk} \dlap \tE^{jc} F^k_{bc} + \im \dlap \tE^{ic} \partial_{(b} \xi_{c)}) \nonumber \\ + \frac{\im}{2} N^a \left( (\partial_t - N^c \partial_c) \xi_b + \partial_b \xi_0 - N^c \partial_b \xi_c - 2 \im \dlap \tE^{ic} F^i_{bc} \right) = 0.
\end{align}

The term in the brackets on the second line vanishes by the evolution equation for $\xi_a$, and multiplying by $\utE^i_a$ we find

\begin{align}
    F^i_{0a} = - N^b F^i_{ab} + \im\epsilon^{ijk} \dlap \tE^{jb} F^k_{ab} - \im \dlap \tE^{ib} \partial_{(a} \xi_{b)}.
\end{align}

To put this in the form of an evolution equation, we expand the first two curvatures using $F^i_{0a} = \partial_t A^i_a - \partial_a A^i_0 + \epsilon^{ijk} A^j_0 A^k_a = \partial_t A^i_a - \partial^A_a A^i_0$ and similarly $F^i_{ab} = \partial^A_a A^i_b - \partial_b A^i_a$, such that

\begin{align}
    (\partial_t - N^b \partial_b) A^i_a = \partial^A_a A^i_0 - N^b \partial^A_a A^i_b + \im \epsilon^{ijk} \dlap \tE^{jb} F^k_{ab} - \im \dlap \tE^{ib} \partial_{(a} \xi_{b)}.
\end{align}
The only remaining component is $\mu,\nu = 0,0$, for which we find
\begin{align}
    \dlap \tE^{i0a} F^i_{0a} + \im \dlap \tg^{00} \partial_t \xi_0 + \frac{\im}{2} \dlap \tg^{0a} (\partial_t \xi_a + \partial_a \xi_0) = 0,
\end{align}

which, using the decomposition, becomes

\begin{align}
    \dlap \tE^{ia} F^i_{0a} - \im (\partial_t + N^a \partial_a) \xi_0 + \frac{\im}{2} N^a ( \partial_t \xi_a - \partial_a \xi_0) = 0.
\end{align}
Rearranging this for $(\partial_t - N^a \partial_a) \xi_0$ gives
\begin{align}
    (\partial_t - N^a \partial_a)\xi_0 = - N^a \partial_a \xi_0 - (\dlap^2 \tE^{ia} \tE^{ib} - N^a N^b) \partial_a \xi_b \nonumber \\ -2 \im \dlap N^a \tE^{ib} F^i_{ab} + \epsilon^{ijk} \dlap^2 \tE^{ia} \tE^{jb} F^k_{ab}.
\end{align}

The final evolution equation comes from the Lorenz condition, \cref{eq:num-plebanski-lorenz-gauge-components}. Expanding this into 3+1 components gives

\begin{align}
    \tg^{00} \partial_t A^i_0 + \tg^{0a} (\partial_t A^i_a + \partial_a A^i_0) + \tg^{ab} \partial_a A^i_b + \frac{\im}{2} \tE^{i0a} ( \partial_t \xi_a - \partial_a \xi_0) + \frac{\im}{2} \tE^{iab} \partial_a \xi_b = 0.
\end{align}

Using the decomposition and collecting terms, we find

\begin{align}
    -\partial_t A^i_0 + N^a (\partial_t A^i_a + \partial_a A^i_0) + (\dlap^2 \tE^{ja} \tE^{jb} - N^a N^b) \partial_a A^i_b + \frac{\im}{2} \tE^{ia}( \partial_t \xi_a - \partial_a \xi_0) \nonumber \\ + \frac{1}{2} \epsilon^{ijk} \dlap \tE^{ja} \tE^{kb} \partial_a \xi_b + \frac{\im}{2} \tE^{ia} N^b \partial_a \xi_b + \frac{\im}{2} \tE^{ib} N^a \partial_b \xi_a = 0.
\end{align}

Substituting the previous evolution equations for $A^i_a$ and $\xi_a$ and solving for $(\partial_t -N^a \partial_a) A^i_0$ gives our final connection evolution equation,

\begin{align}
    (\partial_t - N^a \partial_a) A^i_0 = & \ (\dlap^2 \tE^{ja} \tE^{jb} - N^a N^b) \partial_a A^i_b + N^a \partial^A_a A^i_0 - \im \epsilon^{ijk} \dlap \tE^{ja} N^b F^k_{ab} \nonumber \\ &\ + \dlap^2 \tE^{ia} \tE^{jb} F^j_{ab} - \im \dlap \tE^{ia} \partial_a \xi_0 - \im N^a \dlap \tE^{ib} \partial_{[a} \xi_{b]} \nonumber \\ &\ + \frac{1}{2} \epsilon^{ijk} \dlap^2 \tE^{ja} \tE^{kb} \partial_a \xi_b.
\end{align}

This completes the decomposition of the connection equations of motion.

\subsection{Conformal Frame Equations of Motion}

In this subsection we derive the equations of motion for the conformally related fields. Instead of decomposing \cref{eq:nonlin-conformal-comp-evo-sys}, it is easier to use the translation from the original fields to the conformal ones, together with the original field evolution equations.

We start with the conformal 3-frame, $H^{ia}$, for which

\begin{align}
    (\partial_t - N^b \partial_b) H^{ia} & = \dlap (\partial_t - N^b \partial_b)\tE^{ia} + \tE^{ia} (\partial_t - N^b \partial_b) \dlap \nonumber \\ & = -  \epsilon^{ijk} (A^j_0 - N^b A^j_b) \dlap \tE^{ka} - \im \epsilon^{ijk} \dlap \tE^{kb} \partial_b^A (\dlap \tE^{ja}) - \dlap \tE^{ib} \partial_b N^a + \dlap \tE^{ia} \partial_b N^b \nonumber \\ & \quad + \im (c_1 + c_2) \dlap^2 \tE^{ia} \tE^{jb} \partial_b \chi^j + \frac{c_2 - c_1}{2 \im c_2} \left( \epsilon^{ijk} \dlap^2 \tE^{ja} \partial^A_b \tE^{kb} + (c_1 + c_2) \dlap^2 \tE^{ja} \tE^{ib} \partial_b \chi^j \right. \nonumber \\ & \quad \left. - (3 c_2 + c_1) \dlap^2 \tE^{ja} \tE^{jb} \partial_b \chi^i \right) - \dlap \tE^{ia} \partial_b N^b - 2\im c_2 \dlap^2 \tE^{ia} \tE^{jb} \partial_b \chi^j - \dlap \tE^{ia} (\xi_0 - N^b \xi_b) \nonumber \\
    & = -\epsilon^{ijk} (A^j_0 - N^b A^j_b) H^{ka} - H^{ia} (\xi_0 - N^b \xi_b) - \im \epsilon^{ijk} H^{kb} \partial_b^A H^{ja} - H^{ib} \partial_b N^a + \frac{c_2 - c_1}{2\im c_2} \left( 2 c_2 H^{ia} H^{jb} \partial_b \chi^j \right. \nonumber \\ & \quad \left. + \epsilon^{ijk} H^{ja} \partial^A_b H^{kb} - \epsilon^{ijk} H^{ja} H^{kb} \partial_b n + (c_1 + c_2) H^{ja} H^{ib} \partial_b \chi^j - (3c_2 + c_1) H^{ja} H^{jb} \partial_b \chi^i \right).
\end{align}

If we expand $\partial^A \chi^i = \partial \chi^i + \epsilon^{ijk} A^j \chi^k$ in the terms on the r.h.s.\ of the last equality, they can be combined into $\Omega^i_a$ and $\Omega^i_0$. Doing this gives

\begin{align}
    (\partial^\Omega_t - N^b \partial^\Omega_b) H^{ia} & = -\im \epsilon^{ijk} H^{kb} \partial_b^\Omega H^{ja} - H^{ib} \partial_b N^a + \frac{c_2-c_1}{2\im c_2} \left( \epsilon^{ijk} H^{ja} \partial_b^A H^{kb} - \epsilon^{ijk} H^{ja} H^{kb} \partial_b n +  2 c_2 H^{ia} H^{jb} \partial_b \chi^j \right. \nonumber \\ & \quad \left. + (c_1 + c_2) H^{ja} H^{ib} \partial_b \chi^j - (3 c_2 + c_1) H^{ja} H^{jb} \partial_b \chi^i \right).
\end{align} 

The first term in the brackets retains the derivative $\partial^A$ because it does not separate neatly into $\Omega^i$ and $\omega$; instead, we regard $A^i = A^i(\Omega,\omega)$ via the translation given in the main text.

The remaining frame evolution equations are easier to translate. For the lapse equation we simply find

\begin{align}
    (\partial_t - N^a \partial_a) n = & -\partial_a N^a - 2 \im c_2 H^{ia} \partial_a \chi^i - \xi_0 + N^a \xi_a.
\end{align}

For the shift, we have

\begin{align}
    (\partial_t - N^b \partial_b) N^a = & - H^{ib} \partial_b^A H^{ia} - (c_2 - c_1) \epsilon^{ijk} H^{ja} H^{kb} \partial_b \chi^i + H^{ia} H^{ib} \xi_b \nonumber \\ = & -H^{ib} \partial^\Omega_b H^{ia} - (c_2 - c_1) \epsilon^{ijk} H^{ja} H^{kb} \partial_b \chi^i.
\end{align}

Here the connections $A^i$ and $\xi$ have been replaced with $\Omega^i$. Lastly, for the gauge fields $\chi^i$ we find

\begin{align}
    (\partial_t - N^a \partial_a) \chi^i = \frac{1}{2\im c_2} \partial_a^A H^{ia} - \frac{1}{2 \im c_2} H^{ia} \partial_a n + \frac{c_1 + c_2}{2 \im c_2} \epsilon^{ijk} H^{ja} \partial_a \chi^k.
\end{align}

Again, the connection terms are written assuming that $A^i = A^i(\Omega,\omega)$, since they do not simplify further. This completes the conformal frame 3+1 decomposition. 

\subsection{Conformal Connection Equations of Motion}

To decompose the first equation on the third line in \cref{eq:nonlin-conformal-comp-evo-sys} we first contract the $\mu,\nu$ indices with $\Sigma^{k\mu\sigma} \asd^j{}_\sigma{}^\nu$ and, using the algebra of the $\Sigma^i$, find that this is equivalent to

\begin{align}
    \asd^{j\mu\nu} F{(\Omega)}^i_{\mu\nu} = 0.
\end{align}

No equations are lost, since $\Sigma^{i\mu\sigma} \asd^j{}_\sigma{}^\nu$ forms a complete basis for the symmetric tracefree $4\times 4$ tensors. To start decomposing $\asd^i$, it is easiest to relate it to the already computed self-dual 2-forms $\sd^i$. Recall that the bar operator represents minus the complex conjugate, i.e.\ $\asd^i = -(\sd^*)^i$. This allows us to define $\tbE^{i\mu\nu} = \frac{1}{2} \teps^{\mu\nu\rho\sigma} \asd^i_{\rho\sigma} = - (\tE^*)^{i\mu\nu}$. Therefore, the decomposition is

\begin{align}
    \tbE^{i0a} = \tbE^{ia}, \quad \tbE^{iab} = - \im \epsilon^{ijk} \dlap \tbE^{ja} \tbE^{kb} + \tbE^{ia} N^b - \tbE^{ib} N^a,
\end{align}

with $\tbE^{ia} = -(\tE^*)^{ia}$. Applying this to the connection equation of motion produces

\begin{align}
    \tbE^{j\mu\nu} F{(\Omega)}^i_{\mu\nu} = 2 \tbE^{jb} F{(\Omega)}^i_{0b} - \im \epsilon^{jkl} \dlap \tbE^{kb} \tbE^{lc} F{(\Omega)}^i_{bc} + 2 \tbE^{jb} N^c F{(\Omega)}^i_{bc} = 0.
\end{align}

Using the inverse of the anti-self-dual triad, $\bar{\utE}^j_a$, we convert the index $j$ to $a$,

\begin{align}
    \frac{1}{2} \bar{\utE}^j_a \tbE^{j\mu\nu} F{(\Omega)}^i_{\mu\nu} = F{(\Omega)}^i_{0a} - \frac{\im}{2} \dlap \bar{\utE}^j_a \epsilon^{jkl}\tbE^{kb} \tbE^{lc} F{(\Omega)}^i_{bc} + N^b F{(\Omega)}^i_{ab} = 0.
\end{align}

We then use the identity

\begin{align}
    \epsilon^{jkl} \tbE^{kb} \tbE^{lc} = \det(\tbE) \teps^{dbc} \bar{\utE}^j_d \label{eq:num-matrix-identity-on-bartE}
\end{align}

and solve for $F{(\Omega)}^i_{0a}$ to obtain 

\begin{align}
    F{(\Omega)}^i_{0a} = \frac{\im}{2} \dlap \det(\tbE) \bar{\utE}^j_a \bar{\utE}^j_d  \teps^{dbc} F{(\Omega)}^i_{bc} + N^b F{(\Omega)}^i_{ba}.
\end{align}

We see that the spatial metric has appeared, $\gamma_{ab} = -\det(\tbE) \bar{\utE}^j_a \bar{\utE}^j_d$, and it can be replaced with its definition in terms of the self-dual triads, $\gamma_{ab} = \det(\tE) \utE^j_a \utE^j_b$ (note the minus sign appearing due to the bar operation being minus the conjugate). Equivalently, this replacement is allowed as a consequence of the reality conditions. The evolution equation can then be written in terms of purely self-dual objects:

\begin{align}
    F{(\Omega)}^i_{0a} = -\frac{\im}{2} \dlap \det(\tE) \utE^j_a \utE^j_d  \teps^{dbc} F{(\Omega)}^i_{bc} + N^b F{(\Omega)}^i_{ba}.
\end{align}

Using the analogue of \cref{eq:num-matrix-identity-on-bartE} for $\tE$ and expanding the first and last curvatures into derivatives of $\Omega^i$ gives

\begin{align}
    (\partial_t - N^b \partial_b) \Omega^i_a = \partial^\Omega_a \Omega^i_0 - N^b \partial^\Omega_a \Omega^i_b - \frac{\im}{2} \dlap \utE^j_a \epsilon^{jkl} \tE^{kb} \tE^{lc} F{(\Omega)}^i_{bc}.
\end{align}

Finally, using $H^{ia} = \dlap\tE^{ia}$ and $H^i_a = \utE^i_a/\dlap$, we have the equation of motion for $\Omega^i_a$ written using only objects in the 12-component sector,

\begin{align}
    (\partial_t - N^b \partial_b) \Omega^i_a = \partial^\Omega_a \Omega^i_0 - N^b \partial^\Omega_a \Omega^i_b - \frac{\im}{2} H^j_a \epsilon^{jkl} H^{kb} H^{lc} F{(\Omega)}^i_{bc}.
\end{align}

The second equation on the third line in \cref{eq:nonlin-conformal-comp-evo-sys} can be densitised using the inverse metric and then decomposed into 3+1 form as follows:

\begin{align}
    \sqrt{\det(H)} g_H^{\mu\nu} \partial_\mu \Omega^i_\nu = -\partial_t \Omega^i_0 + N^a \left( \partial_t \Omega^i_a + \partial_a \Omega^i_0 \right) + \left(H^{ja} H^{jb} - N^a N^b\right) \partial_a \Omega^i_b = 0.
\end{align}

Substituting the equation of motion for $\Omega^i_a$ gives the evolution equation for $\Omega^i_0$,

\begin{align}
    (\partial_t - N^a \partial_a) \Omega^i_0 & = N^a \partial_a \Omega^i_0 + ( H^{ja} H^{jb} - N^a N^b ) \partial_a \Omega^i_b \nonumber \\ & - \frac{\im}{2} N^a H^j_a \epsilon^{jkl} H^{kb} H^{lc} F{(\Omega)}^i_{bc} + \epsilon^{ijk} N^a \Omega^j_a \Omega^k_0.
\end{align}

Next, we tackle the fourth line in \cref{eq:nonlin-conformal-comp-evo-sys}. In this case it is easier to decompose the second equation first; multiplying by $\sqrt{\det(H)}$ and decomposing the frame objects gives

\begin{align}
    \im H^{i a} (\partial_t \omega_a - \partial_a \omega_0) + (-\epsilon^{ijk} H^{ja} H^{kb} - \im N^a H^{ib} + \im N^b H^{ia}) \partial_a \omega_b \nonumber \\  - (-\omega_0 \Omega^i_0 + N^a \omega_a \Omega^i_0 + N^a \omega_0 \Omega^i_a + H^{ja} H^{jb} \omega_a \Omega^i_b - N^a N^b \omega_a \Omega^i_b) = 0.
\end{align}

We then perform two manipulations: first, we multiply by $H^i_{a'}$ and relabel $a' \rightarrow a$; second, we use $\omega_N = \omega_0 - N^a \omega_a$ and $\Omega^i_N = \Omega^i_0 - N^a \Omega^i_a$. What remains is an evolution equation for $\omega_a$,

\begin{align}
    (\partial_t - N^b \partial_b) \omega_a = \partial_a \omega_0 + N^b \partial_a \omega_b - \im \epsilon^{ijk} H^i_a H^{jb} H^{kc} \partial_b \omega_c + \im H^i_a \omega_N \Omega^i_N - \im H^i_a H^{jb} H^{jc} \omega_b \Omega^i_c.
\end{align}

The final equation, the first on the same line, decomposes as

\begin{align}
    -\partial_t \omega_0 + N^a \partial_a \omega_0 + N^a( \partial_t - N^b \partial_b) \omega_a + H^{ia} H^{ib} \partial_a \omega_b + H^{ja} H^{jb} \left( \frac{1}{4} \omega_a \omega_b - \Omega^i_a \Omega^i_b \right) - \frac{1}{4} \omega_N \omega_N + \Omega^i_N \Omega^i_N = 0.
\end{align}

Substituting the evolution equation for $\omega_a$ and rearranging gives the final evolution equation for $\omega_0$,

\begin{align}
    (\partial_t - N^a \partial_a )\omega_0 =\ & N^a \partial_a \omega_0 + N^a N^b \partial_a \omega_b - \im \epsilon^{ijk} N^a H^i_a H^{jb} H^{kc} \partial_b \omega_c + H^{ja} H^{jb} \partial_a \omega_b \nonumber \\ & \ + H^{ja} H^{jb} \left( \frac{1}{4} \omega_a \omega_b - \Omega^i_a \Omega^i_b \right) + \im N^a H^i_a \omega_N \Omega^i_N - \im N^a H^i_a H^{jb} H^{jc} \omega_b \Omega^i_c - \frac{1}{4} \omega_N \omega_N + \Omega^i_N \Omega^i_N.
\end{align}

\section{Eigenvalue Analysis}\label{apx:eigenvalue-analysis}

Here we briefly cover the analysis of the complex eigenvalues and deduce the parameter space for which the eigenvalues have non-positive real parts. Taking the eigenvalues directly from the main text and reparametrising with $z = \frac{\alpha}{2}$ and $\tilde\rho = \frac{1}{2}\left(1+\frac{c_1}{c_2}\right)$, we find the eigenvalues

\begin{align}
    \lambda_\pm = z \pm \sqrt{z^2-k^2}, \quad \lambda_{\pm\pm} = z \pm \sqrt{z^2 \pm 2 \im \tilde\rho \sqrt{k^2} z - k^2}.
\end{align}

Note that $\tilde\rho$ is related to the parameter $\hat\rho = c_1/c_2$ of the main text by $\tilde\rho = (1+\hat\rho)/2$. We require $\re(\lambda_\pm) \leq 0$ and $\re(\lambda_{\pm\pm}) \leq 0$. These conditions imply

\begin{align}
    \re(\lambda_+) + \re(\lambda_-) \leq 0, \quad \Rightarrow \quad \re(z) \leq 0.
\end{align}

Here $\sqrt{x}$ denotes the principal square root of $x$, for which

\begin{align}
    \re(\sqrt{x+\im y}) = \sqrt{\frac{x + \sqrt{x^2 + y^2}}{2}}.
\end{align}

Decomposing $z = x + \im y$, we can write

\begin{align}
    \re(\sqrt{z^2-k^2}) = \sqrt{\frac{x^2-y^2-k^2 + \sqrt{(x^2-y^2-k^2)^2 + 4 x^2 y^2}}{2}}.
\end{align}

Therefore,

\begin{align}
    x + \sqrt{\frac{x^2-y^2-k^2 + \sqrt{(x^2-y^2-k^2)^2 + 4 x^2 y^2}}{2}} \leq 0, \\
    x - \sqrt{\frac{x^2-y^2-k^2 + \sqrt{(x^2-y^2-k^2)^2 + 4 x^2 y^2}}{2}} \leq 0.
\end{align}

We replace the first appearance of $x$ in each line with $x = - |x|$, since we know that $x$ must be non-positive. Rearranging these two inequalities gives

\begin{align}
 + \sqrt{\frac{x^2-y^2-k^2 + \sqrt{(x^2-y^2-k^2)^2 + 4 x^2 y^2}}{2}} \leq |x|, \\
 - \sqrt{\frac{x^2-y^2-k^2 + \sqrt{(x^2-y^2-k^2)^2 + 4 x^2 y^2}}{2}} \leq |x|.   
\end{align}

Since the principal square root always has non-negative real part, the second condition is automatically satisfied. For the first inequality, we can use the fact that $ a \leq b \Leftrightarrow a^2 \leq b^2$ when $b$ is a positive real number and the sign of $a$ is known. This applies here because $a$, being the principal square root of a non-negative real number, is itself non-negative. Using this fact, we obtain

\begin{align}
    \frac{x^2-y^2-k^2 + \sqrt{(x^2-y^2-k^2)^2 + 4 x^2 y^2}}{2} \leq x^2.
\end{align}

Rearranging further gives

\begin{align}
    \sqrt{(x^2-y^2-k^2)^2 + 4 x^2 y^2} \leq x^2 + y^2 + k^2.
\end{align}

Again, since both sides are positive, we can square to find that

\begin{align}
    0 \leq 4 k^2 x^2,
\end{align}

which is always true. Therefore, we have shown that $\re(z) \leq 0 \Rightarrow \re(\lambda_\pm) \leq 0$.

For the remaining eigenvalues, we must make some assumptions about $\tilde\rho$ to make progress. In particular, we analyse the case $\tilde\rho \in \R$. First, we make the replacement

\begin{align}
    u_\pm = z \pm \im \tilde\rho \sqrt{k^2}
\end{align}

and define a new related eigenvalue,

\begin{align}
    \hat{\lambda}_{\pm\pm} = u_\pm \pm \sqrt{u_\pm^2 + (\tilde\rho^2 - 1)k^2}.
\end{align}

These are constructed such that $\re(\lambda_{\pm\pm}) = \re(\hat{\lambda}_{\pm\pm})$. Repeating the previous analysis but replacing $k^2$ with $(1-\tilde\rho^2)k^2$, we find that the requirement becomes

\begin{align}
    \sqrt{(x^2-y^2-(1-\tilde\rho^2) k^2)^2 + 4 x^2 y^2} \leq x^2 + y^2 + (1-\tilde\rho^2)k^2.
\end{align}

There are two cases. If $1-\tilde\rho^2 < 0$, the right-hand side can be arbitrarily negative (since $k^2$ can be arbitrarily large); the inequality is then violated, and the real parts of some eigenvalues have the wrong sign. We are therefore left with the case $1-\tilde\rho^2 \geq 0$. Both sides of the inequality are then positive, and squaring gives

\begin{align}
    0 \leq 4 (1-\tilde\rho^2) k^2 x^2.
\end{align}

This is always true, since $1-\tilde\rho^2$ is non-negative. To see what the condition on $\tilde\rho$ implies for $c_1, c_2$, we use the fact that

\begin{align}
    1-\tilde\rho^2 \geq 0, \quad \Rightarrow \quad 1 \geq \tilde\rho^2, \quad \Rightarrow -1 \leq \tilde\rho \leq 1.
\end{align}

Substituting the definition of $\tilde\rho$ in terms of $c_1, c_2$, we find

\begin{align}
    -1 \leq \frac{1}{2} \left( 1 + \frac{c_1}{c_2} \right) \leq 1, \quad \Rightarrow \quad -3 \leq \frac{c_1}{c_2} \leq 1.
\end{align}

Similarly, for $z$ we can write the non-positive real part condition using $\alpha$ instead. The final condition becomes

\begin{align}
    \re(\alpha) \leq 0, \quad -3 \leq \frac{c_1}{c_2} \leq 1.
\end{align}

\section{Numerical Plots}\label{apx:numerical-plots}

    This appendix contains plots referenced in the main text. The set-up and discussion for each plot are given in \cref{sec:numerical-results}. Each plot shows the same four constraints versus simulation time. The top-left panel shows the reality conditions, i.e.\ the $L^2$ norm over all the imaginary components of the metric constructed from the frame. The top-right panel shows the $L^2$ norm of the Gauss constraints, and the bottom-left panel the same for the momentum constraints. Finally, the bottom-right panel shows the $L^2$ norm of the Hamiltonian constraint.

    \begin{figure}[H]
        \centering
        \includegraphics[width=0.6\textwidth]{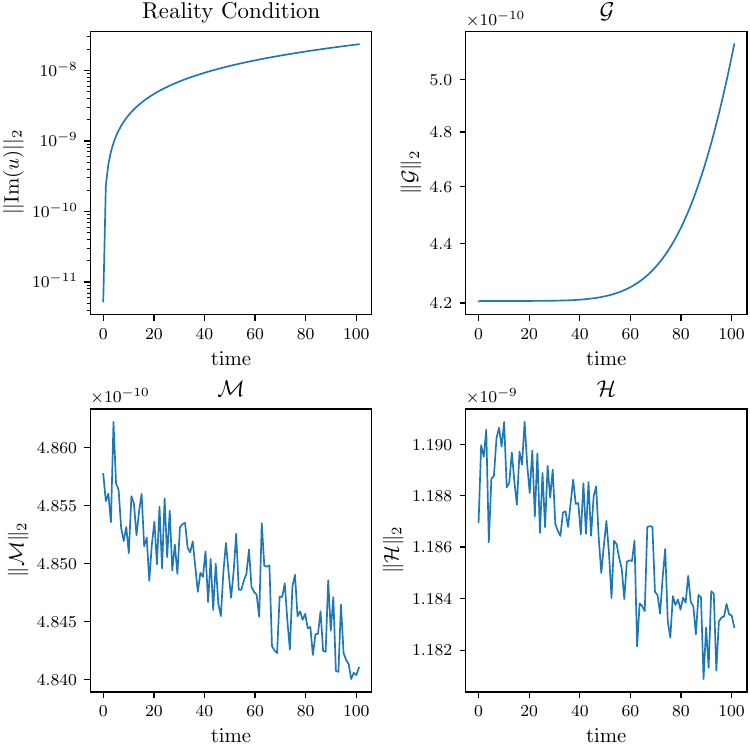}
        \caption{Plot of the $L^2$ norm of the constraints versus time for the original evolution system up to $T=100$.}\label{fig:org-normal-plot}
    \end{figure}

    \begin{figure}[H]
        \centering
        \includegraphics[width=0.6\textwidth]{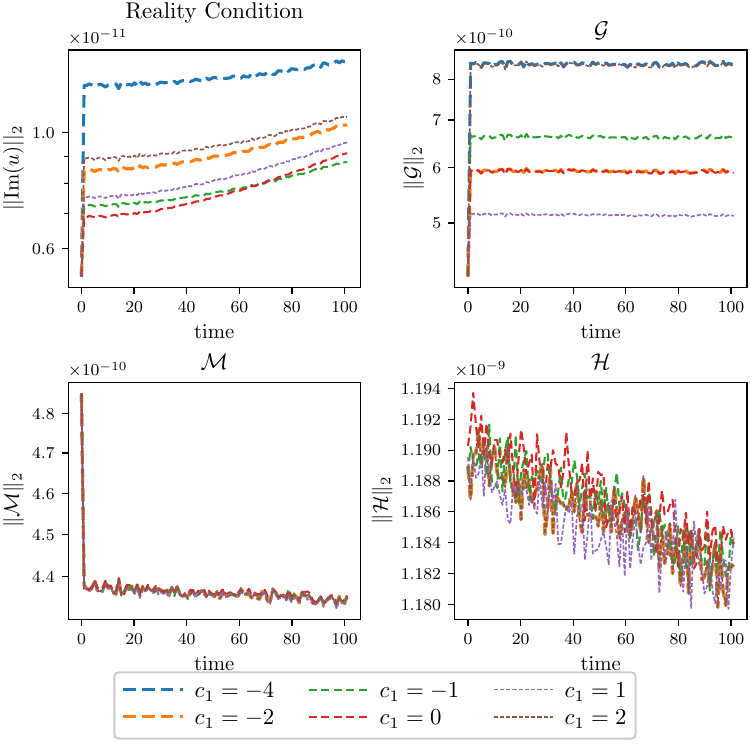}
        \caption{Plot of the $L^2$ norm of the constraints for the hyperbolic evolution system with parameters $c_1 = -4,-2,-1,0,1,2$ and $(c_2,\alpha,\beta) = (1,0,0)$. The simulations were run up to $T = 100$.}\label{fig:hyp-c12}
    \end{figure}

    \begin{figure}[H]
        \centering
        \includegraphics[width=0.6\textwidth]{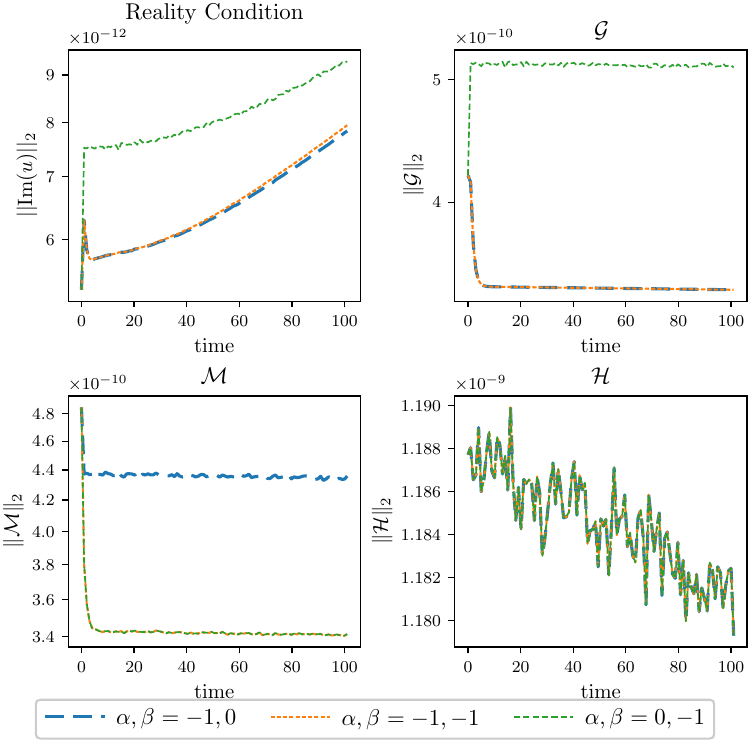}
        \caption{Plot of the $L^2$ norm of the constraints for the stable cases of the $\alpha,\beta$ parameter search for the hyperbolic evolution system. Only stable parameter choices are shown, namely $(\alpha,\beta) = (-1,0),(-1,-1),(0,-1)$. The remaining parameters are fixed at $c_1,c_2 = 1,1$, and the simulations were run up to $T = 100$.}\label{fig:hyp-ab-con}
    \end{figure}

    \begin{figure}[H]
        \centering
        \includegraphics[width=0.6\textwidth]{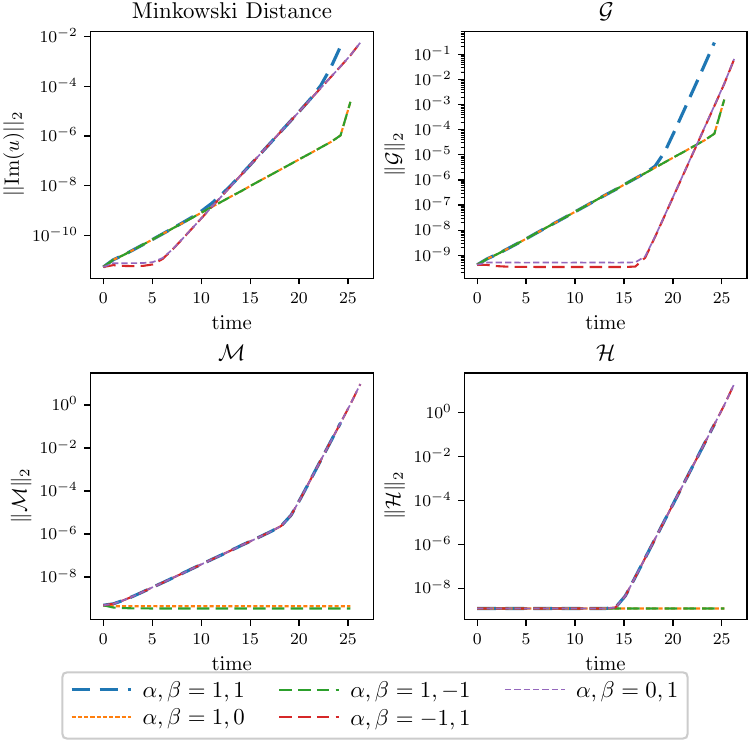}
        \caption{Plot of the $L^2$ norm of the constraints for the unstable cases of the $\alpha,\beta$ parameter search for the hyperbolic evolution system. The unstable parameters chosen were $(\alpha,\beta) = (1,1),(1,0),(1,-1),(-1,1),(0,1)$, and the remaining parameters were $c_1,c_2 = 1,1$. These simulations crashed at around $T = 25$ and did not reach the allotted run time of $T=100$.}\label{fig:hyp-ab-div}
    \end{figure}

    \begin{figure}[H]
        \centering
        \includegraphics[width=0.6\textwidth]{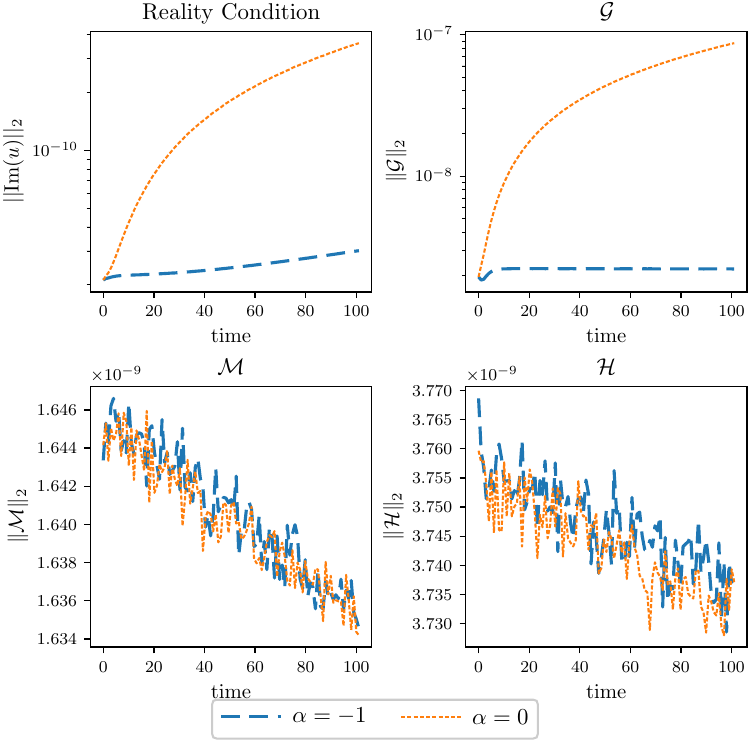}
         \caption{Plot of the $L^2$ norm of the constraints versus simulation time for the conformal evolution system with stable choices for the damping parameter. The coefficients in this case are $\alpha=-1,0$ and $c = 1$. Simulations were performed up to $T=100$.}\label{fig:conf-a-con}
        \end{figure}    

    \begin{figure}[H]
        \centering
        \includegraphics[width=0.6\textwidth]{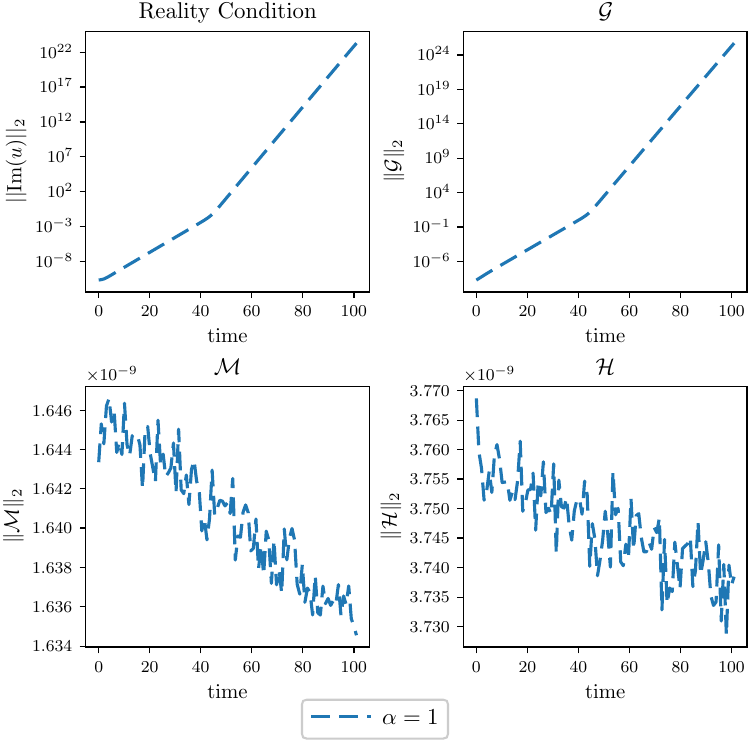}
         \caption{Plot of the $L^2$ norm of the constraints versus simulation time for the conformal evolution system with unstable choices for the damping parameter. The coefficients in this case are $\alpha=1$ and $c = 1$. The simulation was performed up to $T=100$ and did not crash.}\label{fig:conf-a-div}
    \end{figure}
    
    \begin{figure}[H]
        \centering
        \includegraphics[width=0.6\textwidth]{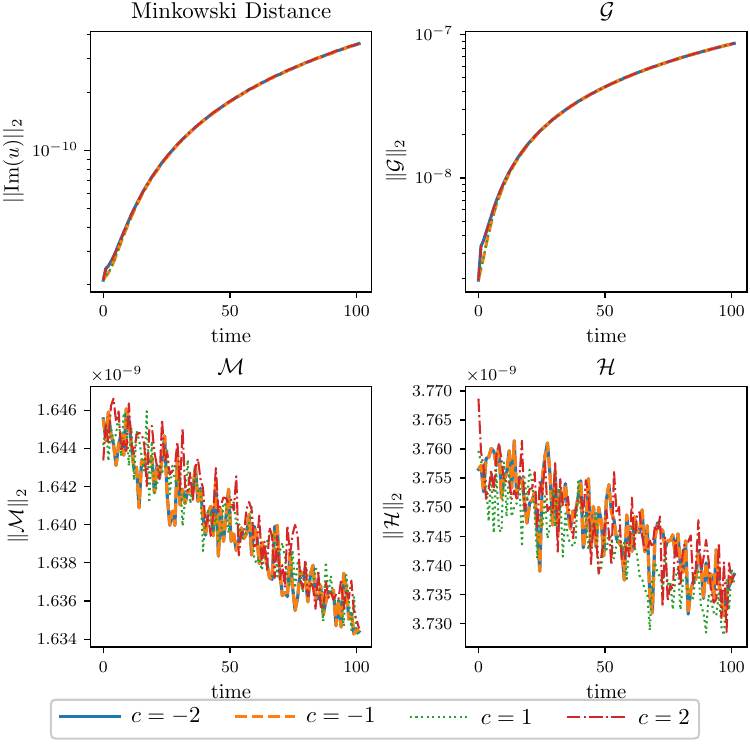}
        \caption{Plot of the $L^2$ norm of the constraints versus simulation time for the $c$ parameter search for the conformal evolution system. The values chosen were $c=-2,-1,1,2$, and simulations were performed up to $T=100$.}\label{fig:conf-c-plot}
    \end{figure}

\printbibliography

\end{document}